\documentclass[fleqn,usenatbib]{mnras}

\usepackage{newtxtext,newtxmath}

\usepackage[T1]{fontenc}

\DeclareRobustCommand{\VAN}[3]{#2}
\let\VANthebibliography\thebibliography
\def\thebibliography{\DeclareRobustCommand{\VAN}[3]{##3}\VANthebibliography}

\usepackage{graphicx}	
\usepackage{amsmath}	
\usepackage{amsfonts} 
\usepackage{algorithm}
\usepackage{algpseudocode}
\usepackage{array}
\usepackage{textcomp}
\usepackage{stfloats}
\usepackage{url}
\usepackage{verbatim}
\usepackage{graphicx}
\usepackage{caption}
\usepackage{subcaption}
\usepackage{subcaption}
\usepackage[dvipsnames]{xcolor}
\usepackage{adjustbox}
\usepackage{hyperref}
\usepackage{booktabs}

\definecolor{linkcolor}{RGB}{92,92,192}
\definecolor{blackcolor}{RGB}{0,0,0}
\hypersetup{colorlinks=true,linkcolor=blackcolor,citecolor=linkcolor,urlcolor=linkcolor}

\usepackage{tikz}
\usepackage{pgfplots}

\usepackage{algpseudocode}
\usepackage{algorithmicx}

\usepackage{bm}

\algblock[Block]{Block}{EndBlock}
\algblockdefx[Block]{Block}{EndBlock}%
[1]{{#1}}%
[1]{{#1}}
\newcommand{\Given}[1]{\State{\bf given} {#1}}

\newcommand{\RepeatFor}[1]{\Repeat {\bf~for} {#1}}

\newcommand{\RR}{\ensuremath{\mathbb{R}}}

\newcommand{\CC}{\ensuremath{\mathbb{C}}}

\title[AIRI: variations and robustness]{The AIRI plug-and-play algorithm for image reconstruction in radio-interferometry: variations and robustness}

\author[M. Terris et al.]{
Matthieu Terris,$^{*1}$
Chao Tang,$^{*1,2}$
Adrian Jackson,$^{2}$
and Yves Wiaux$^{1}$\thanks{E-mail: y.wiaux@hw.ac.uk. $^*$ Equal contribution.}
\\
$^{1}$Institute of Sensors, Signal, and Systems, Heriot-Watt University, Currie, Edinburgh EH14 4AS, UK.\\
$^{2}$EPCC, University of Edinburgh, Potterrow, Edinburgh EH8 9BT, UK
}

\date{Accepted XXX. Received YYY; in original form ZZZ}

\pubyear{2024}

\begin{document}
\label{firstpage}
\pagerange{\pageref{firstpage}--\pageref{lastpage}}
\maketitle

\begin{abstract}
Plug-and-Play (PnP) algorithms are appealing alternatives to proximal algorithms when solving inverse imaging problems. By learning a Deep Neural Network (DNN) denoiser behaving as a proximal operator, one waives the computational complexity of optimisation algorithms induced by sophisticated image priors, and the sub-optimality of handcrafted priors compared to DNNs. 
Such features are highly desirable in radio-interferometric (RI) imaging, where precision and scalability of the image reconstruction process are key. 
In previous work, we introduced AIRI, PnP counterpart to the  unconstrained variant of the SARA optimisation algorithm, 
relying on a forward-backward algorithmic backbone. Here, we introduce variations of AIRI towards a more general and robust PnP paradigm in RI imaging.  Firstly, we show that the AIRI denoisers can be used without any alteration to instantiate a PnP counterpart to the constrained SARA optimisation algorithm itself, 
relying on a  primal-dual forward-backward algorithmic backbone, thus extending the remit of the AIRI paradigm. Secondly, we show that AIRI algorithms are robust to strong variations in the nature of the training dataset, with denoisers trained on medical images yielding similar reconstruction quality to those trained on astronomical images. Thirdly, we develop a functionality to quantify the model uncertainty introduced by the randomness in the training process. We validate the image reconstruction and uncertainty quantification functionality of AIRI algorithms against the SARA family and CLEAN, both in simulation and on real data of the ESO 137-006 galaxy acquired with the MeerKAT telescope. AIRI code is available in the \href{https://basp-group.github.io/BASPLib/}{BASPLib} code library on GitHub.

\end{abstract}


\section{Introduction}

Synthesis imaging by radio interferometry (RI) in astronomy leverages antenna arrays to observe the sky with high resolution and sensitivity \citep{thompson2017interferometry}. 
A new generation of RI telescopes is currently emerging, whose science goals range from studying cosmic magnetism, dark matter, and dark energy, to understanding the structure and evolution of stars and galaxies. Targeting ever increasing resolution and dynamic range to resolve highly complex structure distributions, new arrays feature ever larger antenna counts. They are consequently characterised by unprecedented volumes of observed data, and image sizes. The flagship Square Kilometre Array (\href{https://www.skao.int/}{SKA}) will soon come online \citep{scaife2020big}, the South African \href{https://www.sarao.ac.za/science/meerkat/}{MeerKAT} array and the Australian Pathfinder \href{https://www.atnf.csiro.au/projects/askap/index.html}{ASKAP} two major precursors.

In this new era, the Fourier inverse problem for image formation from visibility data is extremely challenging, with RI imaging algorithms required to deliver a new regime of joint precision and scalability. The CLEAN algorithm holds a near-total monopoly in the field \citep{hogbom1974aperture, thompson2017interferometry, offringa2014wsclean}, owing its success to its simplicity and associated scalability. CLEAN is a highly specialised precursor to the ``Matching Pursuit'' algorithm \citep{mallat1993matching}, iteratively identifying image components from data residuals back-projected in the pixel domain \citep{hogbom1974aperture,schwab1984relaxing}, or some other multi-scale representation \citep{cornwell2008multiscale}. It can loosely be categorised among the variational inference techniques \citep{marsh1987objective}. Yet it requires manual fine-tuning, limits resolution and dynamic range by design to compensate for a simplistic regularisation approach, lacks versatility to handle complex emission and complex calibration effects, and a proper uncertainty quantification functionality.

Building from early work framed in the context of the theory of compressed sensing \citep{wiaux2009compressed}, advanced optimisation, sampling, and deep learning approaches have been developed \citep{kazemi2011radio, garsden2015lofar, dabbech2015moresane, junklewitz2016resolve, repetti2017non, repetti2019scalable, cai2018uncertainty, arras2021comparison, gheller2022convolutional, connor2022deep, dabbech2022first, terris2023image, wilber2023scalableI, wilber2023scalableII, roth2023bayesian, aghabiglou2023ultra, aghabiglou2024r2d2, liaudat2023scalable, dia2023bayesian}. 
Few of those techniques have made their way through from extensive methodological study to validation on challenging real data applications. Firstly, the variational Bayes approach underpinning the RESOLVE algorithm has shown potential to deliver superior precision and robustness to CLEAN \citep{arras2021comparison, roth2023bayesian}, but at the expense of a high computational cost affecting efficiency and scalability. Secondly, modern variational inference techniques using sparsity-based regularisation have been shown to deliver remarkable precision, propelled by advanced proximal optimisation algorithms. These range from the original SARA, which relies on a  primal-dual forward-backward (PDFB) algorithmic backbone to solve an optimisation problem with a constrained data-fidelity term \citep{carrillo2012sparsity, onose2016scalable, thouvenin2023parallel}, 
to uSARA, which relies on a forward-backward (FB) algorithmic backbone to solve an optimisation problem with an unconstrained data-fidelity term
\citep{repetti2020forward, dabbech2022first, wilber2023scalableI}. Thirdly, a ``plug-and-play'' version of uSARA was introduced, dubbed AIRI, which consists in substituting the sparsity-promoting regularisation operator with a learned Deep Neural Networks (DNNs) denoiser. AIRI was demonstrated to further enhance imaging precision over uSARA. 
It also brings a significant acceleration owing to the fast inference of its DNN denoiser \citep{terris2023image, wilber2023scalableII}

In this work,  we introduce variations of AIRI towards a more general and robust PnP paradigm for RI imaging, showing that: (i) AIRI denoisers can be plugged without any alteration in other algorithmic backbones than FB, in particular in PDFB to instantiate a PnP counterpart to SARA itself; (ii) AIRI algorithms are robust to strong variations in the nature of the training dataset; (iii) AIRI algorithms can be endowed with a functionality to quantify the uncertainty inherent to its learned regularisation model.

The remainder of this paper is organised as follows. In Section~\ref{section:proposed_methodology}, we detail the proposed approach unifying constrained and unconstrained approaches for PnP and optimization approaches. In Section~\ref{section:synthetic_resulst}, we validate the image reconstruction of AIRI algorithms against state-of-the-art algorithms on simulated measurements. In Section~\ref{section:real_data}, we showcase the performance of the proposed algorithm on real measurements of the ESO 137-006 galaxy acquired with the MeerKAT telescope \citep{jonas2016meerkat}.

\section{Proposed methodology} \label{section:proposed_methodology}

In this section, we recall the RI inverse problem and introduce our constrained PnP algorithm, dubbed cAIRI. We next detail our methodology for  
curating image datasets
with high dynamic range.

\subsection{The RI inverse problem}

\begin{figure}
\centering
\adjustbox{valign=t}{\begin{subfigure}[b]{0.14\textwidth}
\includegraphics[width=\textwidth]{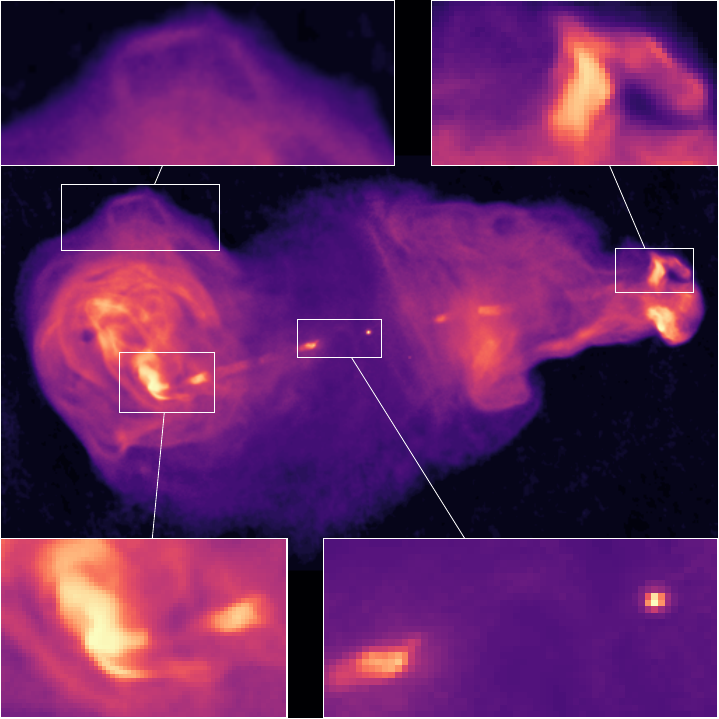}
\vspace{-0.92em}
\caption{}
\end{subfigure}}
\hfill
\adjustbox{valign=t}{\begin{subfigure}[b]{0.16\textwidth}
\includegraphics[width=\textwidth]{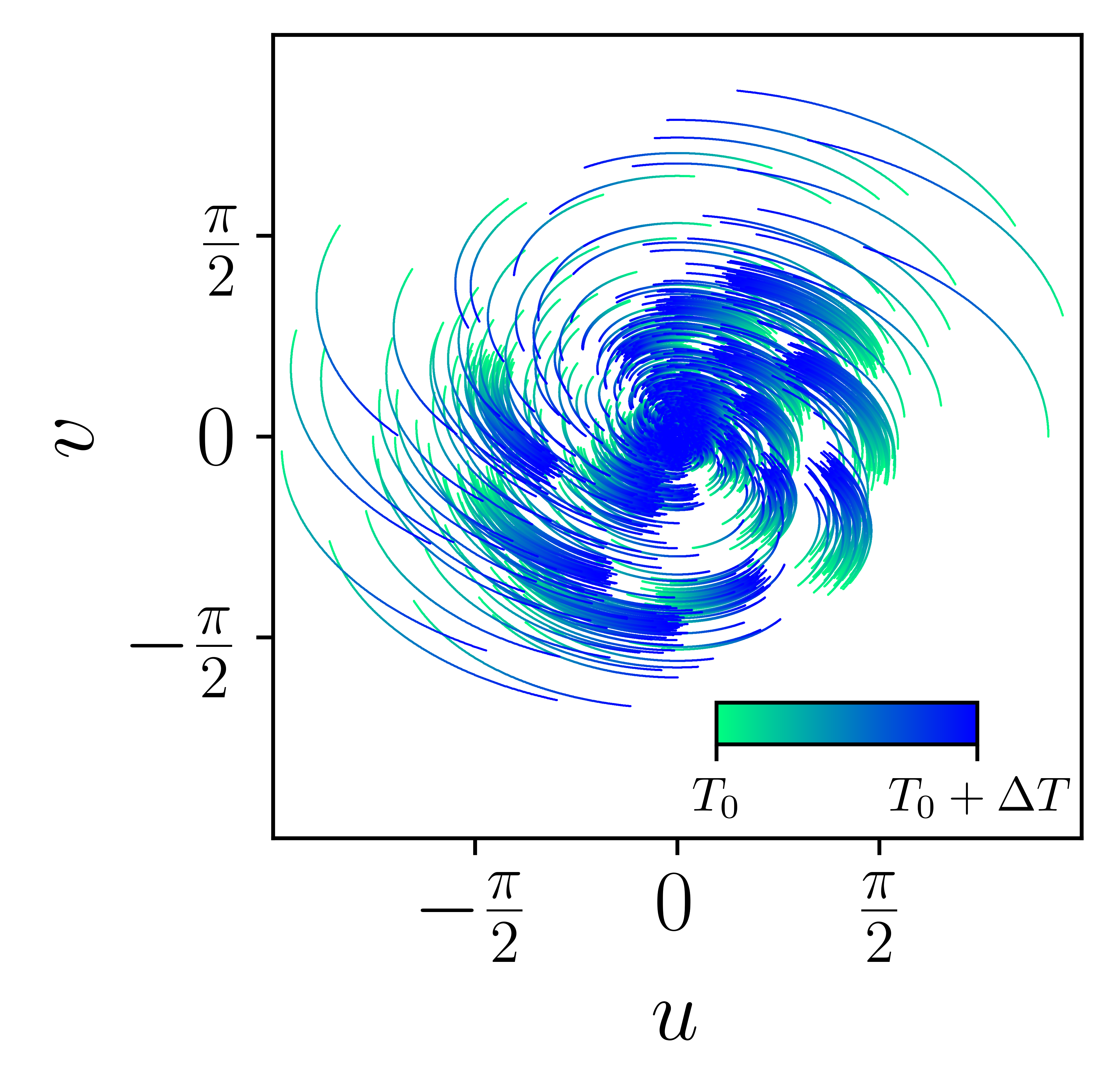}
\vspace{-2em}
\caption{}
\end{subfigure}}
\hfill
\adjustbox{valign=t}{\begin{subfigure}[b]{0.14\textwidth}
\includegraphics[width=\textwidth]{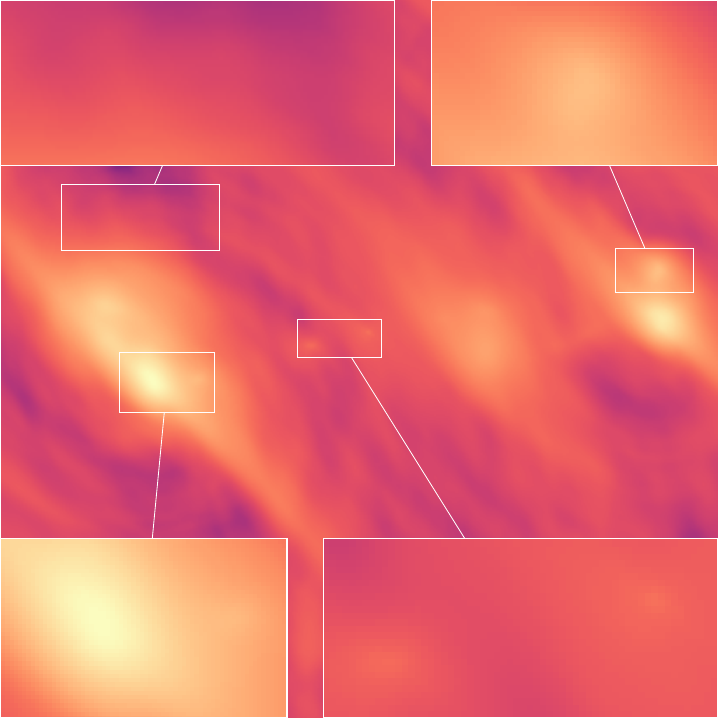}
\vspace{-0.92em}
\caption{}
\end{subfigure}}
\caption{Illustration of a simulated RI imaging problem. (a) shows a groundtruth radio galaxy, displayed in logarithmic scale. (b) shows the sampled points in the spatial Fourier domain. At each time $T_0\leq t \leq T_0+\Delta T$, each pair of antennas samples a single point, during a total acquisition time $\Delta T$. As the acquisition time increases, the locus of the points moves in the Fourier domain due to the Earth's rotation. The colorbar illustrates this effect. (c) shows the back-projected measurements in the image domain.}
\label{fig:illustration}
\end{figure}

Focusing on monochromatic intensity imaging on a small field of view (FoV), and in the absence of atmospheric and instrumental perturbations, each pair of antennas of an RI array acquires a noisy Fourier component of the intensity image to be formed, called a visibility. The associated Fourier mode (also called $uv$-point) is given by the projection of the corresponding baseline, expressed in units of the observation wavelength, onto the plane perpendicular to the line of sight \citep{thompson2017interferometry}. Gathering measurements from all antenna pairs throughout the duration of acquisition, an RI array thus samples an incomplete coverage of the spatial Fourier domain of the image of interest. A discrete formulation of the 
underpinning 
inverse problem, aiming to restore the target image $\overline{\bm{x}}\in \RR^{n}$ from the measured complex visibilities  $\bm{y} \in \CC^{m}$, reads:
\begin{equation}
    \bm{y} = \bm{\Phi} \overline{\bm{x}}+\bm{e},
\label{eq:inv_pb_gen}
\end{equation}
where $\bm{\Phi} = \bm{\mathrm{GFZ}}\in \CC^{m \times n}$ is the measurement operator, $\bm{\mathrm{G}} \in \CC^{m \times d}$ is a sparse interpolation matrix, encoding the non-uniform Fourier transform, $\bm{\mathrm{F}} \in \CC^{d\times d}$ is the 2D Discrete Fourier Transform, $\bm{\mathrm{Z}} \in \RR^{d\times n}$ is a zero-padding operator, incorporating the correction for the convolution performed through the operator $\bm{\mathrm{G}}$, and $\bm{e}\in \CC^{m}$ is a realization of some i.i.d. Gaussian random noise, with mean $0$ and standard deviation $\eta>0$. Backprojecting problem \eqref{eq:inv_pb_gen} in the image domain gives the alternative formulation
\begin{equation}
    \operatorname{Re}\{\bm{\Phi}^\dagger \bm{y}\}= \operatorname{Re}\{\bm{\Phi}^\dagger\bm{\Phi}\} \overline{\bm{x}}+\operatorname{Re}\{\bm{\Phi}^\dagger \bm{e}\},
\label{eq:invpbim}
\end{equation}
where $(\cdot)^\dagger$ denotes the complex conjugate transpose and $\operatorname{Re}\{\cdot\}$ the real part. $\operatorname{Re}\{\bm{\Phi}^\dagger \bm{y}\}$ is known as the dirty image, $\operatorname{Re}\{\bm{\Phi}^\dagger\bm{\Phi}\}$ is the operator representing the convolution of $\overline{\bm{x}}$ by the point spread function, also known as the dirty beam, and $\operatorname{Re}\{\bm{\Phi}^\dagger \bm{e}\}$ is the noise backprojected in the image domain. An illustration of these concepts are provided in Fig.~\ref{fig:illustration}.

\subsection{Unconstrained and constrained approaches}

A common approach for solving the ill-posed inverse problem \eqref{eq:inv_pb_gen} consists in reformulating it as a minimization problem yielding an estimate $\widehat{\bm{x}}$ of $\overline{\bm{x}}$ as
\begin{equation}
    \widehat{\bm{x}} = \underset{\bm{x}}{\arg\min} f(\bm{x})+\lambda r(\bm{x}),
\label{eq:minimization}
\end{equation}
\noindent where $\lambda>0$ is a regularisation parameter. In turn, appropriate choices for $f$, $r$ and $\lambda$ ensure the quality of the reconstructed image $\widehat{\bm{x}}$.
Given the Gaussian nature of the noise, there are primarily two possible strategies for the choice of $f$. 
\begin{algorithm}[t]
\caption{AIRI algorithm}
\begin{algorithmic}[1]
\small
\Given{$0<\gamma<2/L$, denoiser $\operatorname{D}$, $\bm{x}_0\in\RR^n$, $\xi>0$}
\RepeatFor{$k=0,1,\ldots$}
\State {$
\bm{x}_{k+1} = \operatorname{D}(\bm{x}_k-\gamma \nabla f(\bm{x}_k))
$}
\Until{{{$\|x_{k+1}-x_{k}\|/\|x_{k+1}\|<\xi$}}}
\State \Return {$x_{k+1}$}
\end{algorithmic}
\label{algo:pnp_fb}
\end{algorithm}
A first possible choice consists in the least squared error, \emph{i.e.}
\begin{equation}
f(\bm{x}) = \frac{1}{2}  \|\bm{\mathrm{\Phi}} \bm{x} - \bm{y} \|^2,
\label{eq:f_smooth}
\end{equation}
and in this case, algorithms for solving \eqref{eq:minimization} typically take the form of Algorithm~\ref{algo:pnp_fb}, where $L = \| \bm{\Phi} \|_S^2$ is the spectral norm of the measurement operator, $\gamma$ is the step size, $(\cdot)^\dagger$ denotes the Hermitian adjoint and $\operatorname{D}=\operatorname{prox}_{\gamma \lambda r}$ is the proximity operator\footnote{The proximity operator of a convex function $r$ is defined as $\operatorname{prox}_{r}(\bm{y}) = \underset{\bm{x}}{\operatorname{argmin}}\, r(\bm{x})+\frac{1}{2}\|\bm{x}-\bm{y}\|^2$.} of $\gamma \lambda r$. In particular, the uSARA imaging algorithm
\citep{terris2023image, repetti2020forward} 
writes as Algorithm~\ref{algo:pnp_fb} for a 
handcrafted
sparsity-inducing prior $r$.  
A second strategy consists in adopting a constrained approach by minimizing $r(\bm{x})$ subject to $\|\bm{\mathrm{\Phi}}\bm{x}-y\|\leq \varepsilon$, which can be reformulated as \eqref{eq:minimization} with
\begin{equation}
    f(\bm{x}) = \iota_{\mathcal{B}(\bm{y}, \varepsilon)}( \mathrm{\bm{\Phi}} \bm{x}),
\label{eq:f_indicator}
\end{equation}
where $\iota_{\mathcal{B}(\bm{y}, \varepsilon)}(\cdot)$ denotes the indicator\footnote{The indicator $\iota_C$ of a closed convex set $C$ is the function defined as $\iota_C( \bm{x} )=0$ if $ \bm{x} \in C$ and $\iota_C( \bm{x} )=+\infty$ otherwise.} function of $\mathcal{B}(\bm{y}, \varepsilon)$, the $\ell_2$ ball of radius $\varepsilon$ and centred in $\bm{y}$. The value of $\varepsilon$ can be estimated with the bound on a $\chi^2$ distribution with $2m$ degrees of freedom, \emph{i.e.} $\varepsilon^2 = (2m + 4\sqrt{m})\eta^2$ \citep{carrillo2012sparsity}. This problem leads to different algorithms, a classical instance being given in Algorithm~\ref{algo:pnp_pd}, where $\operatorname{D}=\operatorname{prox}_{\gamma r}$.
\begin{algorithm}[t]
\caption{cAIRI algorithm}
\begin{algorithmic}[1]
\small
\Given{$0<\gamma<1/L$, denoiser $\operatorname{D}$, $\bm{x}_0\in\RR^n$, $\xi>0$}
\RepeatFor{$k=0,1,\ldots$}
\State{$\bm{x}_{k+1} = \operatorname{D}(\bm{x}_k-\gamma {\rm Re}\{\mathrm{\bm{\Phi}}^\dagger \bm{u}_k\})$}
\State{$\bm{v}_{k} = \bm{u}_k+\mathrm{\bm{\Phi}}(2\bm{x}_{k+1}-\bm{x}_k)$}
\State{$\bm{u}_{k+1}=\bm{v}_k- \operatorname{prox}_{f}(\bm{v}_k)$}
\Until{{{$\|\bm{x}_{k+1}-\bm{x}_{k}\|/\|\bm{x}_{k+1}\|<\xi$}}}
\State \Return {$\bm{x}_{k+1}$}
\end{algorithmic}
\label{algo:pnp_pd}
\end{algorithm}
In particular, the SARA algorithm \citep{onose2016scalable} writes as Algorithm~\ref{algo:pnp_pd} when $\operatorname{D}=\operatorname{prox}_{\gamma r}$ for a well-chosen sparsity-inducing prior function $r$. 

Although Algorithms~\ref{algo:pnp_fb} and \ref{algo:pnp_pd} exhibit distinct algorithmic backbones, these two algorithms can be related within the unifying minimization framework \eqref{eq:minimization} up to the choice of $f$ \citep{combettes2011proximal, pesquet2021learning}.
In \citet{terris2023image}, we proposed to adopt a PnP approach and replaced $\operatorname{D}$ in Algorithm~\ref{algo:pnp_fb} with a denoising DNN and dubbed the resulting algorithm AIRI. In this work, we propose to extend this approach to the constrained framework and introduce constrained AIRI (cAIRI) Algorithm~\ref{algo:pnp_pd}, the constrained variant of AIRI. 
Importantly, these two algorithms, as well as uSARA and SARA, fall in the same general framework propelled by the minimization framework \eqref{eq:minimization}. Furthermore, AIRI and cAIRI can be seen as generalizations of the state-of-the-art uSARA and SARA algorithms, but relying on denoising DNNs $\operatorname{D}$ with higher expressive power than sparsity inducing priors underpinning uSARA and SARA.

\subsection{Generating synthetic low-dynamic range datasets}
\label{subsection:general_exponentiation}

We define the dynamic range of an image as the ratio of pixel intensities between brightest and faintest features. For natural images, this ratio tends to be low. In the absence of groundtruth RI image datasets, \citet{terris2023image} proposed to generate a low-dynamic range dataset from Optical Astronomical Imaging (OAI) images to serve as a basis for training PnP denoisers (as well as end-to-end DNNs for benchmarking) to solve the inverse problem \eqref{eq:inv_pb_gen}.

In order to investigate the influence of the nature of the images used in the training of the denoiser, we also propose to generate a low-dynamic range dataset from magnetic resonance imaging (MRI) images. The second dataset we examine contains MRI images derived from the fastMRI single-coil knee images \citep{zbontar2018fastmri}. To break symmetries and enlarge the image size, random patches from each MRI slices are extracted, rotated, translated, and concatenated. An edge tapering filter is then applied to eliminate image discontinuities, resulting in new MRI images with varied content.

Both the OAI and MRI-born images were preprocessed with a blind image restoration network (SCUNet, \citealt{zhang2022practical}) to remove noise and artefacts. Pixel-wise soft-thresholding is then applied to remove residual background.
This processing strategy yields 5000 low-dynamic range 
OAI-born and MRI-born groundtruth images of size 512$\times$512 respectively. 

\begin{figure}%
\centering%
\begin{tabular}{@{\hspace{0.\tabcolsep}} c @{\hspace{-0.3\tabcolsep}} c @{\hspace{-0.5\tabcolsep}} c @{\hspace{-0.3\tabcolsep}} c}
& OAID & MRID & \\
\rotatebox[origin=c]{90}{Raw} %
& 
\raisebox{-0.5\height}{
\includegraphics[width=0.11\textwidth]{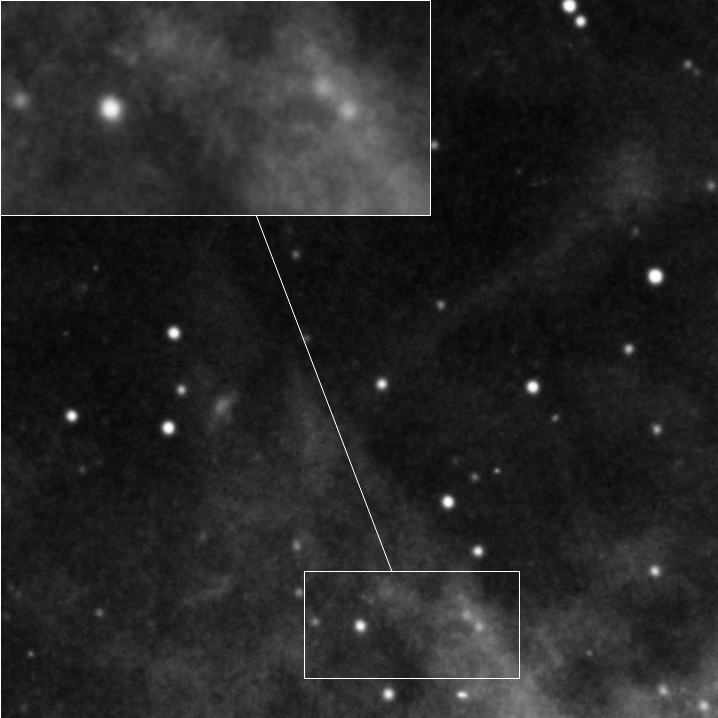}\!
\includegraphics[width=0.11\textwidth]{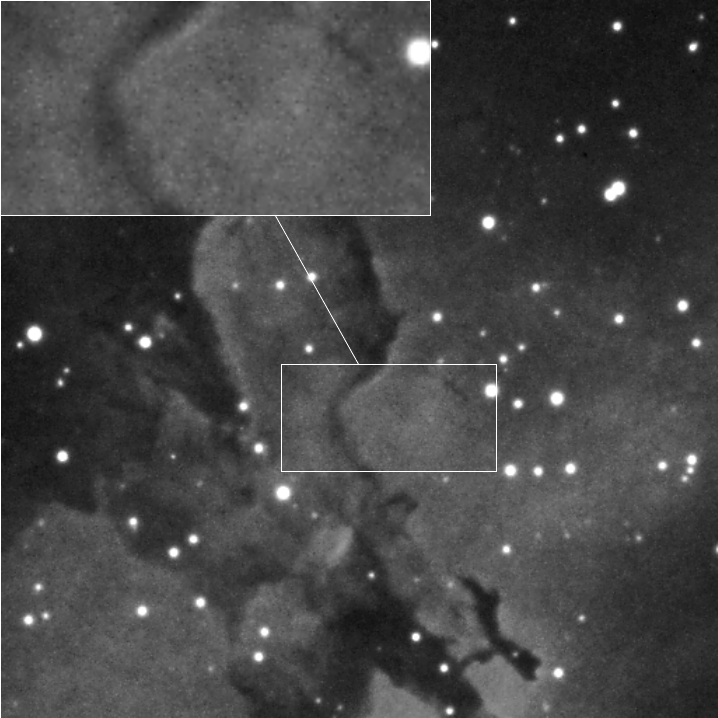}
} %
&
\raisebox{-0.5\height}{
\includegraphics[width=0.11\textwidth]{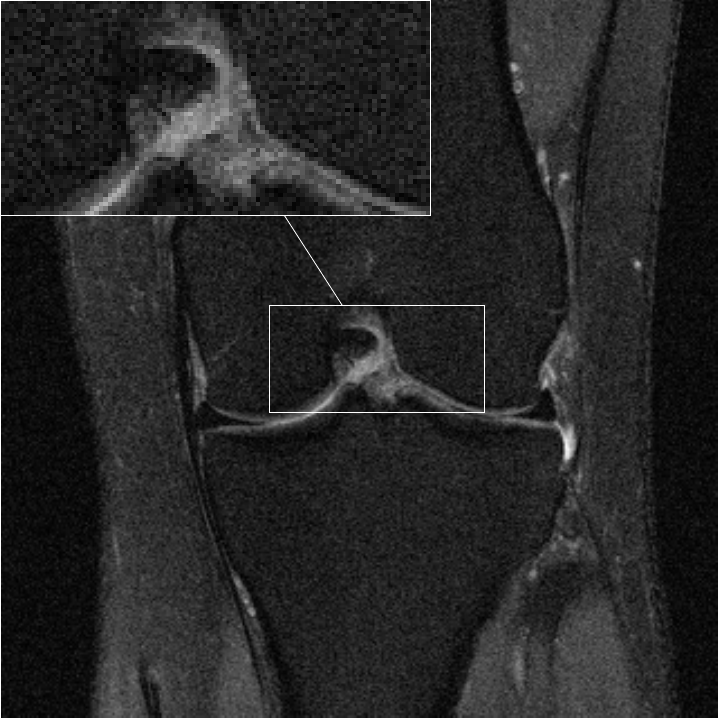}\!
\includegraphics[width=0.11\textwidth]{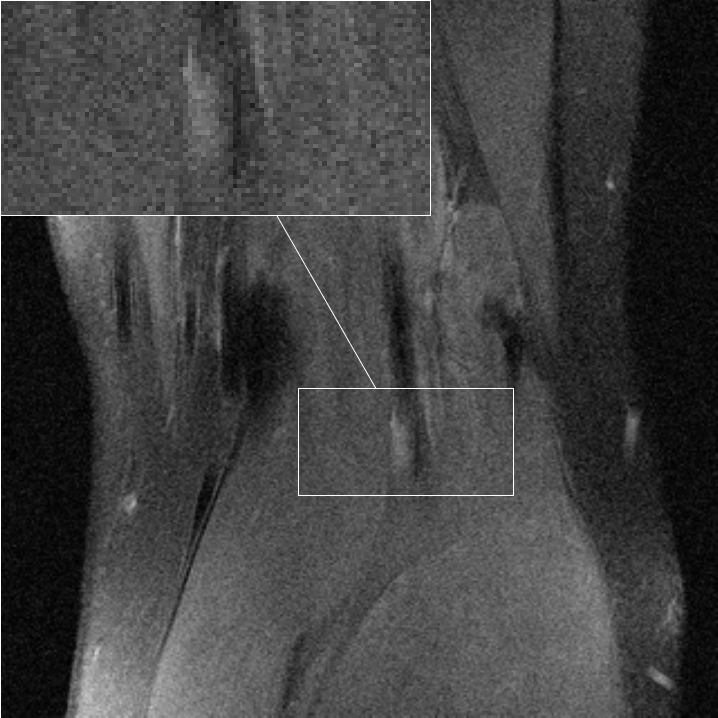}
} %
&
\raisebox{-0.5\height}{\includegraphics[width=0.025\textwidth]{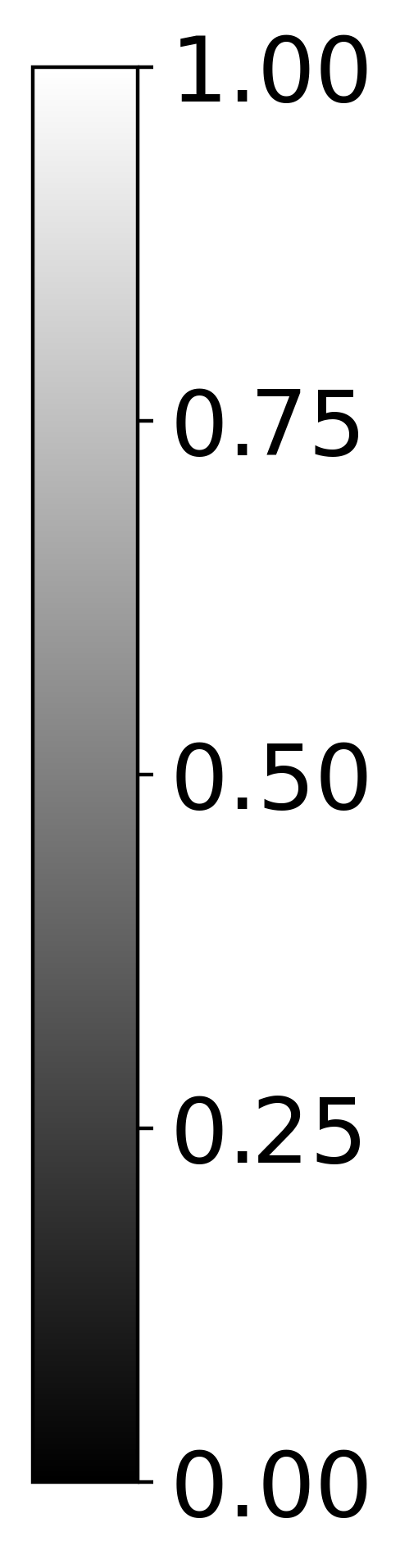}} %
\\
& (a) & (b) & \\
\rotatebox[origin=c]{90}{Preprocessed} %
& 
\raisebox{-0.5\height}{
\includegraphics[width=0.11\textwidth]{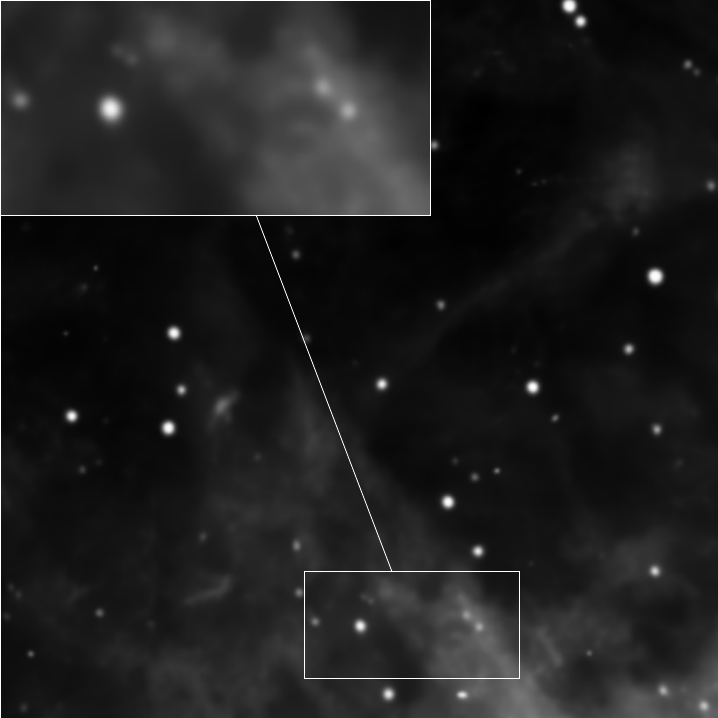}\!
\includegraphics[width=0.11\textwidth]{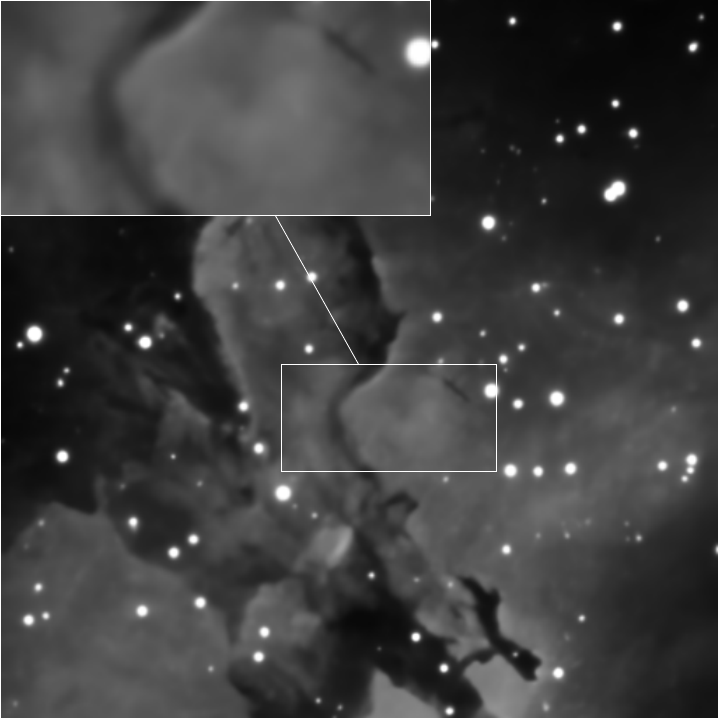}
} %
&
\raisebox{-0.5\height}{
\includegraphics[width=0.11\textwidth]{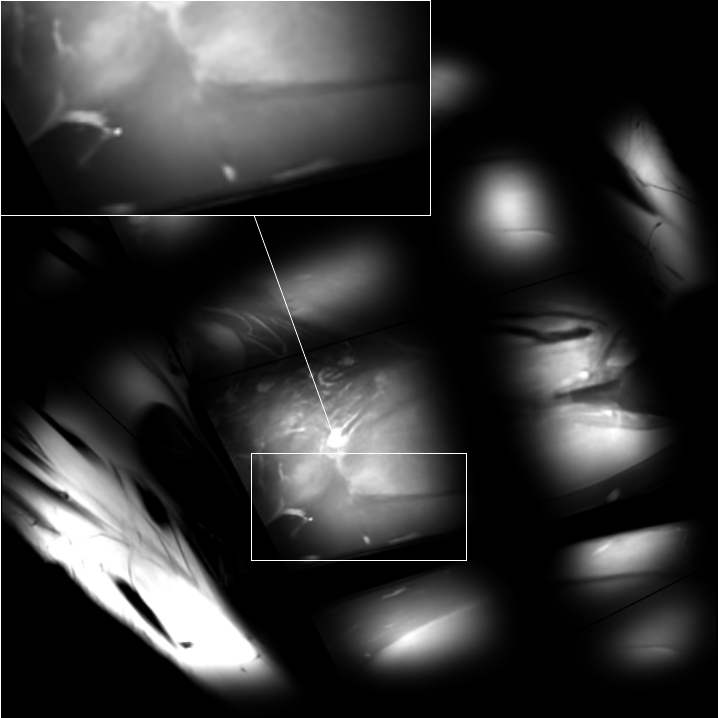}\!
\includegraphics[width=0.11\textwidth]{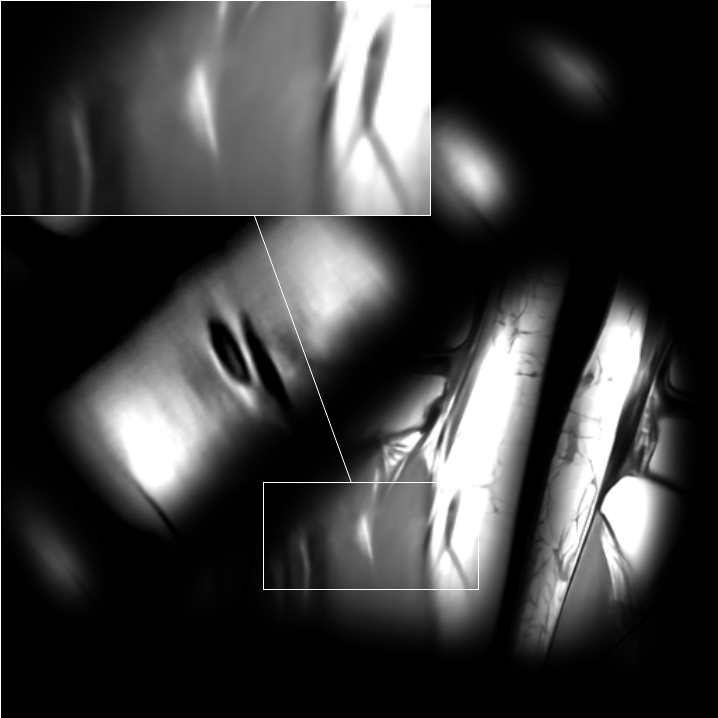}
} %
&
\raisebox{-0.5\height}{\includegraphics[width=0.025\textwidth]{exp_results/512_results/colorbar_ver_lin.png}} %
\\
& (c) & (d) & \\
\rotatebox[origin=c]{90}{Exponentiated} %
& 
\raisebox{-0.5\height}{
\includegraphics[width=0.11\textwidth]{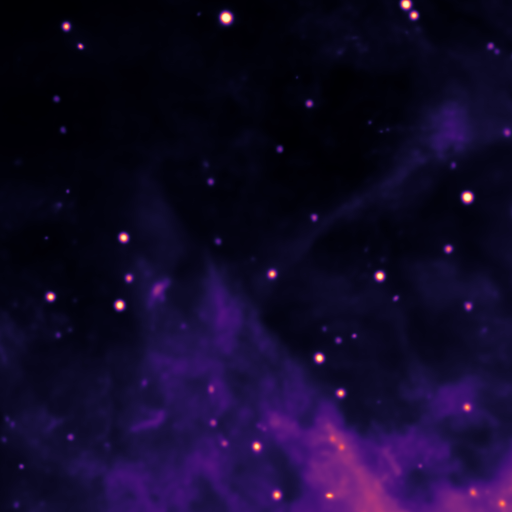}\!
\includegraphics[width=0.11\textwidth]{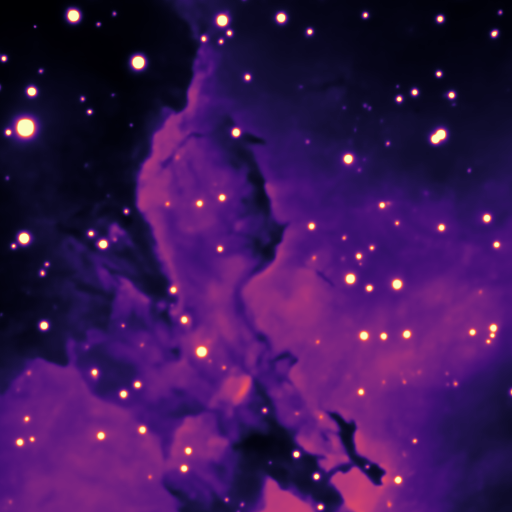}
} %
&
\raisebox{-0.5\height}{
\includegraphics[width=0.11\textwidth]{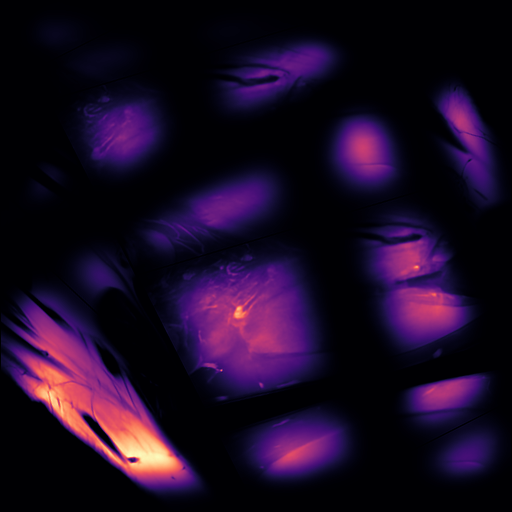}\!
\includegraphics[width=0.11\textwidth]{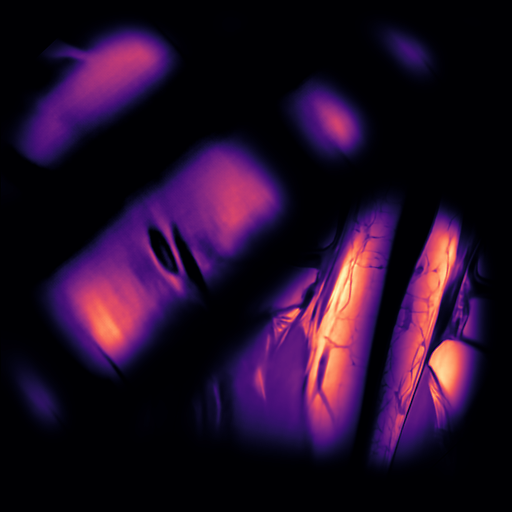}
} %
&
\raisebox{-0.5\height}{\includegraphics[width=0.025\textwidth]{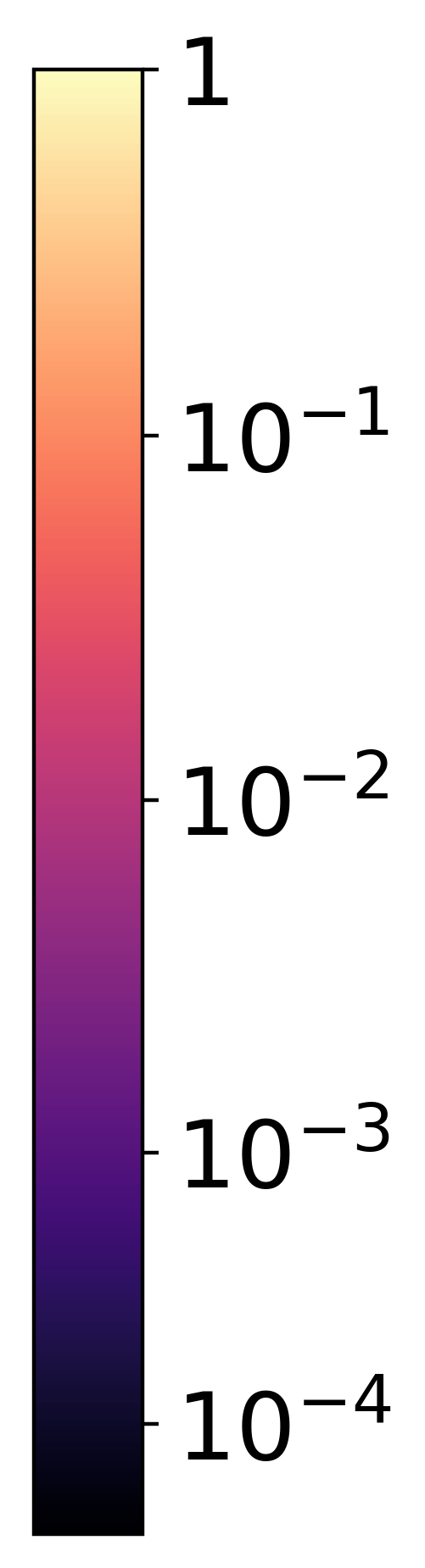}} %
\\
& (e) & (f) & \\
\end{tabular}

\caption{The different preprocessing steps of our optical astronomy image dataset (OAID; left) and MRI dataset (MRID; right). (a) and (b) show raw images; (c) and (d) show images preprocessed with a SCUNet and the soft-thresholding removing the background; (e) and (f) show images after exponentiation. For images in MRID, random translations and rotations are also applied to increase image size and to break symmetries as discussed in section \ref{subsection:general_exponentiation}.}
\label{fig:exponentiated}
\end{figure}

We show in Fig.~\ref{fig:exponentiated} (a)-(d) the raw and preprocessed images from both optical astronomy image dataset (OAID) and MRI dataset (MRID). 
While the semantic content of images strongly differs between the two datasets, both are normalized to a maximum intensity of 1. However, the residual noise levels differ between the two. As a result, the minimum non-zero intensity, denoted $\sigma_0$, also varies between the datasets, leading to different dynamic ranges. For each dataset, $\sigma_0$ is chosen such that the intensities of $1\%$ non-zero pixels in the corresponding dataset are below this value, which are around $0.02$ and $0.01$ for OAID and MRID respectively.

\subsection{Training a shelf of high dynamic range denoisers}
\label{sect:dynamic_range_est}

A distinctive characteristic of radio images is their high dynamic range, with important features exhibiting very faint intensity (for instance radio plumes or radio relics, see \citet{missaglia2019watcat, wittor2021exploring}). Observation with modern instrument will typically target value up to $10^6$ or even higher. Considering without loss of generality images with peak intensity normalised to 1, this implies faint intensities
down to as low as $10^{-6}$. In order to convert the available low-dynamic range images to high dynamic ranges, \citet{terris2023image} proposed a simple  exponentiation procedure.
Given a low-dynamic range image $\bm{u}$, we apply a pixel-wise exponentiation transform, denoted $\operatorname{rexp}_a$:
\begin{equation}
    \operatorname{rexp}_a \colon \bm{u} \mapsto \bm{u}_\text{max}(a^{\bm{u}/\bm{u}_\text{max}}-1)/a. 
    \label{eq:intensity_expo}
\end{equation}
Indeed, it can be shown that this transform increases the dynamic range of the image for sufficiently large values of $a$. 
In this work, we assume a maximum pixel intensity of $1$, i.e.
$\bm{u}_{\text{max}} = 1$.

Next, we link the exponentiation factor $a$ to the target dynamic range $1/\sigma$ by following the approach of \citet{terris2023image}, where the faintest intensity in the exponentiated image, $(a^{\sigma_0} - 1)/a$, is equated to the inverse of the dynamic range $\sigma$.
This yields the following equation to be solved for $a$, given the faintest intensity $\sigma_0$ in the clean, low-dynamic range training dataset, and the inverse of the target dynamic range $\sigma$:
\begin{equation}
    (a \sigma + 1) ^ { 1 / \sigma_0} - a = 0.
    \label{eq:expo_factor}
\end{equation} 
We solve \eqref{eq:expo_factor} numerically for the exponentiation parameter $a$ at different training noise levels of interest to our experiments.

Applying this transform to images in OAID and MRID online during training yields synthetic high dynamic range radio-astronomical training images, as shown in Fig.~\ref{fig:exponentiated} (e)-(f).

Given these high dynamic range datasets, the common approach for training the denoiser $\operatorname{D}$ consists in minimizing a denoising loss with Lipschitz regularization \citep{terris2023image}.
Denoting $\bm{\theta}$ the learnable parameters of $\operatorname{D}$, for a synthetic high dynamic range sample $\bm{u}$, the training loss writes 
\begin{equation}
\begin{aligned}
    \mathcal{L}(\bm{\theta}, \bm{u}) = \underbrace{\|\operatorname{D}_{\bm{\theta}}(\bm{u}+\sigma \bm{w})-\bm{u}\|_1}_{\text{denoising}}+ \,\kappa\underbrace{\operatorname{max}\{ \| \boldsymbol{\nabla} \operatorname{Q}_{\bm{\theta}}(\widetilde{\bm{u}})\|_{\rm{S}}, 1-\delta \}}_{\text{Lipschitz regularization}},
\end{aligned}
\label{eq:training_loss}
\end{equation}
where $\bm{w}$ is the realization of Gaussian random noise with standard deviation 1, $\sigma>0$ denotes the training noise level, $\operatorname{Q}$ is defined as $\operatorname{Q} = 2\operatorname{D}-\operatorname{Id}$,
$\|\bm{\nabla} \operatorname{Q} (\cdot)\|_{\text{S}}$ denotes the spectral norm of the Jacobian of $\operatorname{Q}$, $\kappa,$ and $\delta$ are positive constants, and $\widetilde{\bm{u}}$ is a point sampled at random on the segment $[\bm{u}, \bm{u}+\sigma \bm{w}]$.

Finally, to avoid the necessity of training denoisers for each measurement with different target dynamic ranges,
we build a
shelf of denoisers trained for different noise levels covering the full range of possible dynamic ranges of interest for a class of radio observations.

\subsection{Selecting denoiser from shelf during reconstruction}
\label{subsect:denoiser_selection}

During reconstruction, both the maximum image intensity and the effective noise level in the image domain need to be estimated from the data to provide an estimate of the target dynamic range, necessary for choosing the appropriate trained denoiser based on its training noise level.

Firstly, the maximum intensity of the back-projected image, denoted as $\widetilde{\alpha}$, provides a rough estimation of the sought peak value $\alpha$ of the target image $\overline{\bm{x}}$. Then, following \citet{terris2023image}, the inverse of the signal-to-noise ratio of the observed data, \emph{i.e.} the inverse of the target dynamic range, can be estimated from the properties of $\mathrm{\bm{\Phi}}$ and $\bm{e}$ in \eqref{eq:inv_pb_gen} as:
\begin{equation}
    \sigma_\text{heu} = \frac{\eta}{\widetilde{\alpha}\sqrt{2\| \mathrm{\bm{\Phi}} \|^2_{\rm S}}},
    \label{eq:heuristic}
\end{equation}
where we recall that $\eta$ is the standard deviation of $e$ in \eqref{eq:inv_pb_gen}. 

Next, the network $\operatorname{D}_\sigma$ will be chosen where $\sigma$ is the closest training noise level below $\sigma_{\text{heu}}$.
The activation of the denoiser is thus rescaled with a factor $\beta=\widetilde{\alpha}\sigma_{\text{heu}}/\sigma$ as
\begin{equation}
\label{eq:rescaled_denoiser}
 \operatorname{D}(\cdot) = \beta \operatorname{D}_{\sigma}(\cdot / \beta),
\end{equation}
ensuring both that the input to the network has maximum intensity below $1$ and that the effective standard deviation of the residual noise in the input to the denoiser matches the noise level $\sigma$ for which it was trained.

In practice, the peak value of the back-projected image is a loose upper bound estimation of the real peak value. To address the inaccuracy of the $\sigma_{\text{heu}}$ estimate, we update $\widetilde{\alpha}$ as the maximum intensity of the reconstructed image at each iteration of the PnP algorithm. The value of $\sigma_{\text{heu}}$ is monitored at each iteration to check whether $\operatorname{D}_\sigma$ and the scaling factor $\beta$ need to be updated. 
With the combination of the denoiser shelf and the selection strategy, 
we fully decouple the training and reconstruction.

\subsection{Model uncertainty quantification}
\label{sect:model_uncertainty}

If obtaining an estimate of the solution to a specific model is desirable, it is also of interest to be able to quantify the uncertainty in the recovered solution.
In practice, this uncertainty quantification is often performed by sampling from the posterior linked to $r$ in \eqref{eq:minimization},
\emph{e.g.}~with MCMC methods \citep{laumont2022bayesian, mukherjee2023learned}, but other strategies have been proposed \citep{repetti2019scalable, zhang2019probabilistic, liaudat2023scalable}. In a nutshell, these methods aim at quantifying the aleatoric uncertainty induced by the measurement procedure \eqref{eq:inv_pb_gen} in the reconstructed image.

Similarly to \citet{narnhofer2021bayesian}, we instead propose in this work to investigate the epistemic uncertainty in the reconstruction, induced by variations in the denoiser $\operatorname{D}$. In fact, while the choice of $r$ in \eqref{eq:minimization}
is a cornerstone of the reconstruction quality, its relationship with the denoiser $\operatorname{D}$ in the case of PnP algorithms remains unclear since no closed form for $r$ is available in general\footnote{
Only a maximally monotone operator can be associated to $\operatorname{D}$. These operators can be seen as a generalization of subdifferentials for convex functionals. We refer the reader to \citet{bauschke2011convex} and  \citet{pesquet2021learning} for more details.}, not to mention the high number of parameters and the (by nature) imperfect training procedure of $\operatorname{D}$. 
 
Training the same architecture $\operatorname{D}$ for the same task but in different settings (\emph{i.e.}~with different initialisation, Gaussian noise realisation or patch selection) leads to networks $\operatorname{D}$ with different weights, but consistent denoising performance. In turn, these different denoiser realizations yield distinct prior realizations. More precisely, assume that we train $K \in \mathbb{N}$ denoisers $(\operatorname{D}_k)_{1\leq k \leq K}$ with the same training procedure (loss and datasets) and that the sole difference between these denoisers arises from distinct random realizations influencing their training process. Intuitively, these denoisers yield unique regularizers $(r_k)_{1\leq k\leq K}$, and plugging $\operatorname{D}_k$ within the PnP Algorithms~\ref{algo:pnp_fb} and \ref{algo:pnp_pd} returns $\widehat{\bm{x}}_k$ satisfying \eqref{eq:minimization}. This procedure gives us a set of solutions $\operatorname{Sol} = \{\widehat{\bm{x}}_1, \hdots, \widehat{\bm{x}}_K\}$ to each problem \eqref{eq:inv_pb_gen}. In turn, the (pixel-wise) empirical standard deviation of $\operatorname{Sol}$ yields an estimator of the epistemic uncertainty \citep{lahlou2021deup}.

We note that previous works have studied aleatoric and epistemic uncertainties in the context of end-to-end supervised learning with Bayesian neural networks \citep{kendall2017uncertainties, baumgartner2019phiseg, xue2019reliable,  edupuganti2020uncertainty, kwon2018uncertainty, gao2022bayesian}.

\section{Experimental results on simulated data}
\label{section:synthetic_resulst}

In this section, we show the performance of different algorithms for reconstructing RI images from simulated observations.

\begin{figure*}
\centering
\small
\begin{tabular}{@{\hspace{0.\tabcolsep}}c @{\hspace{0.1\tabcolsep}} c @{\hspace{0.1\tabcolsep}} c @{\hspace{0.1\tabcolsep}} c @{\hspace{0.1\tabcolsep}} c @{\hspace{0.1\tabcolsep}} c @{\hspace{0.1\tabcolsep}} c @{\hspace{0.0\tabcolsep}} l}
\includegraphics[width=0.158\textwidth]{exp_results/512_results/zoom_true.pdf} &
\includegraphics[width=0.158\textwidth]{exp_results/512_results/zoom_dirty.pdf} &
\includegraphics[width=0.158\textwidth]{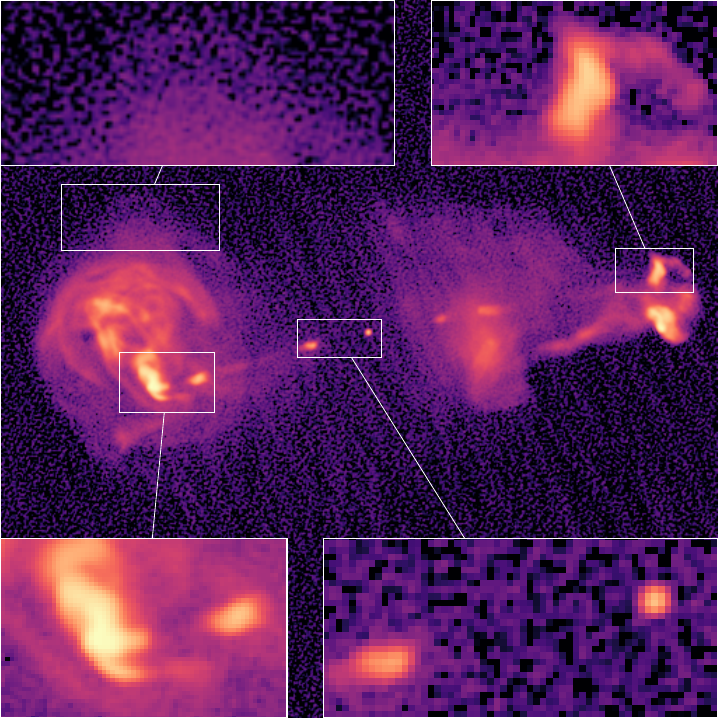} &
\includegraphics[width=0.158\textwidth]{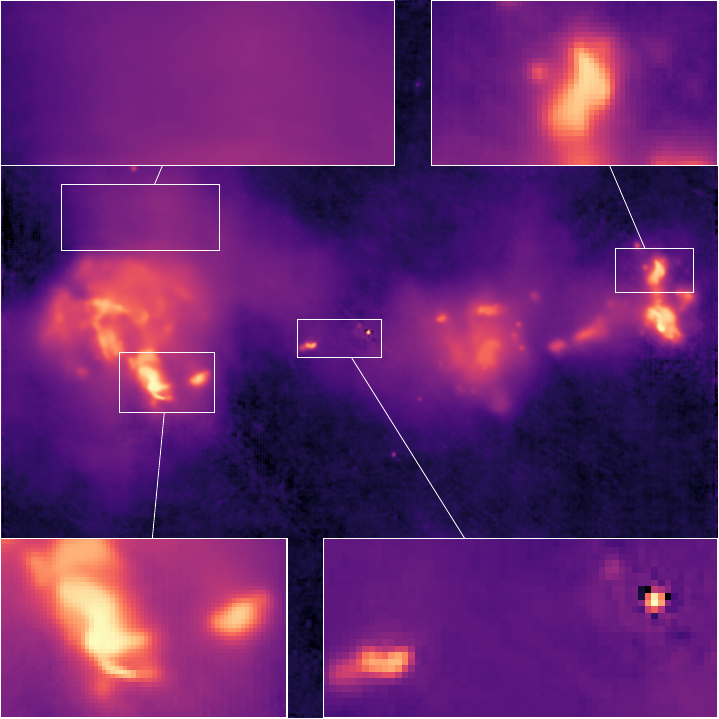} &
\includegraphics[width=0.158\textwidth]{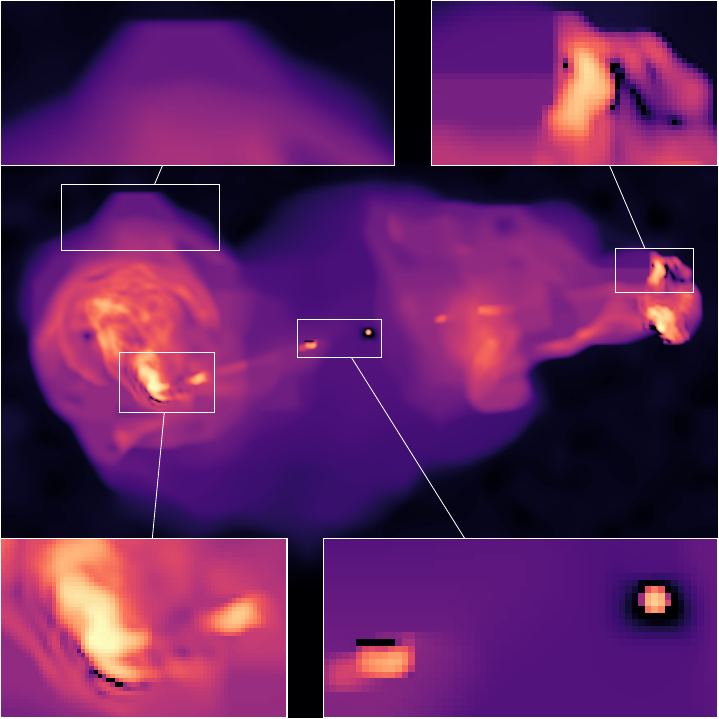} &
\includegraphics[width=0.158\textwidth]{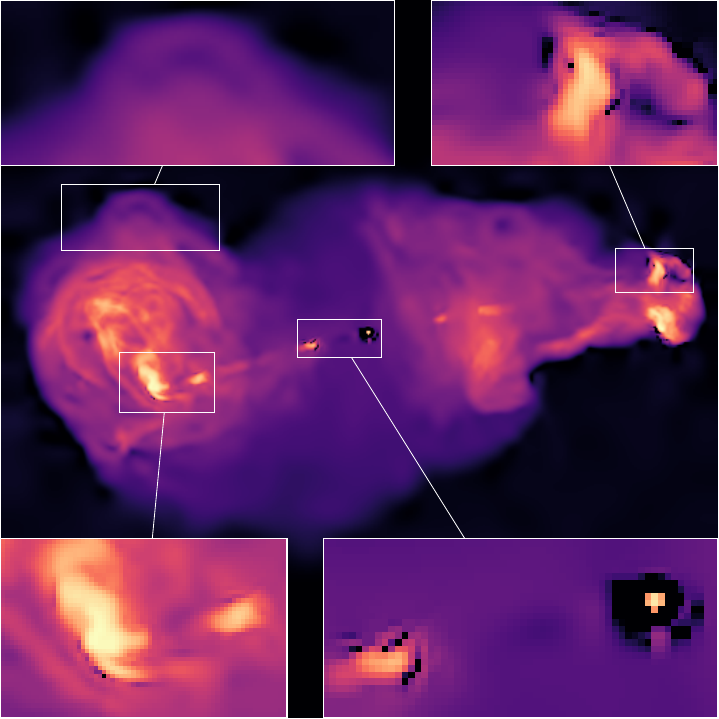} &
\raisebox{0.05\height}[0pt][0pt]{\includegraphics[width=0.04\textwidth]{exp_results/512_results/colorbar_vertical_new.png}} \\
(a) Groundtruth & (b) Back-projection & (c) MS-CLEAN & (d) UNet &  (e) uPnP\textsubscript{BM3D} & (f) uSARA & \\
&  &   $(5.45 {\rm dB}, 7.07 {\rm dB})$  & $(17.02 {\rm dB}, 7.42 {\rm dB})$ & $(20.44 {\rm dB}, 24.01 {\rm dB})$ & $(24.53 {\rm dB}, 24.98 {\rm dB})$ & \\
 \includegraphics[width=0.158\textwidth]{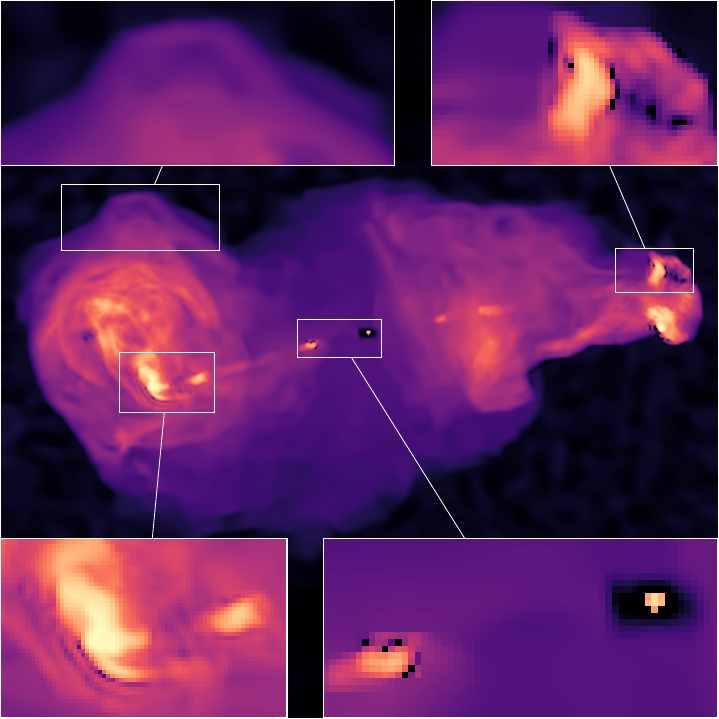} &
\includegraphics[width=0.158\textwidth]{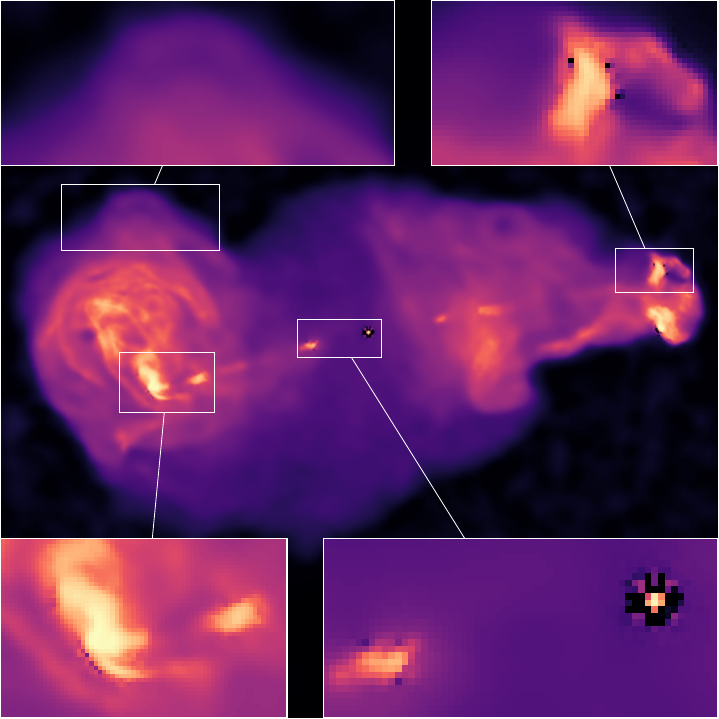} &
\includegraphics[width=0.158\textwidth]{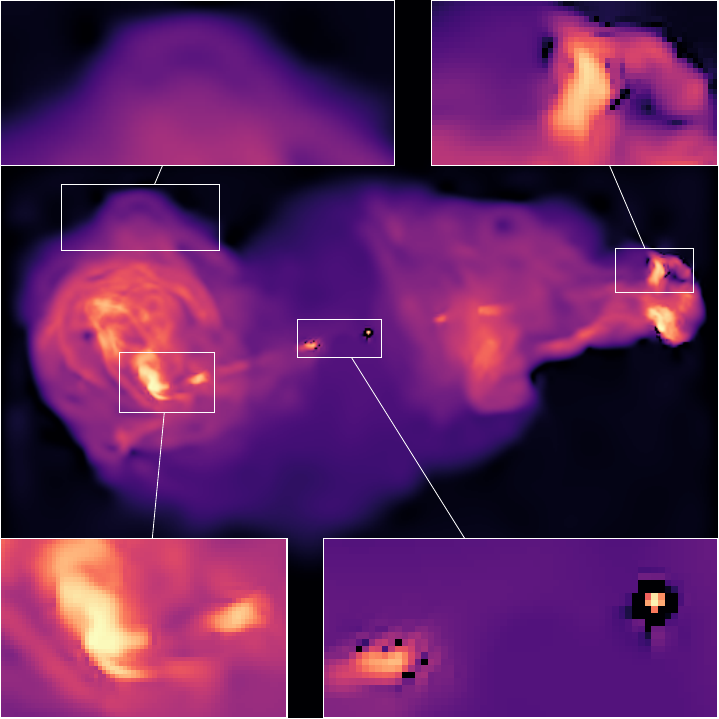} &
\includegraphics[width=0.158\textwidth]{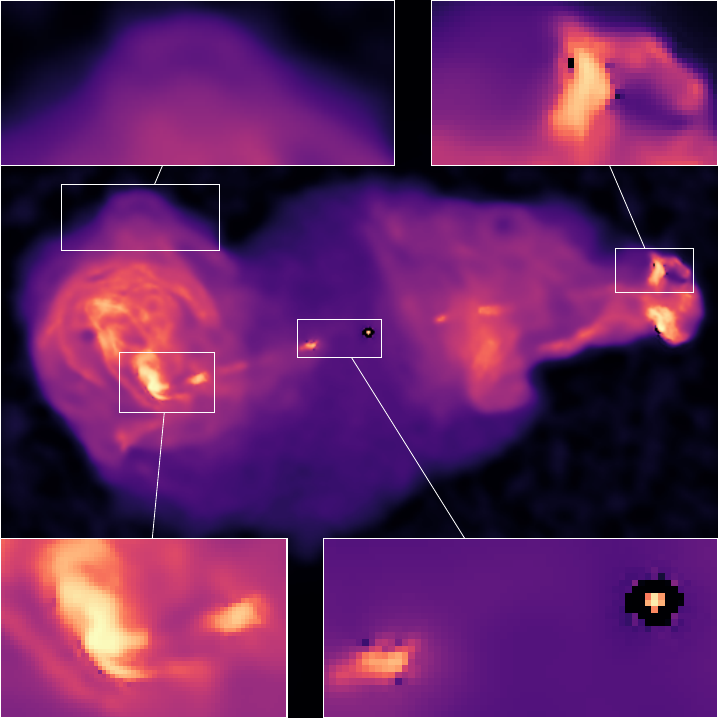} &
\includegraphics[width=0.158\textwidth]{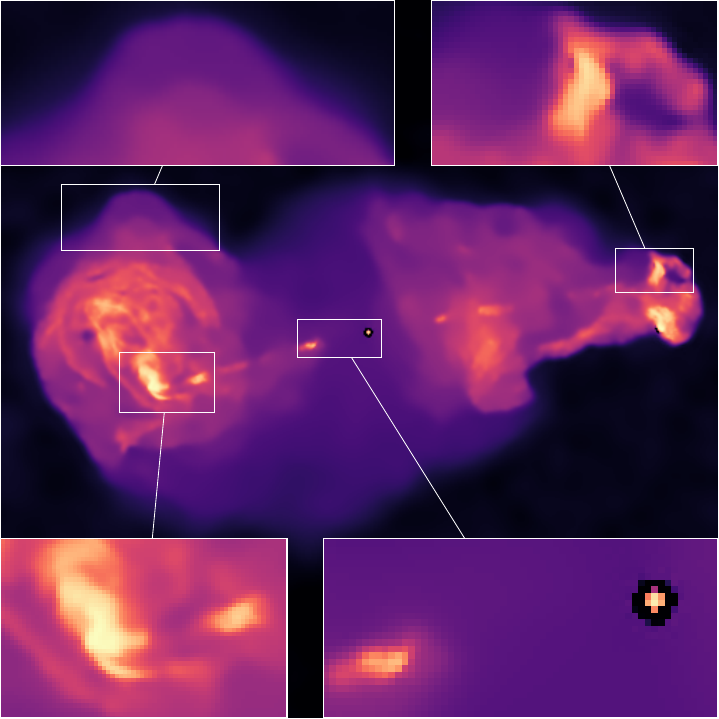} &
\includegraphics[width=0.158\textwidth]{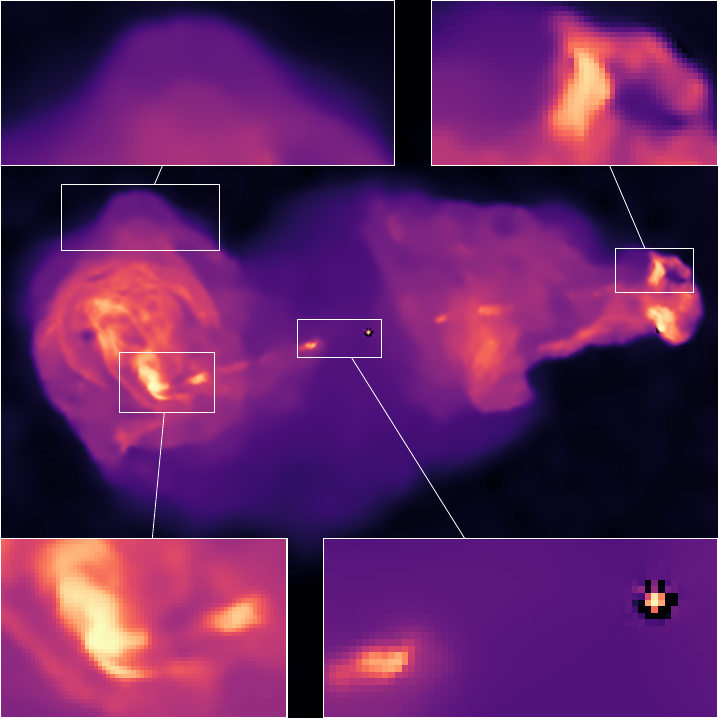} & 
\raisebox{0.05\height}[0pt][0pt]{\includegraphics[width=0.04\textwidth]{exp_results/512_results/colorbar_vertical_new.png}} \\
(g) cPnP\textsubscript{BM3D} & (h) cAIRI\textsubscript{MRID} & (i) SARA  & (j) AIRI\textsubscript{MRID} & (k) AIRI\textsubscript{OAID} & (l) cAIRI\textsubscript{OAID} & \\
 $(23.24 {\rm dB}, 25.08 {\rm dB})$ & $(27.77 {\rm dB}, 25.11 {\rm dB})$  & $(27.74 {\rm dB}, 25.16 {\rm dB})$ &  $(26.51 {\rm dB}, 25.25 {\rm dB})$  & $(26.48 {\rm dB}, 25.79 {\rm dB})$ & $(28.65 {\rm dB}, 26.29 {\rm dB})$ & \\
\includegraphics[width=0.158\textwidth]{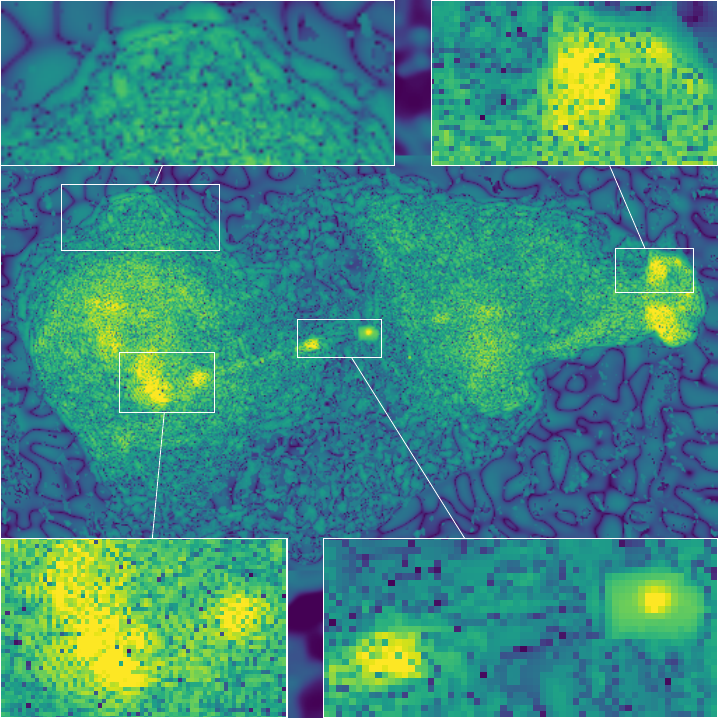} &
\includegraphics[width=0.158\textwidth]{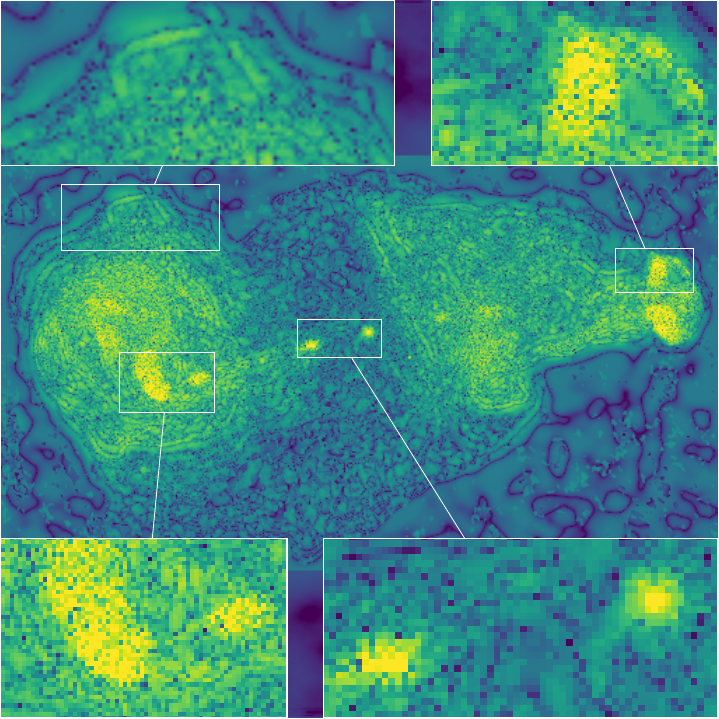} &
\includegraphics[width=0.158\textwidth]{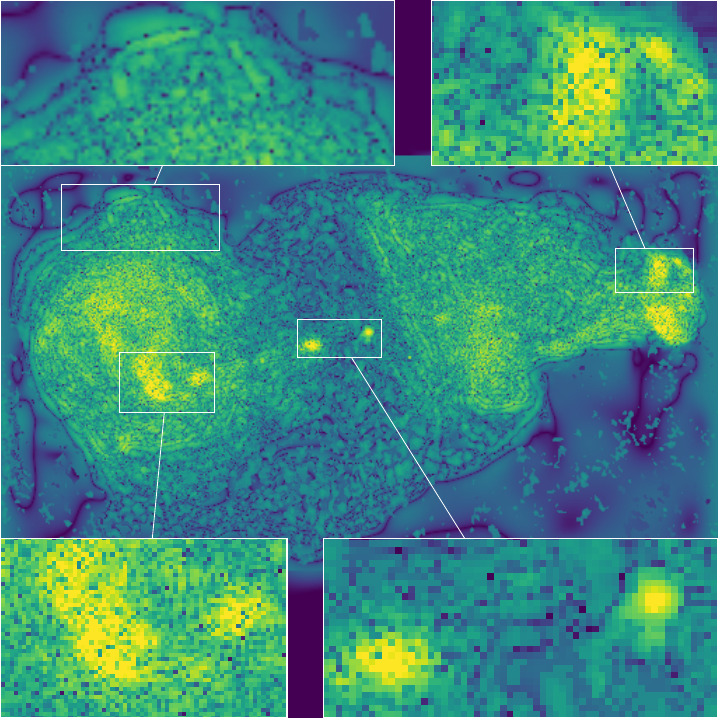} &
\includegraphics[width=0.158\textwidth]{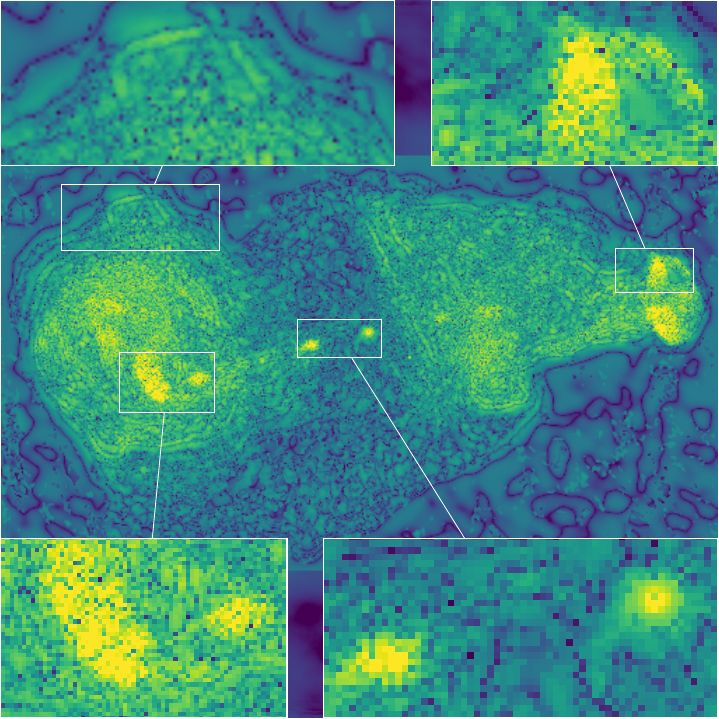} &
\includegraphics[width=0.158\textwidth]{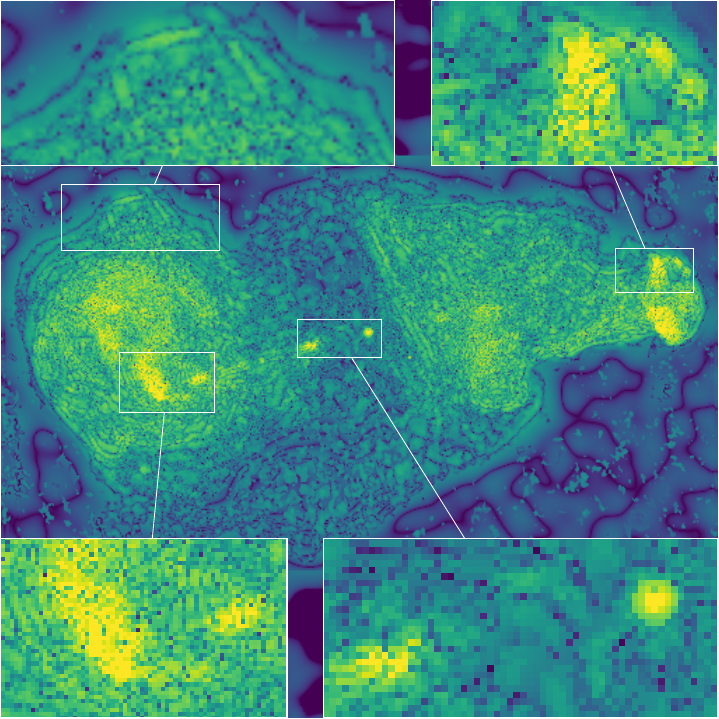} &
\includegraphics[width=0.158\textwidth]{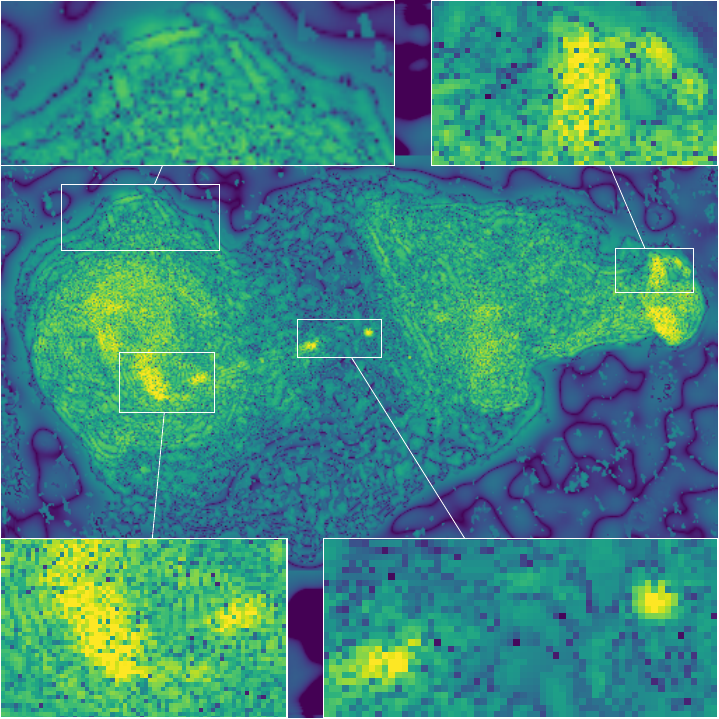} & 
\raisebox{0.05\height}[0pt][0pt]{\includegraphics[width=0.04\textwidth]{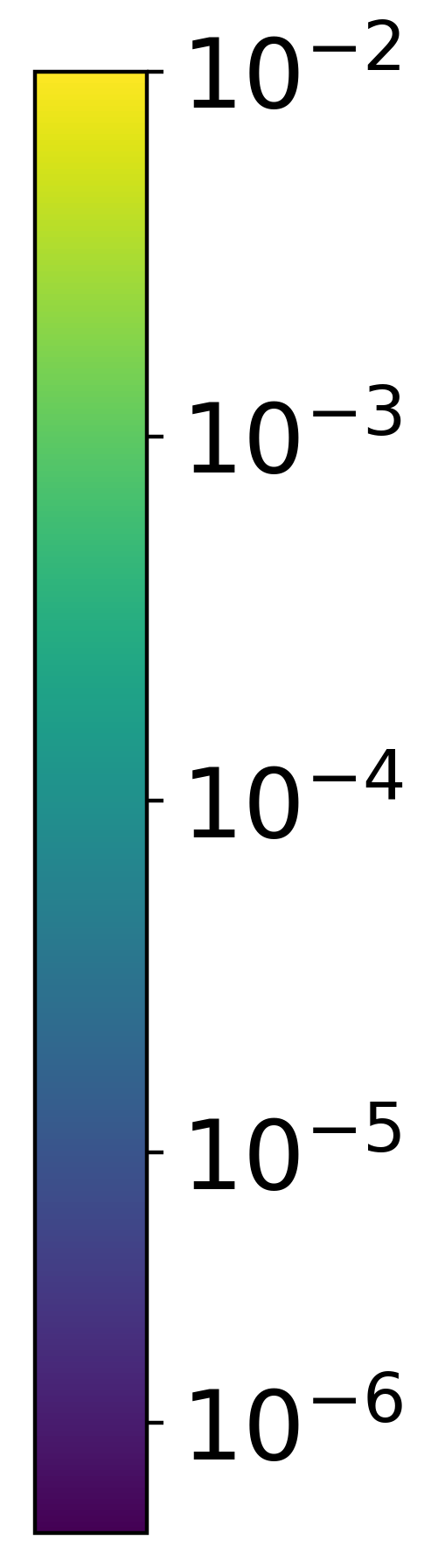}}\\
(m) cPnP\textsubscript{BM3D} error  & (n) cAIRI\textsubscript{MRID} error & (o) SARA error  
& (p) AIRI\textsubscript{MRID} error
& (q) AIRI\textsubscript{OAID} error & (r) cAIRI\textsubscript{OAID} error & \\
\includegraphics[width=0.158\textwidth]{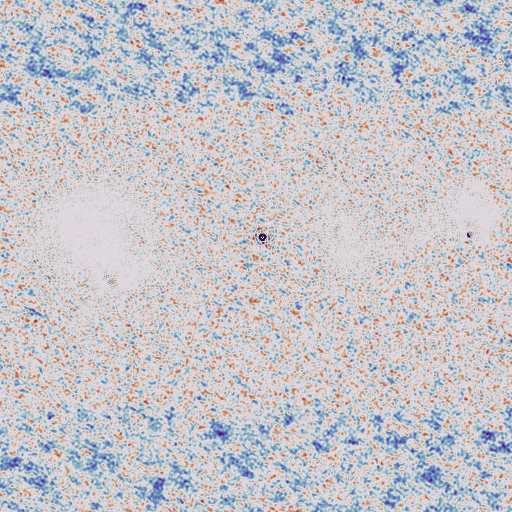} &
\includegraphics[width=0.158\textwidth]{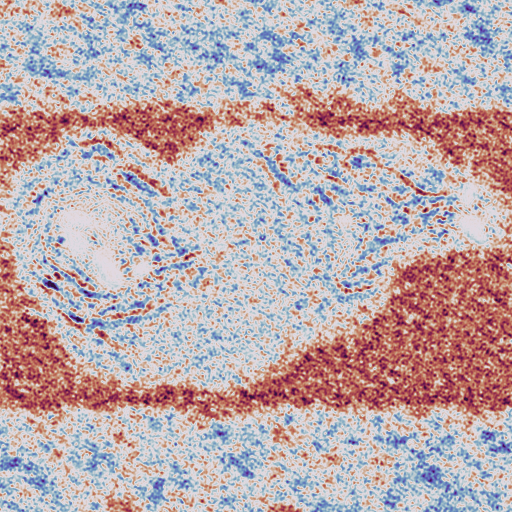} &
 \includegraphics[width=0.158\textwidth]{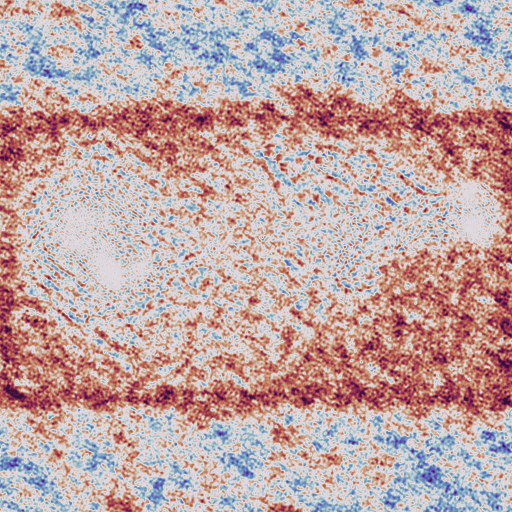} &
\includegraphics[width=0.158\textwidth]{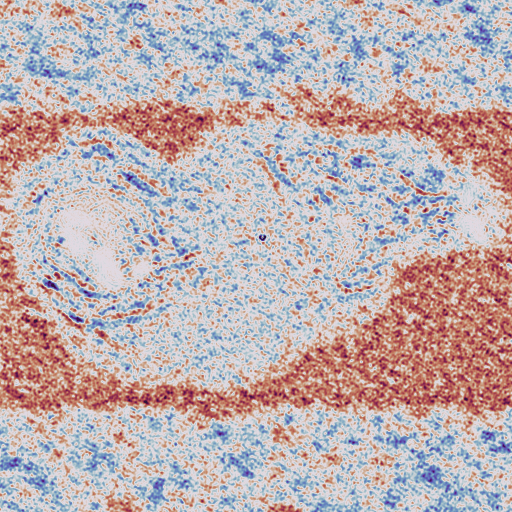} &
\includegraphics[width=0.158\textwidth]{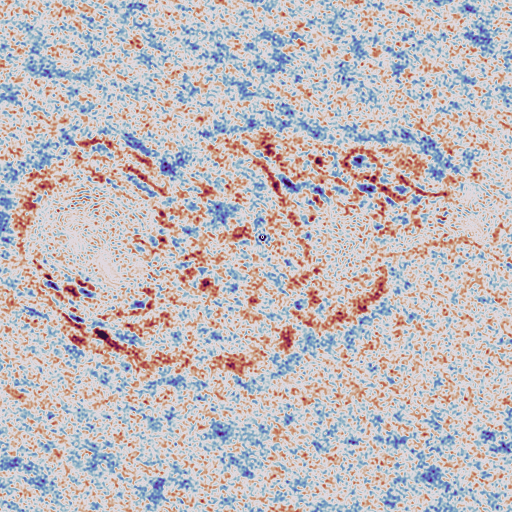} &
\includegraphics[width=0.158\textwidth]{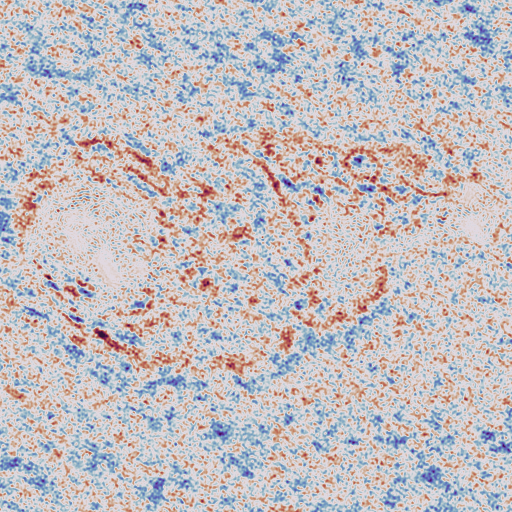} & 
\raisebox{0.05\height}[0pt][0pt]{\includegraphics[width=0.04\textwidth]{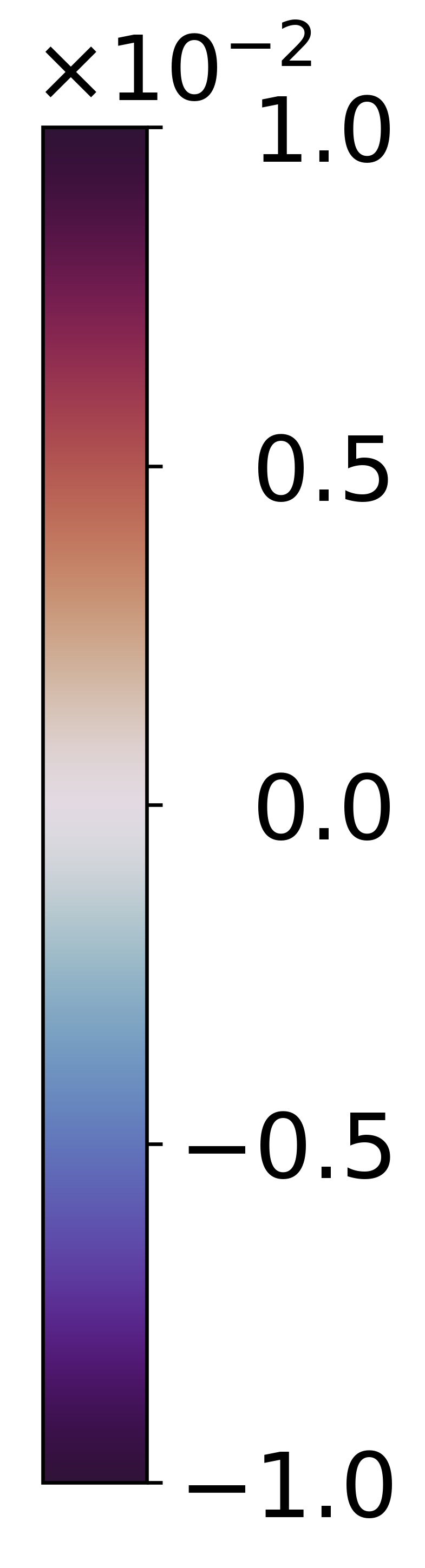}}\\
(s) cPnP\textsubscript{BM3D} residual  & (t) cAIRI\textsubscript{MRID} residual  & (u) SARA residual &  (v) AIRI\textsubscript{MRID} residual & (w) AIRI\textsubscript{OAID} residual & (x) cAIRI\textsubscript{OAID} residual & \\
\end{tabular}
\caption{Results of various algorithms for the RI imaging problem. Top left image (a) shows the groundtruth image, (b) shows the simulated measurements back-projected in the real image domain. Other images on rows 1 and 2 show reconstructions with different imaging algorithms in increasing order of logSNR values. The metrics (SNR, logSNR) associated with each reconstruction are shown beneath each image. 
Row 3 shows error maps for different algorithms in row 2.
Row 4 shows residual dirty images of associated algorithms.
All images in rows 1, 2 and 3 are displayed in logarithmic scale (see colorbar on the right) while the residuals in row 4 are shown in linear scale. Insets show magnified areas of the image.}
\label{fig:3c353_dt8}
\end{figure*}

\subsection{Test dataset and evaluation metrics}

We use 3 radio images (3c353, Hercules A, and Centaurus A, see \citet{terris2023image} for more details) of size $N=$ 512$\times$512 as ground-truth images to simulate testing measurements. Each image has maximum intensity value normalised to 1, and a dynamic range slightly above $10^4$.  
For each image, we generate RI measurements following \eqref{eq:inv_pb_gen} with various observation settings simulating the behaviours of the MeerKAT telescope \citep{jonas2016meerkat}.
Namely, each simulated operator $\mathrm{\bm{\Phi}}$ arises from different pointing direction and observation time.
More specifically, we consider five distinct pointing directions chosen at random, and for each direction, we vary the observation time with $\Delta T$ selected from the set $\{1\text{h}, 2\text{h}, 4\text{h}, 8\text{h}\}$. 
This leads to a total of 4$\times$5 inverse problems for each groundtruth image, \emph{i.e.}~a total of 4$\times$15 inverse problems with $\bm{y}\in\mathbb{C}^m$ such that $m = 4\times 10^5$ for $\Delta T = 1\text{h}$, 
$m = 8\times 10^5$ for $\Delta T = 2\text{h}$, 
$m = 1.6\times 10^6$ for $\Delta T = 4\text{h}$, 
$m = 3.2\times 10^6$ for $\Delta T = 8\text{h}$.
Fig.~\ref{fig:illustration}~(b) shows one of the test sampling patterns for a given pointing direction and observation time $\Delta T = 4\text{h}$.
The dynamic ranges of the observations
fall in the range $[2.5 \times 10^3, 1.4 \times 10^4]$, 
with the average dynamic range for each $\Delta T$ as follows: $2.9 \times 10^3$ for $\Delta T = 1\text{h}$, $3.9 \times 10^3$ for $\Delta T = 2\text{h}$, $5.6 \times 10^3$ $\Delta T = 4\text{h}$, and $1.1 \times 10^4$ for $\Delta T = 8\text{h}$.

To emphasize low intensity features, radio images are usually displayed in logarithmic scale with the following explicit transform $\operatorname{rlog}_a(\bm{x})= \bm{x}_{\text{max}}\log_a(a \bm{x} / \bm{x}_{\text{max}}+1)$, which maps an image $\bm{x}$ with dynamic range $a\gg 1$ and pixel values in $[\bm{x}_\text{max}/a;\bm{x}_\text{max}]$ to a low-dynamic range image $\operatorname{rlog}_a(\bm{x})$, with dynamic range $\log_a(2)$ and pixel values in the range $[\bm{x}_\text{max}\log_a(2);\bm{x}_\text{max}]$.
Therefore, we evaluate the reconstruction accuracy of the high and low intensities with the SNR and logSNR metrics. Given a groundtruth image $\overline{\bm{x}}$ and an estimated image $\widehat{\bm{x}}$, the reconstruction SNR is defined as $\operatorname{SNR}(\overline{\bm{x}}, \widehat{\bm{x}}) = 20\log(\|\overline{\bm{x}}\|/\|\overline{\bm{x}}-\widehat{\bm{x}}\|)$.
Given the high DR set up, in order to evaluate the performance of the algorithms at faint intensities, we consider the logSNR metric \citep{terris2023image} writing $\operatorname{logSNR}(\overline{\bm{x}}, \widehat{\bm{x}}) = \operatorname{SNR}(\operatorname{rlog}_a(\overline{\bm{x}}),\operatorname{rlog}_a(\widehat{\bm{x}}))$.
The images in this article are presented in a logarithmic scale, i.e. for an image $\bm{x}$, visualizations show $\operatorname{rlog}_a(\bm{x})$.
The values of $a$ in $\operatorname{rlog}_a(\cdot)$ should ideally be set to the corresponding dynamic range of each measurement. For the $\operatorname{logSNR}$ metric, we set $a=2.5 \times 10^3$, the lowest dynamic range of all the synthetic measurements.  
Choosing a single value for $a$ preserves the consistency of the $\operatorname{logSNR}$ metric across different reconstructions and $\Delta T$'s. 
For the visualisation of images, a value of $a=10^4$ is chosen, more representative of the dynamic range values for the $\Delta T = 4 \text{h}, 8 \text{h}$ cases displayed.
Because the groundtruth images are normalised, $\bm{x}_{\text{max}}$ in $\operatorname{rlog}_a(\cdot)$ is set to 1 for consistency.
Given $K$ trained models, the model uncertainty is estimated using the empirical unbiased standard deviation (std) metric, defined as $\operatorname{std}(\operatorname{Sol}) =  \left.\sqrt{\sum_{k=1}^K(\widehat{\bm{x}}_k-\operatorname{mean}(\operatorname{Sol}))^2}  \middle/ \sqrt{K-1}\right.$. Additionally, we calculate the relative standard deviation, defined as $\operatorname{std/mean}(\operatorname{Sol}) = \operatorname{std}(\operatorname{Sol})/\operatorname{mean}(\operatorname{Sol})$,  to compare the uncertainty for features with different intensities.
 
To assess the data-fidelity quality, we also provide residual dirty images, defined as $\operatorname{Re} \{ \bm{\Phi}^\dagger(\bm{y}-\bm{\Phi}\bm{x}) \}$ \citep[up to a normalization constant, see][]{terris2023image}.

\subsection{AIRI variants and benchmark algorithms}
In order to investigate the influence of the training dataset on the denoising prior, we propose two versions for both AIRI and cAIRI underpinned by two sets of denoisers trained with different datasets.
We denote by AIRI\textsubscript{OAID} (resp. cAIRI\textsubscript{OAID}) the AIRI (resp. cAIRI) algorithm relying on denoisers trained on the OAID dataset, while AIRI\textsubscript{MRID} (resp. cAIRI\textsubscript{MRID}) relies on denoisers trained on the MRID dataset
(see Fig.~\ref{fig:exponentiated}). 
Following \citep{terris2023image}, random rotations are applied before applying the denoising network, which improves stability of the algorithms \citep{terris2024equivariant}.
The algorithms are implemented with MATLAB.

We compare the performance of the proposed AIRI and cAIRI algorithms with several state-of-the-art RI imaging methods : 
(i) multi-scale CLEAN (MS-CLEAN) implemented in the C++ WSClean software package\footnote{\href{https://gitlab.com/aroffringa/wsclean/}{https://gitlab.com/aroffringa/wsclean/}}\textsuperscript{,}\footnote{The full command is: \texttt{wsclean -multiscale -niter 30000 -weight uniform -gain 0.1 -mgain 0.6 -threshold 0.001 -no-reorder -minuvw-m 0.001 -auto-mask 0.001 -mem}.} \citep{offringa2017optimized}; 
(ii) the UNet from \citep{zhang2020plug} that was trained on the proposed synthetic datasets to solve the RI imaging problem in a supervised end-to-end fashion for each $\Delta T$; 
(iii) uSARA and SARA, the MATLAB implementations of which are available in the \href{https://basp-group.github.io/BASPLib/}{BASPLib} code library. We further test our algorithms with the BM3D denoiser \citep{dabov2007image} in Algorithms~\ref{algo:pnp_fb} and \ref{algo:pnp_pd} instead of the proposed trained denoiser. 
BM3D is a non-trained denoiser and we tuned its denoising noise level to achieve the best possible image quality. However, it does not yield convergent PnP algorithms and suffers from slow inference time. The algorithm with a BM3D denoiser plugged into Algorithm~\ref{algo:pnp_fb} is named uPnP\textsubscript{BM3D} and the one with BM3D plugged into Algorithm~\ref{algo:pnp_pd} is named cPnP\textsubscript{BM3D}.

\subsection{Training details}
\label{susection:training_details}

We use the DnCNN network architecture \citep{zhang2017beyond} with the proposed modifications in \citet{terris2023image}: all batch normalization layers are removed, ReLU layers are replaced by leaky ReLU layers, and an additional ReLU is added to ensure positivity of the reconstructed image.

We train a shelf of denoisers for 8 noise levels covering wide possible dynamic ranges of interest, \emph{i.e. $\{2.5 \times 10^{-6}, 5.0 \times 10^{-6}, 1.0 \times 10^{-5}, 2.0 \times 10^{-5}, 4.0 \times 10^{-5}, 8.0 \times 10^{-5}, 1.6 \times 10^{-4}, 3.2 \times 10^{-4}\}$}. 
This choice of noise levels ensures that $\sigma_{\text{heu}}/\sigma \in [1, 2]$ in \eqref{eq:rescaled_denoiser}.
Each denoiser is trained on randomly cropped patches of size 49$\times$49 of the OAID or MRID datasets, with the training loss \eqref{eq:training_loss}. 
As detailed in \citet{terris2023image}, for each image of the synthetic dataset, a set of randomly cropped patches $\bm{x}_s$ is extracted and an associated high-dynamic range patch $\bm{u}_s$ is computed with $\bm{u}_s = \operatorname{rexp}_a(\bm{x}_s)$ where $a$ is chosen as per \eqref{eq:expo_factor}. 
The Lipschitz regularization parameter $\kappa$ in \eqref{eq:training_loss} is fine-tuned for each denoiser and other training hyperparameters are the same. We refer to the denoisers trained with the non-expansiveness term for RI imaging as AIRI denoisers.

\subsection{Results for image estimation}

To illustrate clearly even the minor differences among different methods, we choose to show the reconstructions of one measurement with $\Delta T = 8\text{h}$ in Fig.~\ref{fig:3c353_dt8}, which contain more details compared to others recovered from measurements with smaller observation time. 
The reconstructions are organised in the ascending order of logSNR values and we only display the error maps and residual dirty images for the 6 best results.
We first observe that MS-CLEAN successfully recovers structures with higher intensities, but not the faint and extended emissions. The central black hole is also less compact compared to the ones of other algorithms. This is as expected due to the convolution with the CLEAN beam and the addition of the residual dirty image.
As was observed in \citet{terris2023image}, UNet recovers the high intensity emissions 
better than MS-CLEAN,  
but struggles at lower intensities (see the extended purple areas): this translates into a relatively good SNR, but lower logSNR. 

For the other algorithms, the constrained ones generally yield better reconstructions than their unconstrained counterpart. uSARA and SARA show faithful reconstructions, despite noticeable wavelet artefacts around the central point source, which is challenging to recover for all algorithms. While no significant difference between the two methods can be noticed in logarithmic scale, 
the higher SNR of the SARA reconstruction implies that it recovers high intensity values better than its unconstrained variant.
BM3D-based PnP algorithms (uPnP\textsubscript{BM3D} and cPnP\textsubscript{BM3D}) exhibit mild 
stripe-like artefacts 
(see top right zoom regions of panels (e) and (g)), 
and the discontinuities around the central black hole form significant rectangular shapes. These artefacts could be induced by the square patches selection and hard thresholding strategies within the BM3D algorithm. We recall that these algorithms were fine-tuned to achive best performance.
PnP algorithms with the proposed denoisers AIRI and cAIRI yield better reconstructions than other methods, with little noticeable visual difference between each method despite variations in the reconstruction metrics. However, 
AIRI\textsubscript{OAID} and AIRI\textsubscript{MRID}
do not recover high intensities as well as their constrained variants cAIRI\textsubscript{OAID} and cAIRI\textsubscript{MRID}. This is consistent with the results obtained with uSARA and SARA. Interestingly, the visual results of the OAID and MRID AIRI algorithms are very similar, with the MRID variants  
exhibiting discontinuities around the bright structures.
This may be due to the 
lack of abrupt intensities changes
in the MRI knee images compared to the astronomical images in the training datasets, as shown in Fig.~\ref{fig:exponentiated}.

Notable differences emerge at the level of residual dirty images, where 
MRID-based algorithms and SARA
exhibit slight background misestimation (panels (t)-(v)).
With the same set of denoisers plugged in, the constrained (panels (t) and (x)) and unconstrained (panels (v) and (w)) algorithms yield similar residual images. Also, the denoisers trained with OAID tend to preserve data fidelity better than the ones trained with MRID.
Meanwhile, the residual dirty image of cPnP\textsubscript{BM3D} (panel (s)) is the closest to pure noise in large, smooth regions compared to others, thus maintaining better data fidelity.
However, its absolute error at relatively brighter and compact structures is also higher than for other algorithms.
This observation also indicates that faint residual images do not necessarily correlate with accurate reconstruction.

\begin{figure}
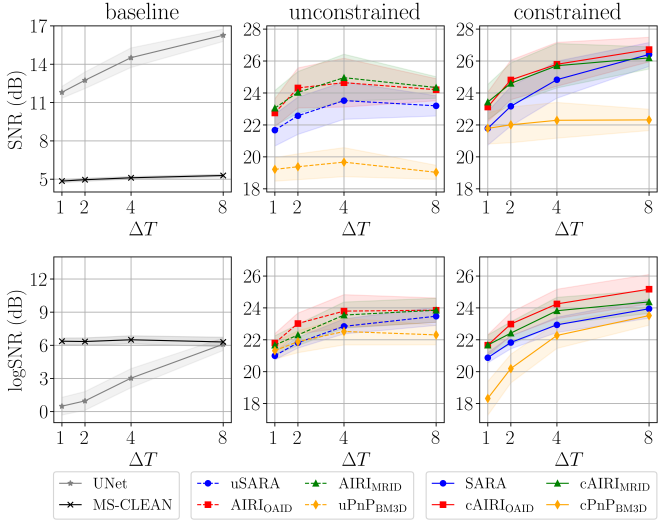

 \centering
 \includegraphics[width=0.49\textwidth]{exp_results/simulation_newscripts/snr_new_scr_2.png} \\
 \includegraphics[width=0.49\textwidth]{exp_results/simulation_newscripts/logsnr_new_scr_2.png}
\caption{
Reconstruction metrics as a function of the observation time. From left to right: i) MS-CLEAN and UNet as baseline algorithms; ii) unconstrained algorithms based on Algorithm~\ref{algo:pnp_fb}; iii) constrained algorithms based on Algorithm~\ref{algo:pnp_pd}.
Top row gives SNR metrics and bottom row gives logSNR metric. Each point is an average value over 15 inverse problems. The shaded areas represent the 95\% confidence intervals.}
\label{fig:metrics}
\end{figure}

Fig.~\ref{fig:metrics} gives the average SNR and logSNR estimated with various algorithms. On these plots, each point is an average over the 15 inverse problems considered for each observation time, error bars representing the 95\% confidence intervals. 
Both uSARA and SARA optimisation algorithms significantly outperform the baseline MS-CLEAN algorithm and the UNet for the two considered metrics. AIRI and cAIRI improve by 1 to 2 dB over their pure optimisation counterparts uSARA and SARA. The constrained variants perform better than the unconstrained ones, particularly on the SNR. In particular, cAIRI outperforms AIRI.  We explain this phenomenon by the good ability of constrained methods to recover high intensity values after fewer iterations than their unconstrained variants.

\begin{figure}
\centering
\small
\begin{tabular}{@{\hspace{0.\tabcolsep}}c @{\hspace{0.1\tabcolsep}}c @{\hspace{0.1\tabcolsep}}c
@{\hspace{0.0\tabcolsep}}l } 
\includegraphics[width=0.145\textwidth]{exp_results/simulation_newscripts/zoom_airi_mrid.pdf} &
\includegraphics[width=0.145\textwidth]{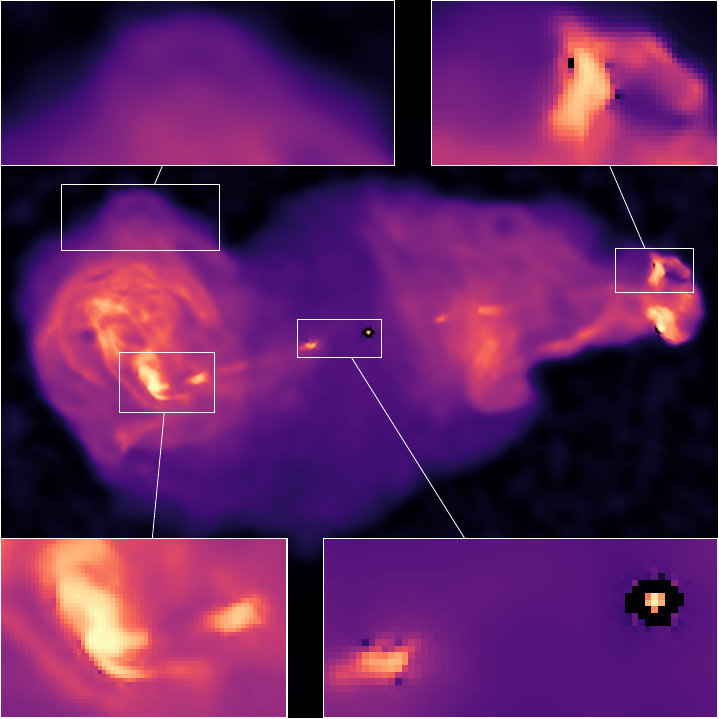} &
\includegraphics[width=0.145\textwidth]{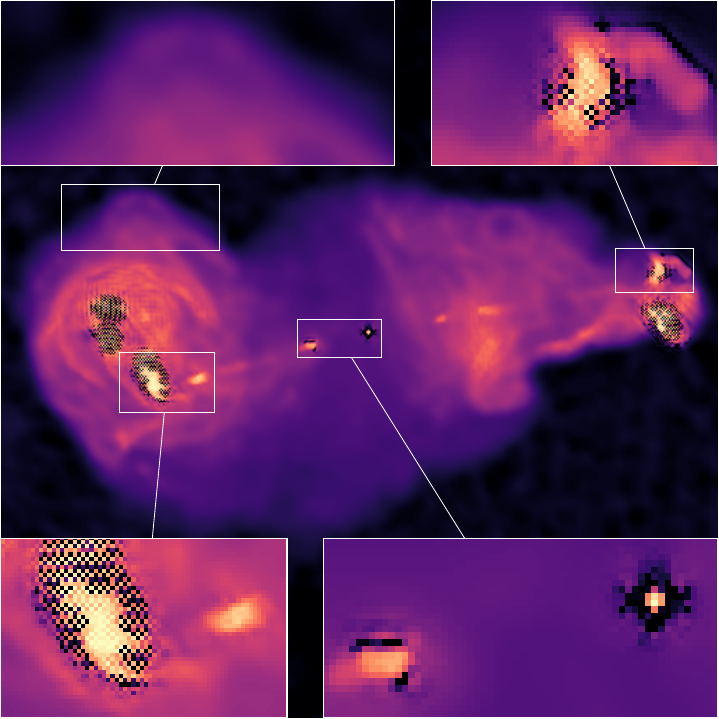} &
\includegraphics[width=0.04\textwidth]{exp_results/512_results/colorbar_vertical_new.png} \\ %
(a) Adaptive Peak & (b) Exact Peak & (c) Dirty Peak & \\
$(26.51 {\rm dB}, 25.25 {\rm dB})$  & $(26.51{\rm dB}, 25.25{\rm dB})$  & $(2.58{\rm dB}, 12.22{\rm dB})$ & \\
\end{tabular}
\caption{AIRI\textsubscript{MRID} reconstructions of 
the same simulated measurement as
in Fig.~\ref{fig:3c353_dt8} with different image peak value $\alpha$ estimation schemes: (a) adaptive peak value estimation; (b) using the exact peak value of the groundtruth (1.00); (c) using the maximum intensity of the back-projected image (61.18). The metrics (SNR, logSNR) associated with each reconstruction are shown beneath each image.}
\label{fig:3c353_dt8_peak}
\end{figure}

In Table~\ref{tab:timings}, we give the computational time of the denoising step for different denoisers. When executed on a GPU, AIRI denoisers demonstrate a 20-fold increase in inference speed compared to the proximity operator in SARA and uSARA. Conversely, the BM3D algorithm exhibits slow inference, raising doubt on its practicality. These results underscore the efficiency of DNN-based denoisers, benefiting seamlessly from GPU acceleration.

\begin{table}
\centering
\small
\begin{tabular}{@{\hskip 0pt}l @{\hskip 8pt} c @{\hskip 5pt} c @{\hskip 5pt} c @{\hskip 5pt} c @{\hskip 0pt}}
 & AIRI (GPU) &  AIRI (CPU)  & SARA  &  BM3D \\ \hline 
Time (s.) & \bf{0.05 $\pm$ 0.02}  &  7.92 $\pm$ 0.52 & 1.31 $\pm$ 1.12 & 15.08 $\pm$ 0.40 \\ 
\hline
\end{tabular}
\caption{Average time (in seconds) taken by the denoising step for solving 60 simulated inverse problems with different denoisers $\operatorname{D}$ when plugged in 
Algorithms~\ref{algo:pnp_fb} and \ref{algo:pnp_pd}.}
\label{tab:timings}
\end{table}

\subsection{Effectiveness of the adaptive peak estimation approach} 
In our simulated experiments, the true peak value in the sought reconstructed image is known (it is equal to 1) and the peak value of the back-projected images vary from 30 to 68 across measurement settings. To validate the adaptive peak estimation scheme proposed in Section~\ref{subsect:denoiser_selection}, we ran the simulated experiments where this approach is compared to two alternatives, where $\widetilde{\alpha}$ is set to either (i) the exact maximum intensity of the groundtruth image, or (ii) the maximum intensity of the back-projected image. 
Results in Fig.~\ref{fig:3c353_dt8_peak} show that the proposed adaptive peak estimation scheme yields reconstructions with the same quality as if the true peak value were known. If the estimated peak value differs too much from the real peak value, like in the situation from Fig.~\ref{fig:3c353_dt8_peak}(c), the reconstruction quality can degrade significantly.
We attribute this suboptimality to the fact that the denoisers are applied to images with peak value far below 1, a range for which they have not been trained.

\subsection{Robustness to denoiser realisation and model uncertainty}
\label{subsection:robustness}

To investigate the robustness of the proposed approach, and following the discussion in Section~\ref{sect:model_uncertainty}, we trained $K=15$ denoisers per training dataset (i.e., OAID and MRID) under identical experimental conditions, but with different random seeds for the (pseudo) random processes during training. Each denoiser was then used in Algorithm~\ref{algo:pnp_fb} and Algorithm~\ref{algo:pnp_pd}, resulting in two sets of 15 solutions, one for each training dataset. For more details, see Section~\ref{sect:model_uncertainty}.

\begin{figure}
\small
\centering
\begin{tabular}{@{\hspace{0.\tabcolsep}} c @{\hspace{0.3\tabcolsep}} c @{\hspace{0.\tabcolsep}} c @{\hspace{0.\tabcolsep}} c @{\hspace{-0.3\tabcolsep}} c}

\rotatebox[origin=c]{90}{cAIRI\textsubscript{OAID}} %
 & 
 \raisebox{ %
 -0.5\height %
 }{%
 \includegraphics[width=0.14\textwidth]{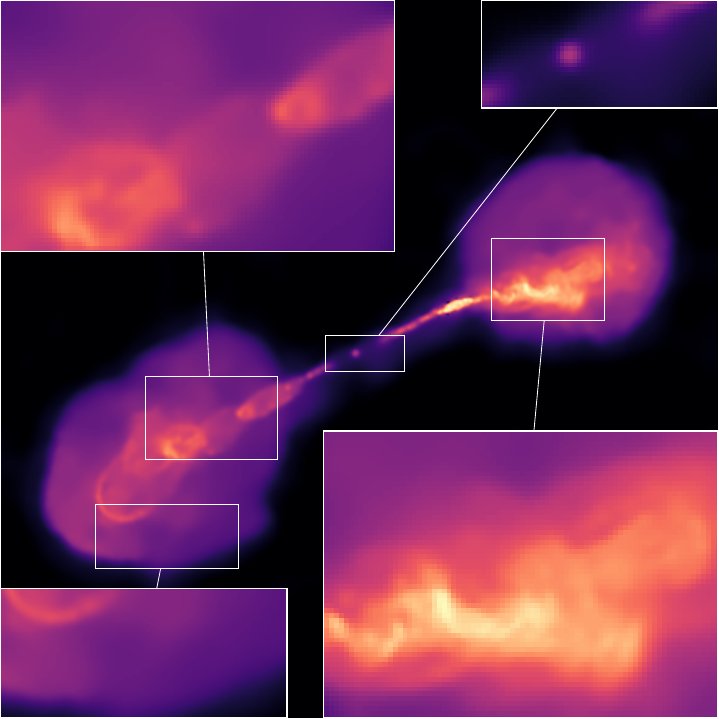} %
 } %
 &
 \raisebox{ %
 -0.5\height %
 }{%
\includegraphics[width=0.14\textwidth]{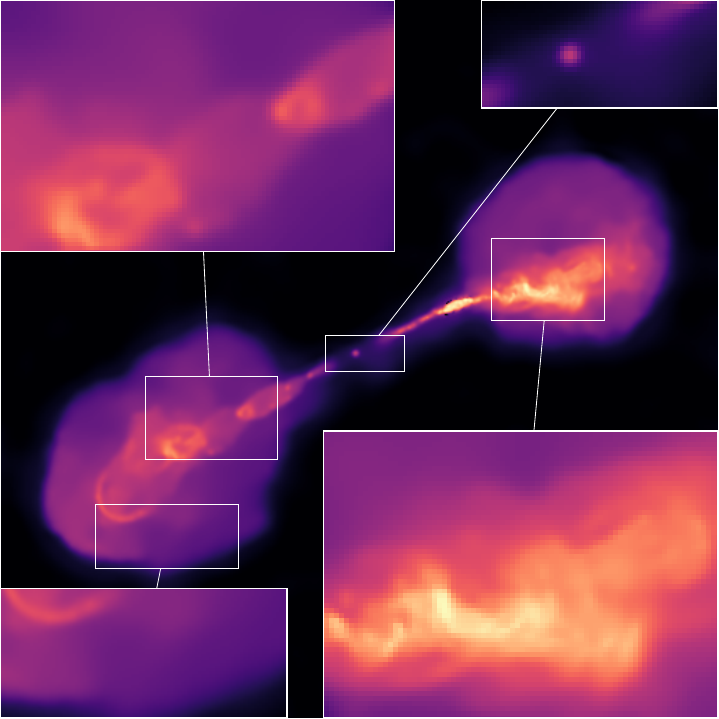} %
} %
&
 \raisebox{ %
 -0.5\height %
 }{%
\includegraphics[width=0.14\textwidth]{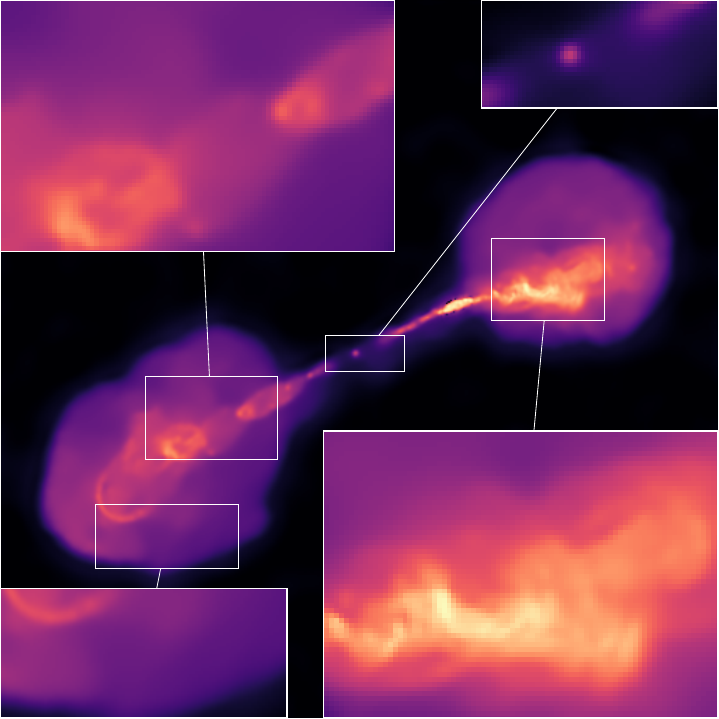} %
} %
&
 \raisebox{ %
 -0.5\height %
 }{%
\includegraphics[width=0.03\textwidth]{exp_results/512_results/colorbar_vertical_new.png} %
} %
\\
& (a) $\widehat{x}_1$ & (b) $\widehat{x}_2$ & (c) $\widehat{x}_3$ \\
&  $(27.16 {\rm dB}, 25.44 {\rm dB})$ & $(27.45 {\rm dB}, 24.85 {\rm dB})$ & $(27.20 {\rm dB}, 24.61 {\rm dB})$ & \\
\rotatebox[origin=c]{90}{cAIRI\textsubscript{OAID}} %
& 
 \raisebox{ %
 -0.5\height %
 }{%
\includegraphics[width=0.14\textwidth]{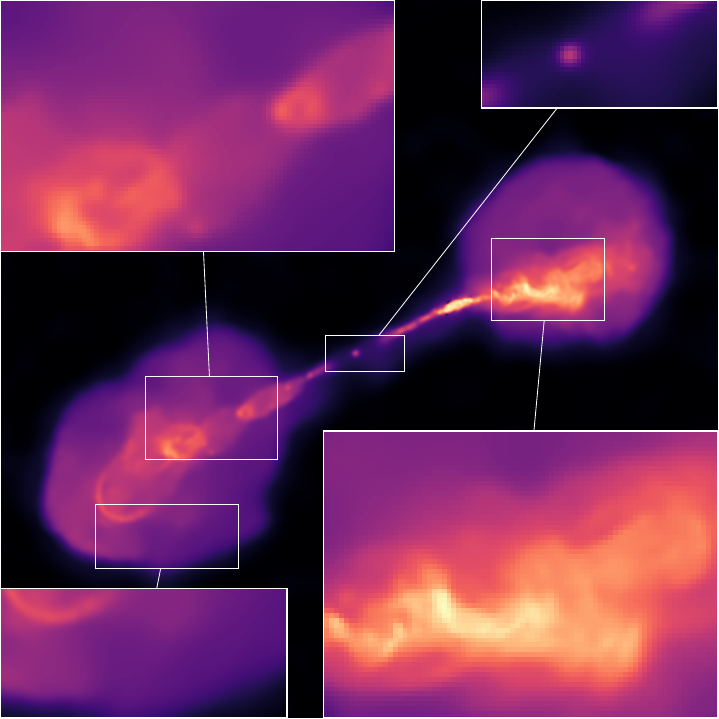} %
} %
&
 \raisebox{ %
 -0.5\height %
 }{%
\includegraphics[width=0.14\textwidth]{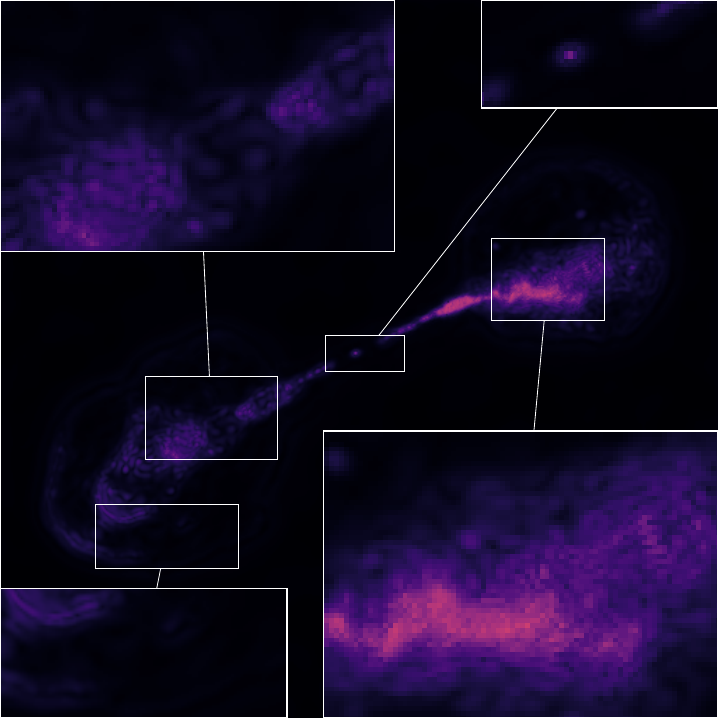} %
} %
&
 \raisebox{ %
 -0.5\height %
 }{%
\includegraphics[width=0.14\textwidth]{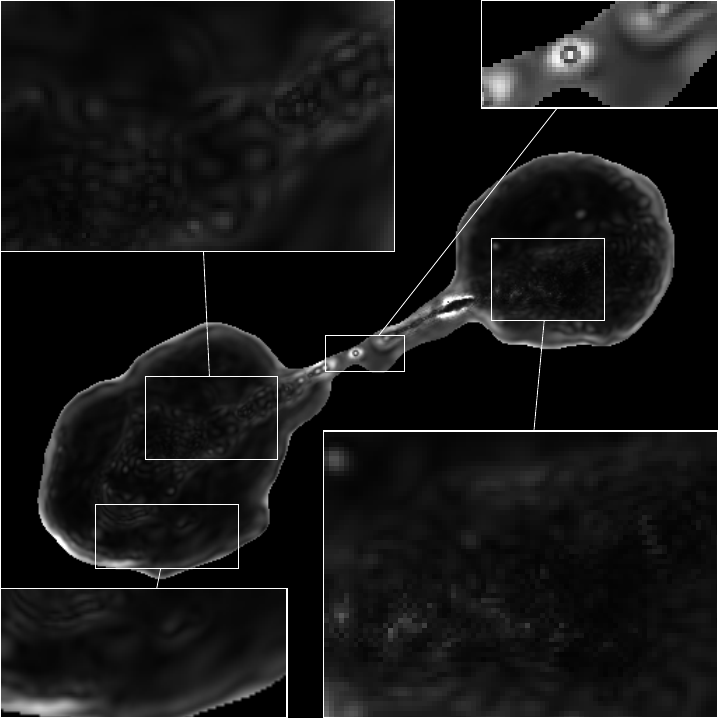} %
} %
& 
 \raisebox{ %
 -0.5\height %
 }{%
 \includegraphics[width=0.03\textwidth]{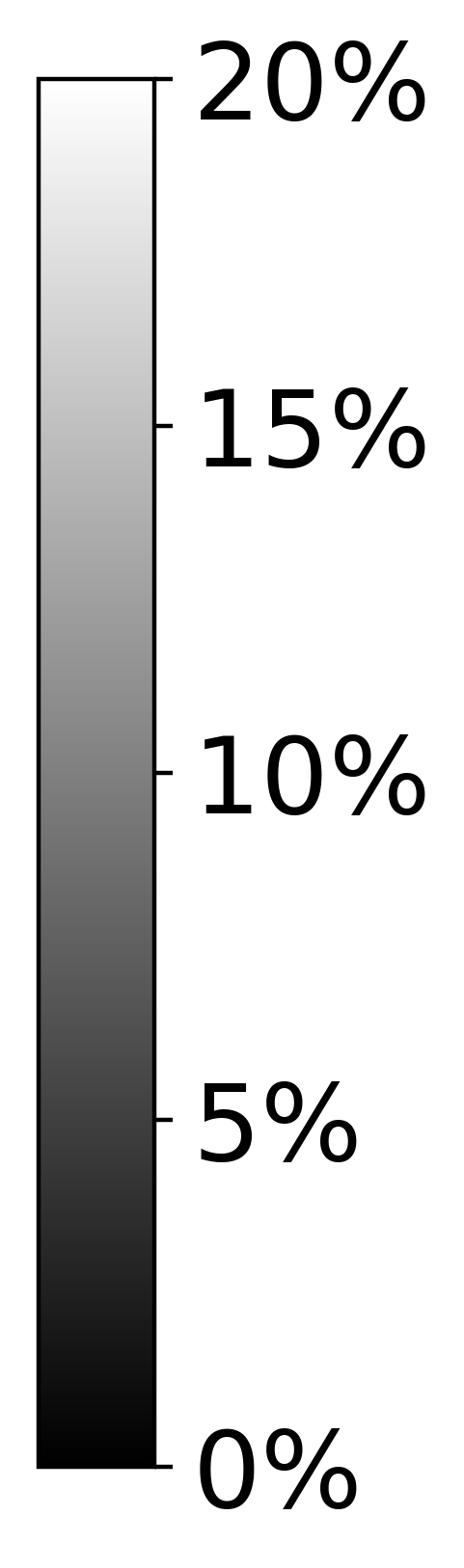} %
} %
\\
& (d) mean image %
& (e) std. image %
& (f) std/mean %
\\
 & $(27.86 {\rm dB}, 25.10 {\rm dB})$ & 
 & & \\

\rotatebox[origin=c]{90}{AIRI\textsubscript{OAID}} %
& 
 \raisebox{ %
 -0.5\height %
 }{%
\includegraphics[width=0.14\textwidth]{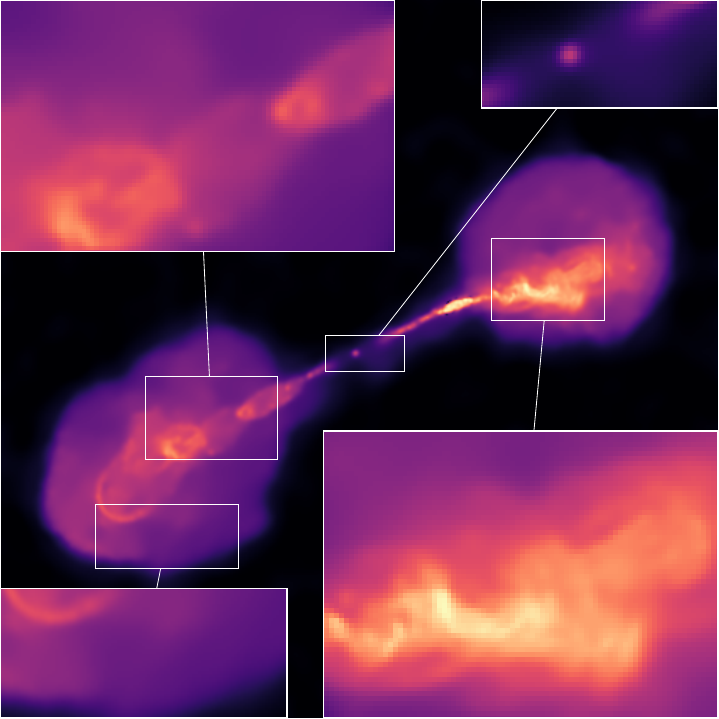} %
} %
&
 \raisebox{ %
 -0.5\height %
 }{%
\includegraphics[width=0.14\textwidth]{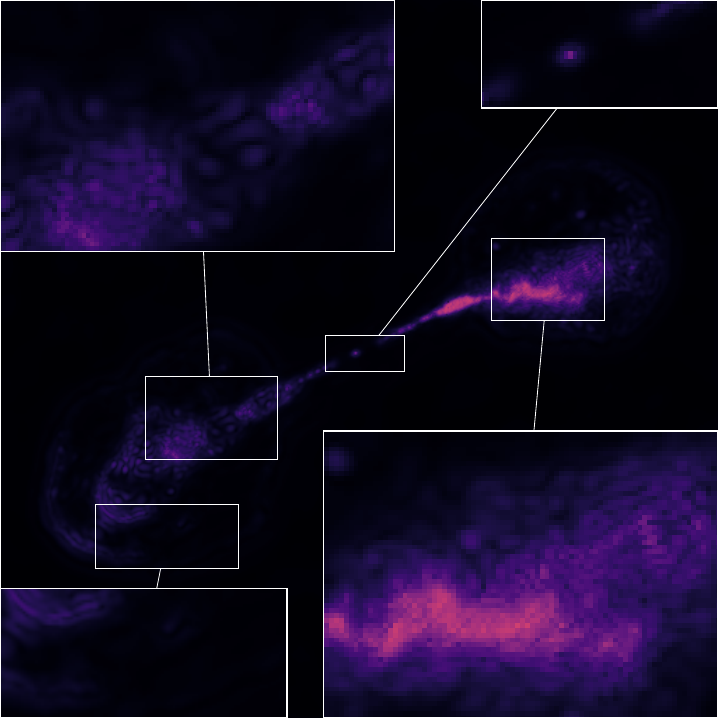} %
} %
&
 \raisebox{ %
 -0.5\height %
 }{%
\includegraphics[width=0.14\textwidth]{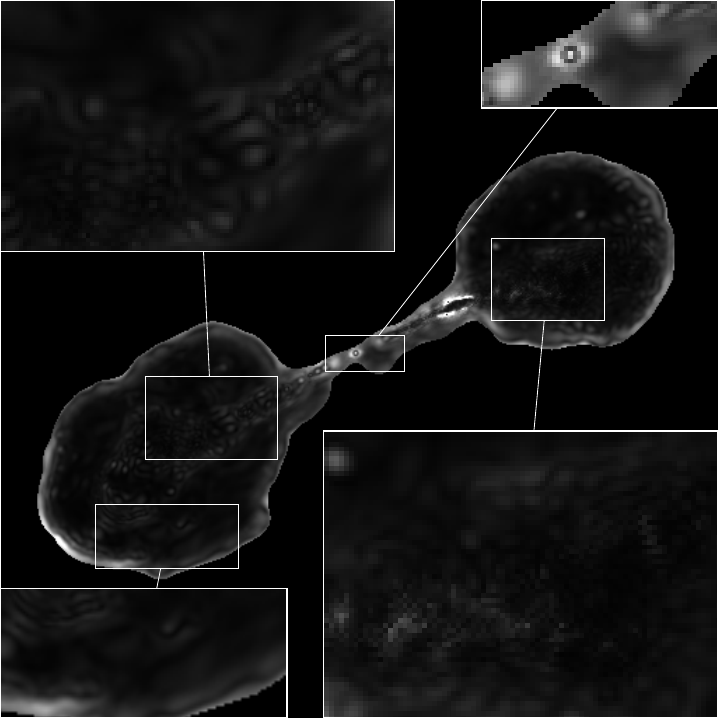} %
} %
& 
 \raisebox{ %
 -0.5\height %
 }{%
\includegraphics[width=0.03\textwidth]{exp_results/robustness_newscripts/per_ver.png} %
} %
\\
& (g) mean image & (h) std. image & (i) std/mean \\
&  $(27.70 {\rm dB}, 25.45 {\rm dB})$ &  & & \\
  
\rotatebox[origin=c]{90}{cAIRI\textsubscript{MRID}} %
& 
 \raisebox{ %
 -0.5\height %
 }{%
\includegraphics[width=0.14\textwidth]{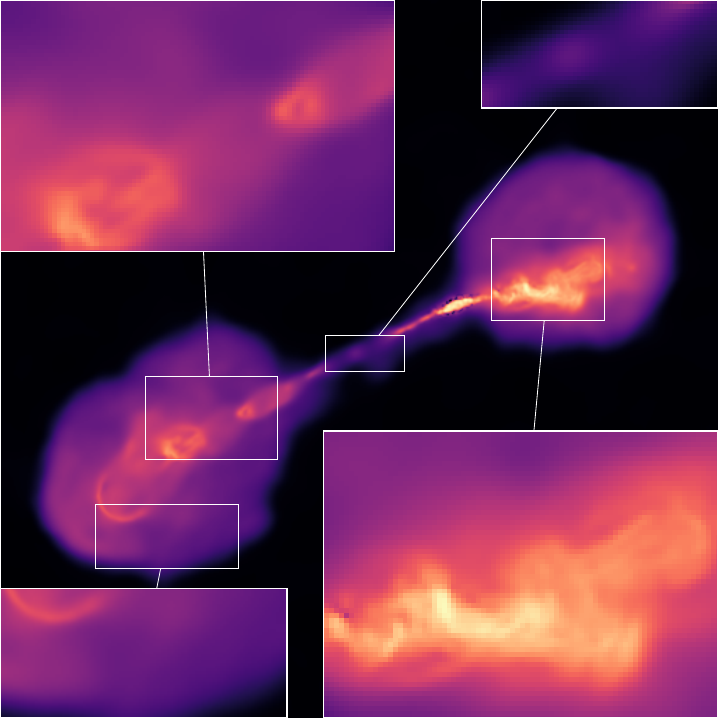} %
} %
&
 \raisebox{ %
 -0.5\height %
 }{%
\includegraphics[width=0.14\textwidth]{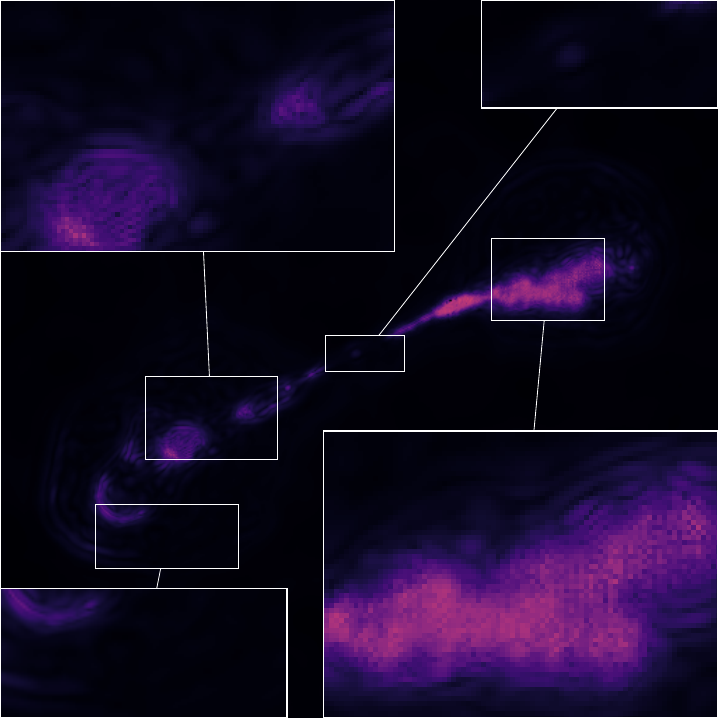} %
} %
&
 \raisebox{ %
 -0.5\height %
 }{%
\includegraphics[width=0.14\textwidth]{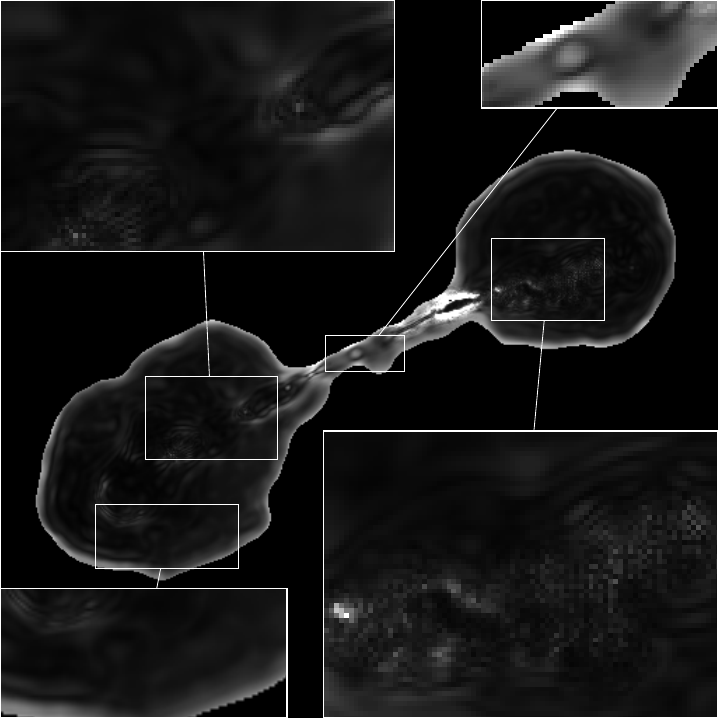} %
} %
& 
 \raisebox{ %
 -0.5\height %
 }{%
\includegraphics[width=0.03\textwidth]{exp_results/robustness_newscripts/per_ver.png} %
} %
\\
& (j) mean image & (k) std. image & (l) std/mean \\
&  $(27.45 {\rm dB}, 25.86 {\rm dB})$ &  & & \\

\rotatebox[origin=c]{90}{AIRI\textsubscript{MRID}} %
& 
 \raisebox{ %
 -0.5\height %
 }{%
\includegraphics[width=0.14\textwidth]{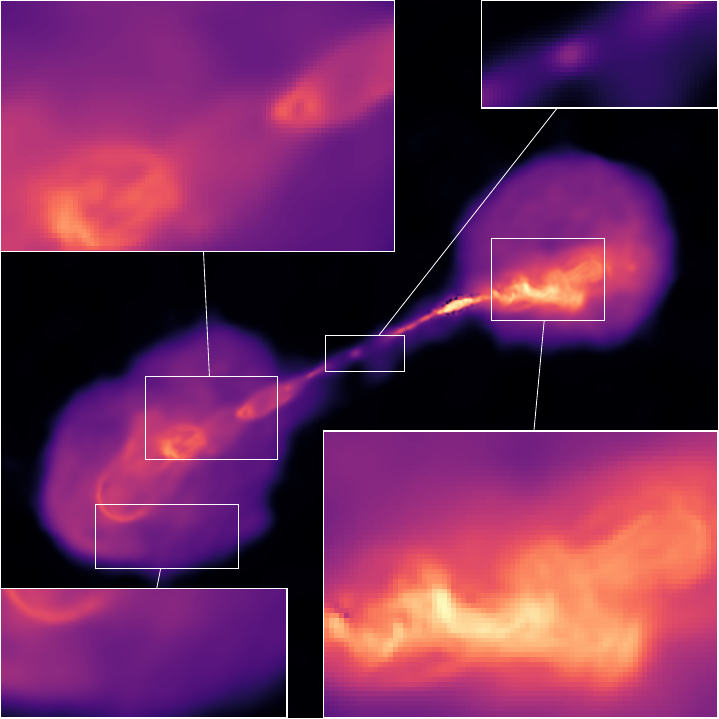} %
} %
&
\raisebox{-0.5\height}{%
\includegraphics[width=0.14\textwidth]{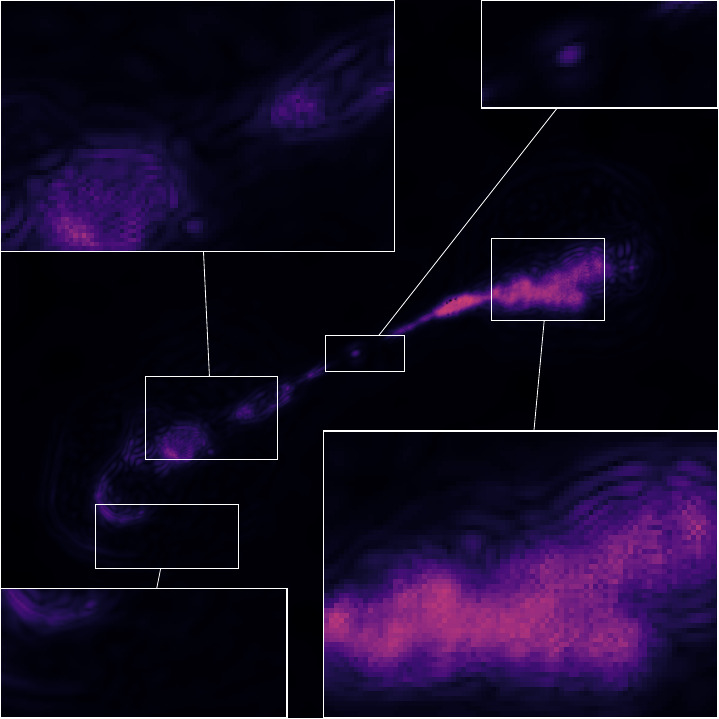} %
} %
&
\raisebox{-0.5\height}{%
\includegraphics[width=0.14\textwidth]{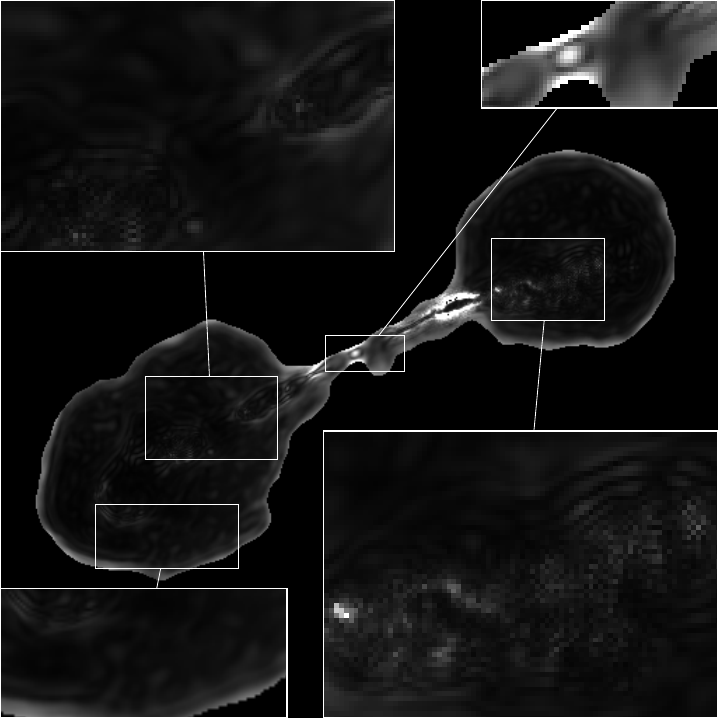} %
} %
& \raisebox{-0.5\height}{%
\includegraphics[width=0.03\textwidth]{exp_results/robustness_newscripts/per_ver.png} %
} %
\\
& (m) mean image & (n) std. image & (o) std/mean \\
&   $(27.46 {\rm dB}, 26.32 {\rm dB})$ &  & & \\
  
\end{tabular}
\caption{Sensitivity of the imaging result to the training conditions of the denoiser realisation. (a)-(c) show image reconstructions with cAIRI\textsubscript{OAID} for 3 denoisers, each trained with a different random seed. 
(d)-(f) show the mean image, standard deviation and relative standard deviation maps for reconstructions obtained with 15 networks trained on the astronomical dataset and plugged in the cAIRI\textsubscript{OAID} algorithm. Next rows show the same results but obtained with other AIRI variants. Insets show magnified areas of the image.
Images (a)-(c), the mean images and standard deviation images share the same colour bar.
}
\label{fig:3c353_uncertainty}
\end{figure}

Fig.~\ref{fig:3c353_uncertainty} shows visual results for Algorithm~\ref{algo:pnp_fb}
and Algorithm~\ref{algo:pnp_pd}
with different denoisers realisations.
We notice that the reconstructions from each random seed, as well as the mean image, show no clear difference, except around the point source at the centre of the image. 

The relative standard deviation, which highlights regions with mean pixel intensities surpassing the estimated noise levels as defined by \eqref{eq:heuristic}, generally demonstrates that the empirical standard deviation remains modest, typically less than 10\% of the mean intensity across most of the image. Notably, this deviates around the central source where the standard deviation increases to about 20\% of the mean intensity. This observation is particularly relevant as the central source estimation is often unsatisfying, most algorithms hallucinating a black ring around it. Our cAIRI algorithm demonstrates better performance in  
estimating central sources, especially when combined with OAID AIRI denoisers, which gives a smaller relative standard deviation around the central sources (top right zoom in Fig.~\ref{fig:3c353_uncertainty} (l).

\begin{figure}
\centering
\includegraphics[width=0.49\textwidth]{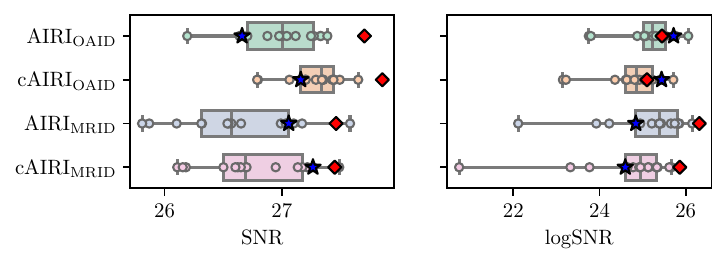}
\caption{Results for 15 different runs of the AIRI and cAIRI algorithms for simulated measurements with $\Delta T=4$h. The red diamond indicates the reconstruction metric of the mean image of the 15 reconstructions. The blue star indicates the models that are used in our deterministic experiments (in Fig.~\ref{fig:3c353_dt8} and~\ref{fig:metrics}).}
\label{fig:metrics_robustness}
\end{figure}

Fig.~\ref{fig:metrics_robustness} gives the reconstruction metrics for 15 denoiser realisations in the context of a single inverse problem at $\Delta T=4\text{h}$, with selected samples depicted in Fig.~\ref{fig:3c353_uncertainty}. The reconstruction quality offered by different denoiser realisations spans about 1.5dB in SNR and 4dB in logSNR.
As illustrated in Fig.~\ref{fig:3c353_uncertainty}, the variations in SNR and logSNR metrics among the reconstructed images are not readily discernible at the image level. 

In general, AIRI tends to provide better logSNR metrics than cAIRI, while the opposite holds for SNR, and reconstructions produced with cAIRI exhibit less variation in SNR compared to those generated with AIRI for the selected measurement. 
Interestingly, the SNR of both the mean images is slightly higher than the mean SNR, indicating that the mean image offers enhanced fidelity. 
Yet, computing this image necessitates multiple denoiser realizations and algorithm runs.

Furthermore, we notice in Fig.~\ref{fig:3c353_uncertainty} that while algorithms relying on OAID-based denoisers tend to provide reconstructions with higher SNR, those relying on MRID-based denoisers tend to provide higher logSNR. Visually, we observe a similar trend as in the previous sections where MRID-based algorithms tend to yield less precise point sources, but these differences are mild.

As the overall variation in reconstruction quality remains small, these results suggest that our AIRI PnP algorithms are resilient to variations inherent to the training process.

\begin{figure*} \centering
\begin{tabular}{c @{\hspace{0.3\tabcolsep}} c}
\includegraphics[width=0.49\textwidth]{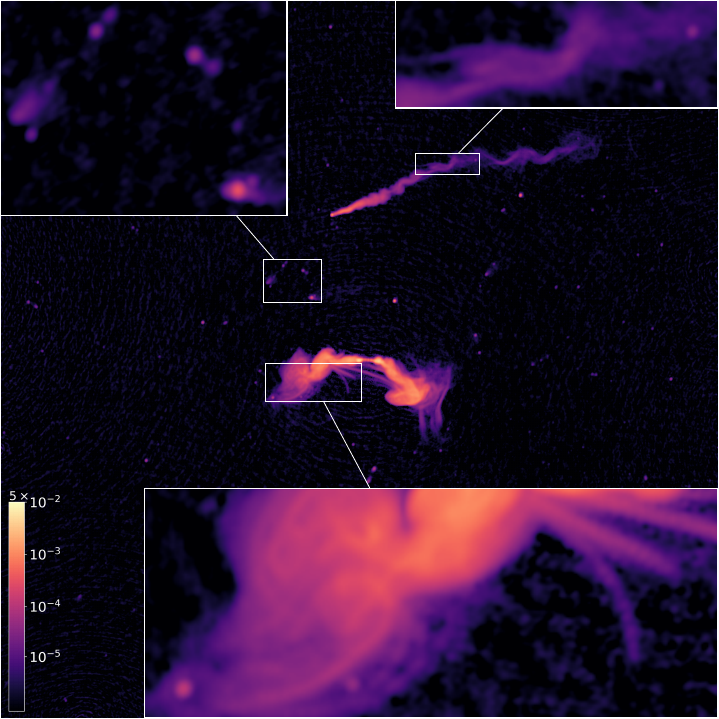} &
\includegraphics[width=0.49\textwidth]{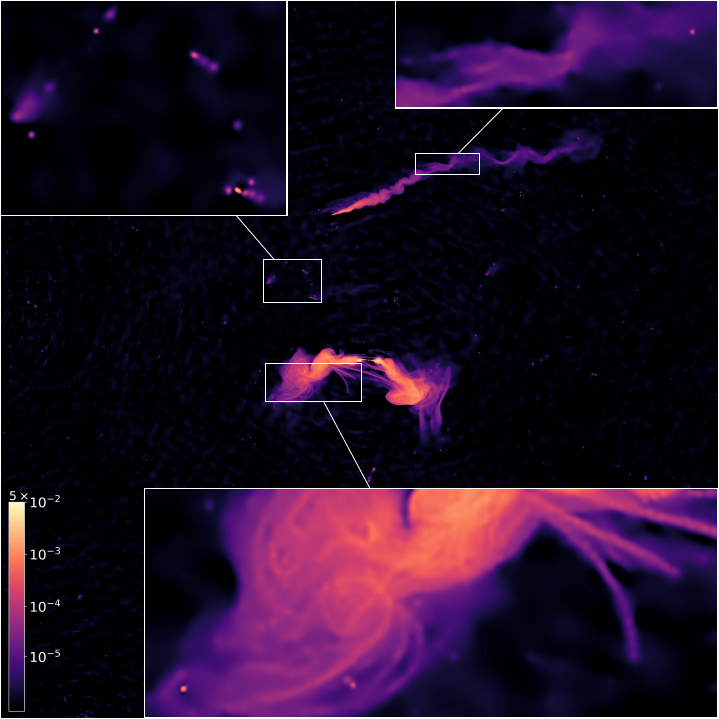} \\
(a) Baseline (MS-CLEAN)  & (b) Proposed (cAIRI\textsubscript{OAID})  \\
\end{tabular}
\caption{
Imaging of ESO 137-006 galaxy and other radio sources of interest from data acquired with the MeerKAT telescope with an image size of 2560$\times$2560 and a FoV of 1.19$\times$1.19 square degree:
(a) imaging result with the baseline MS-CLEAN algorithm; (b) imaging result with the proposed cAIRI\textsubscript{OAID} algorithm.}
\label{fig:meerkat_full_FOV}
\end{figure*}

\begin{figure} \centering
\begin{tabular}{@{\hspace{0.\tabcolsep}} c @{\hspace{0.15\tabcolsep}} c @{\hspace{0.\tabcolsep}}}
\multicolumn{2}{c}{
\includegraphics[width=0.238\textwidth]{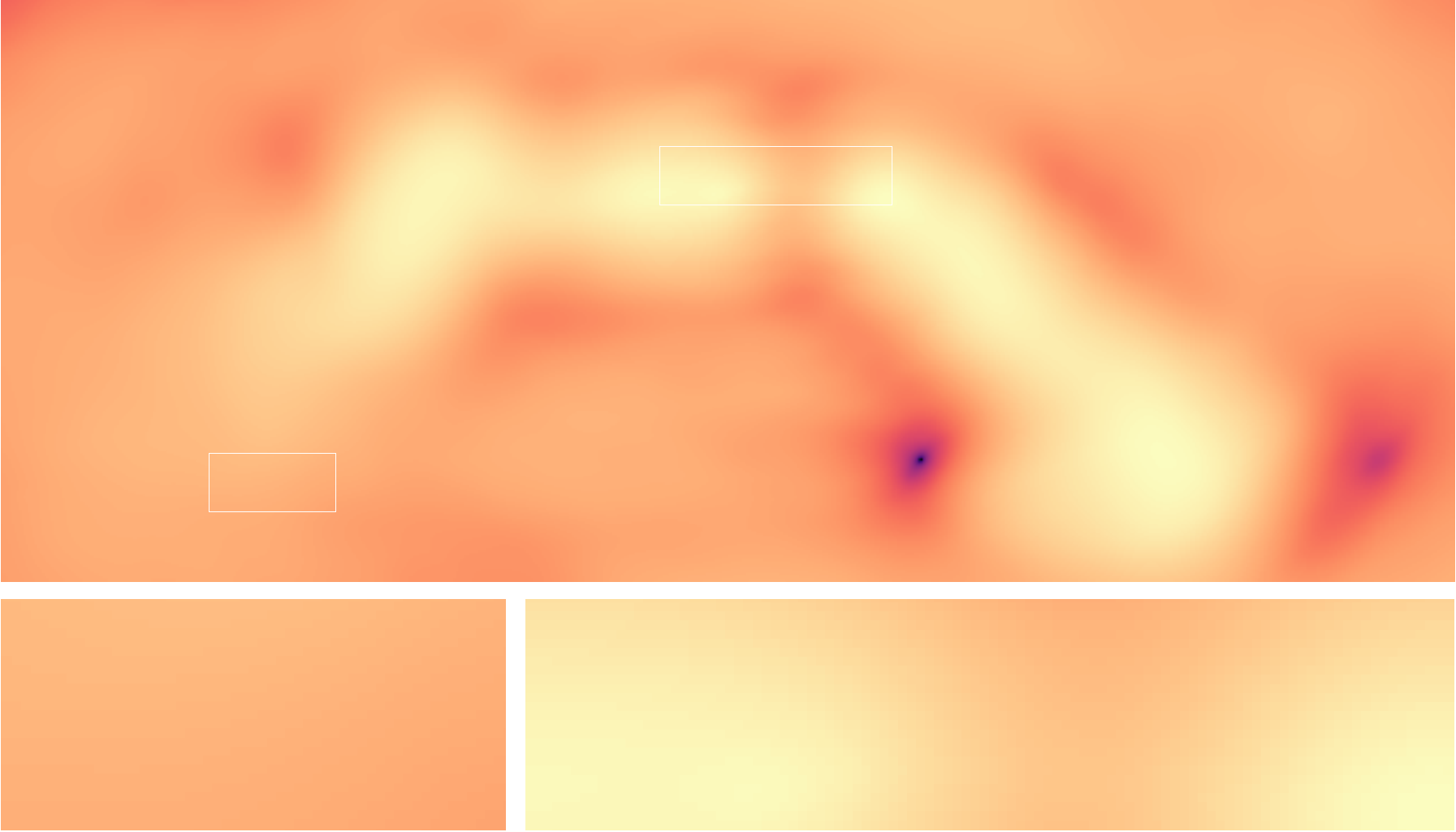}} \\
\multicolumn{2}{c}{(a) Back-projected image} \\
\hspace{0.2\tabcolsep}\includegraphics[height=0.03\textwidth]{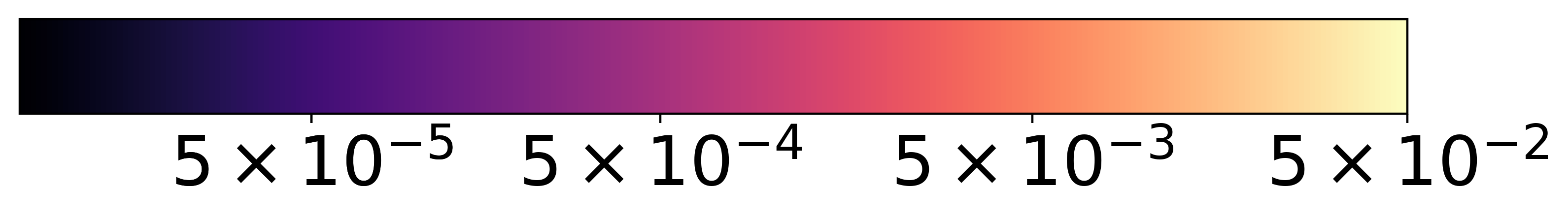} & 
\hspace{0.2\tabcolsep}\includegraphics[height=0.03\textwidth]{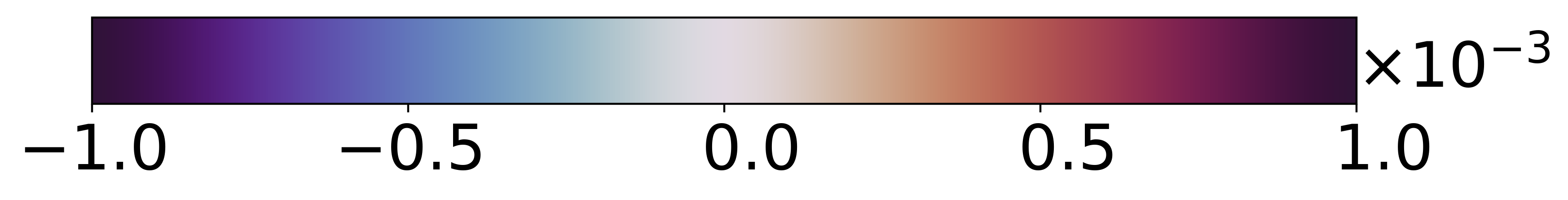}\\
\includegraphics[width=0.238\textwidth]{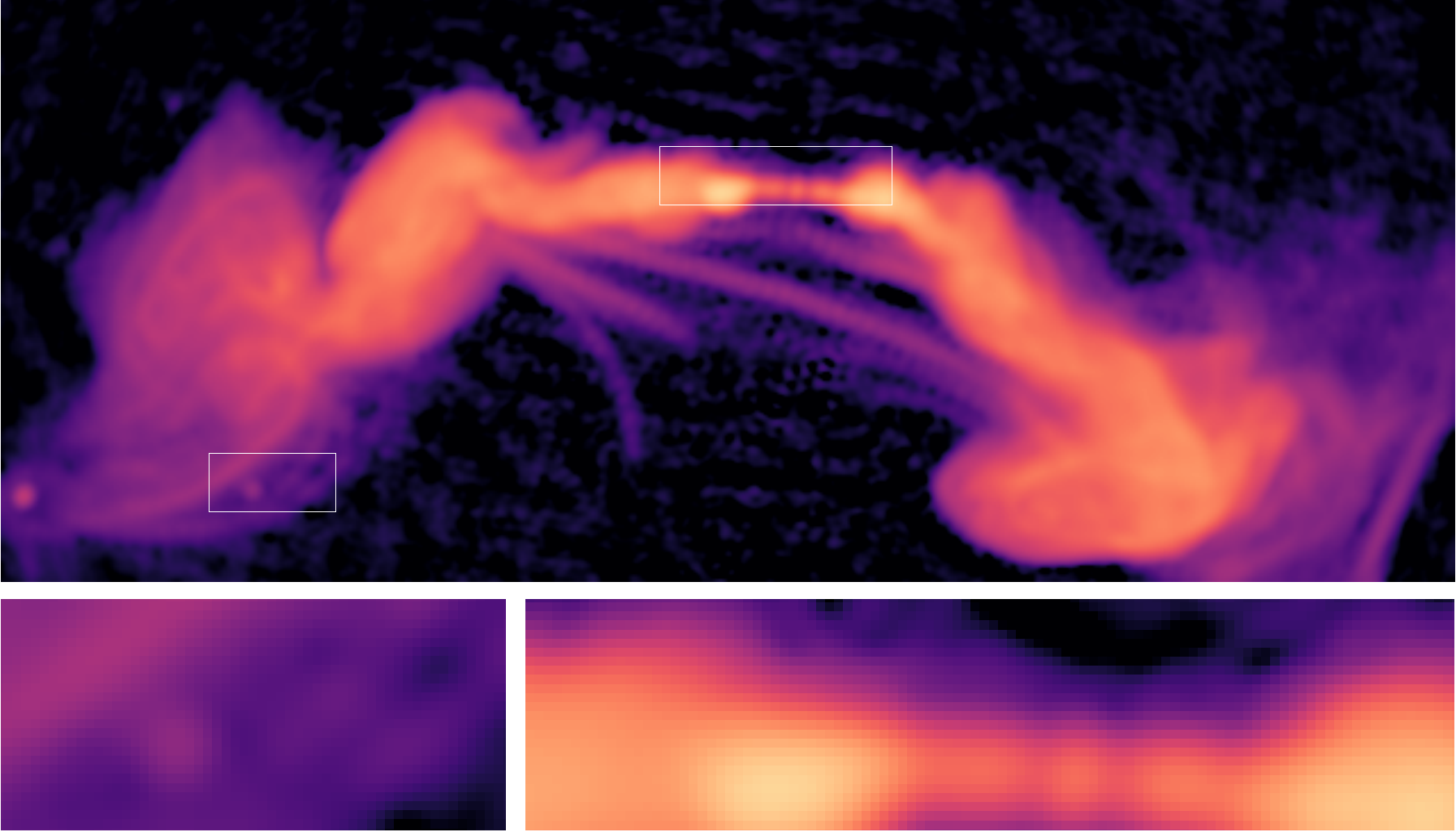} &
\includegraphics[width=0.238\textwidth]{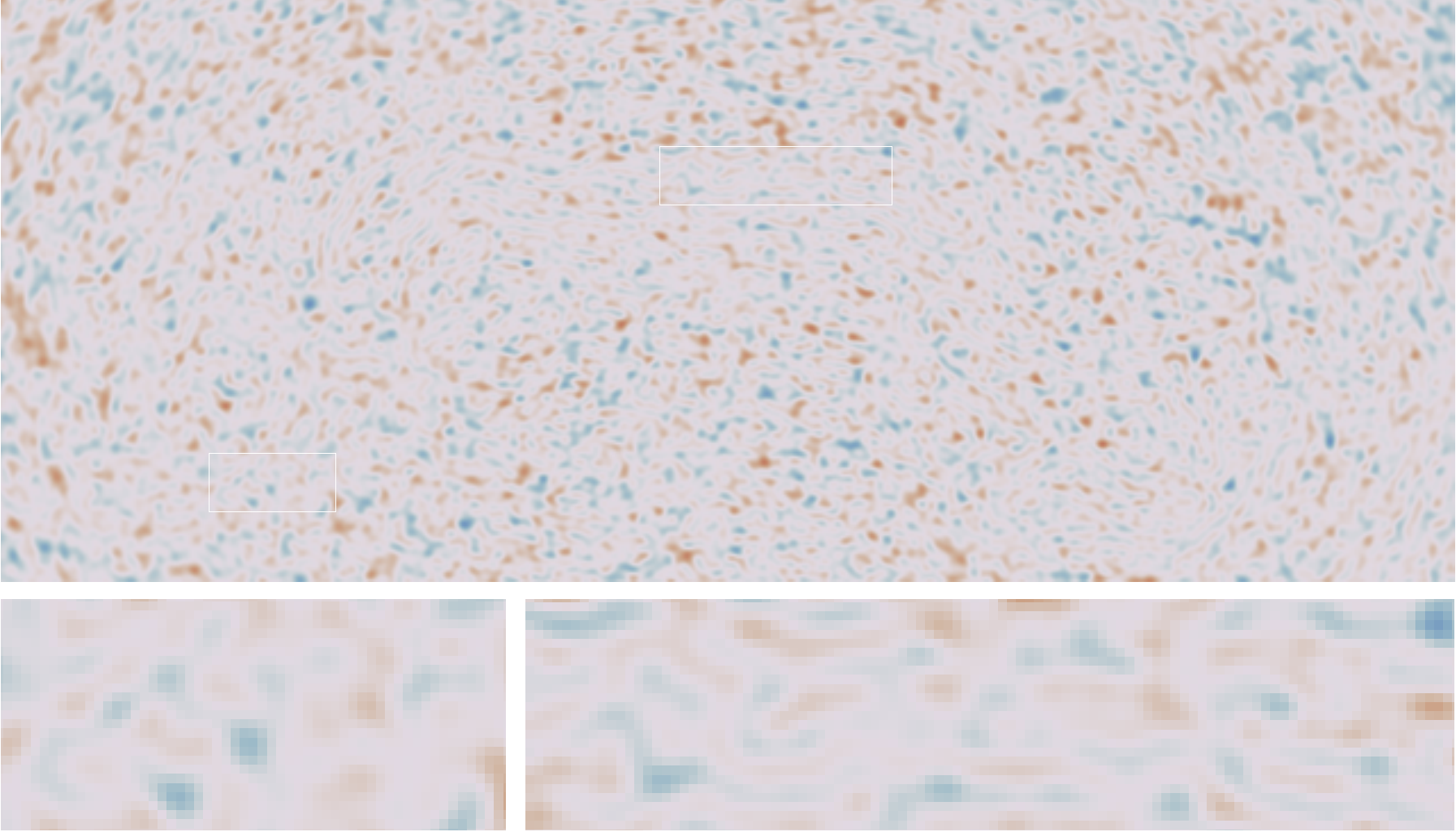} \\
(b) MS-CLEAN & (c) MS-CLEAN Residual \\
\includegraphics[width=0.238\textwidth]{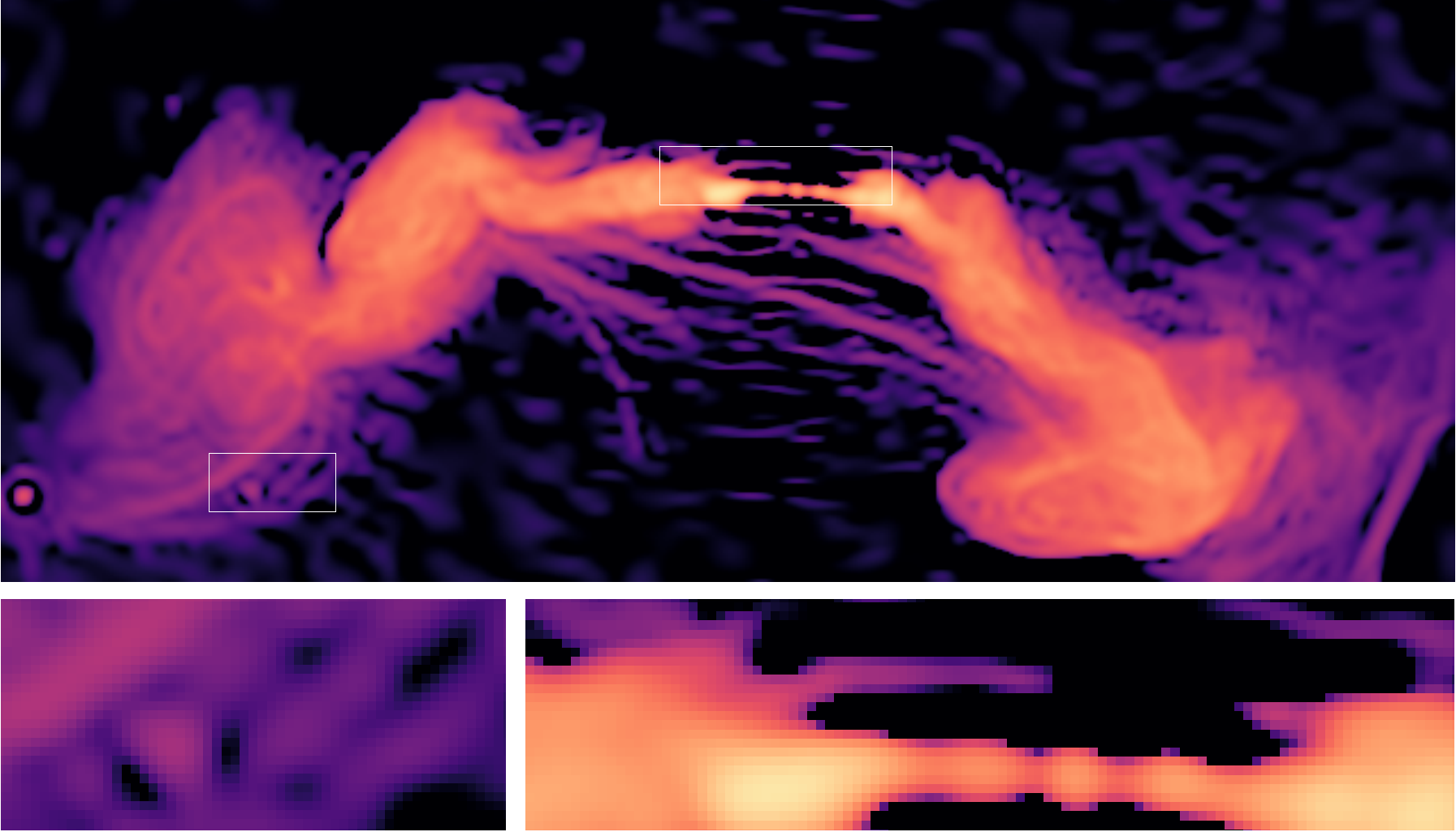} &
\includegraphics[width=0.238\textwidth]{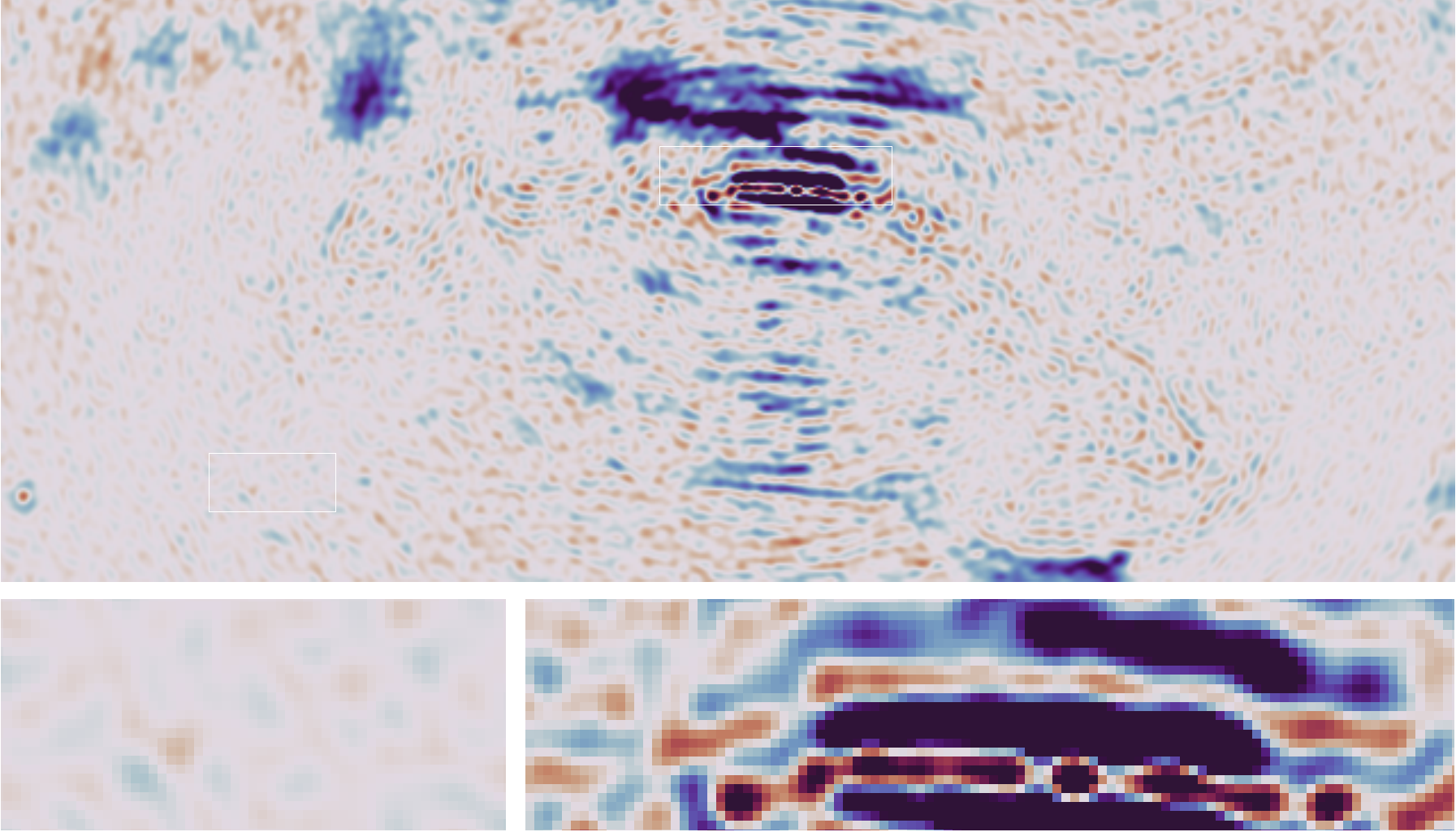} \\
(d) uSARA & (e) uSARA Residual \\
\includegraphics[width=0.238\textwidth]{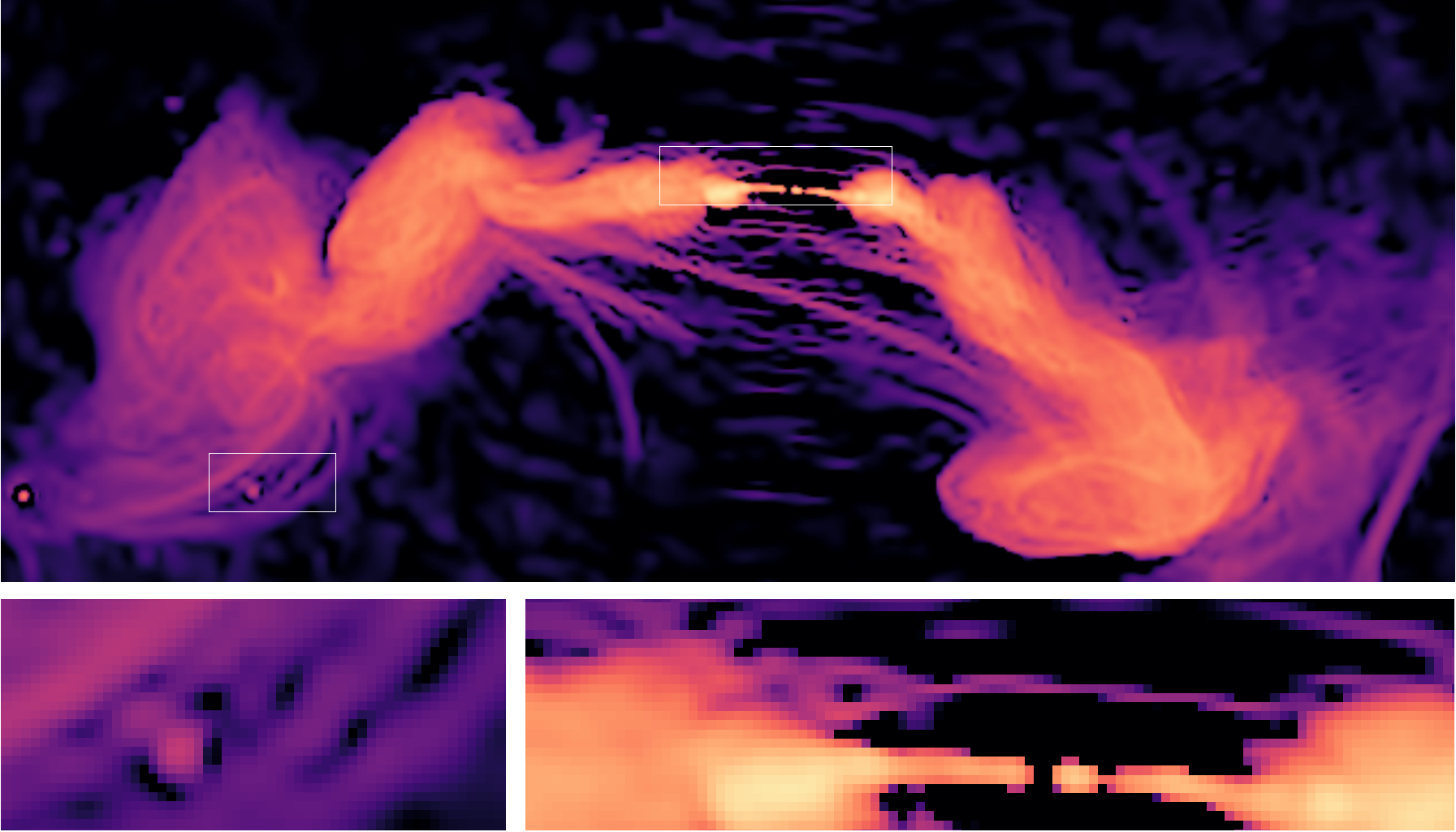} &
\includegraphics[width=0.238\textwidth]{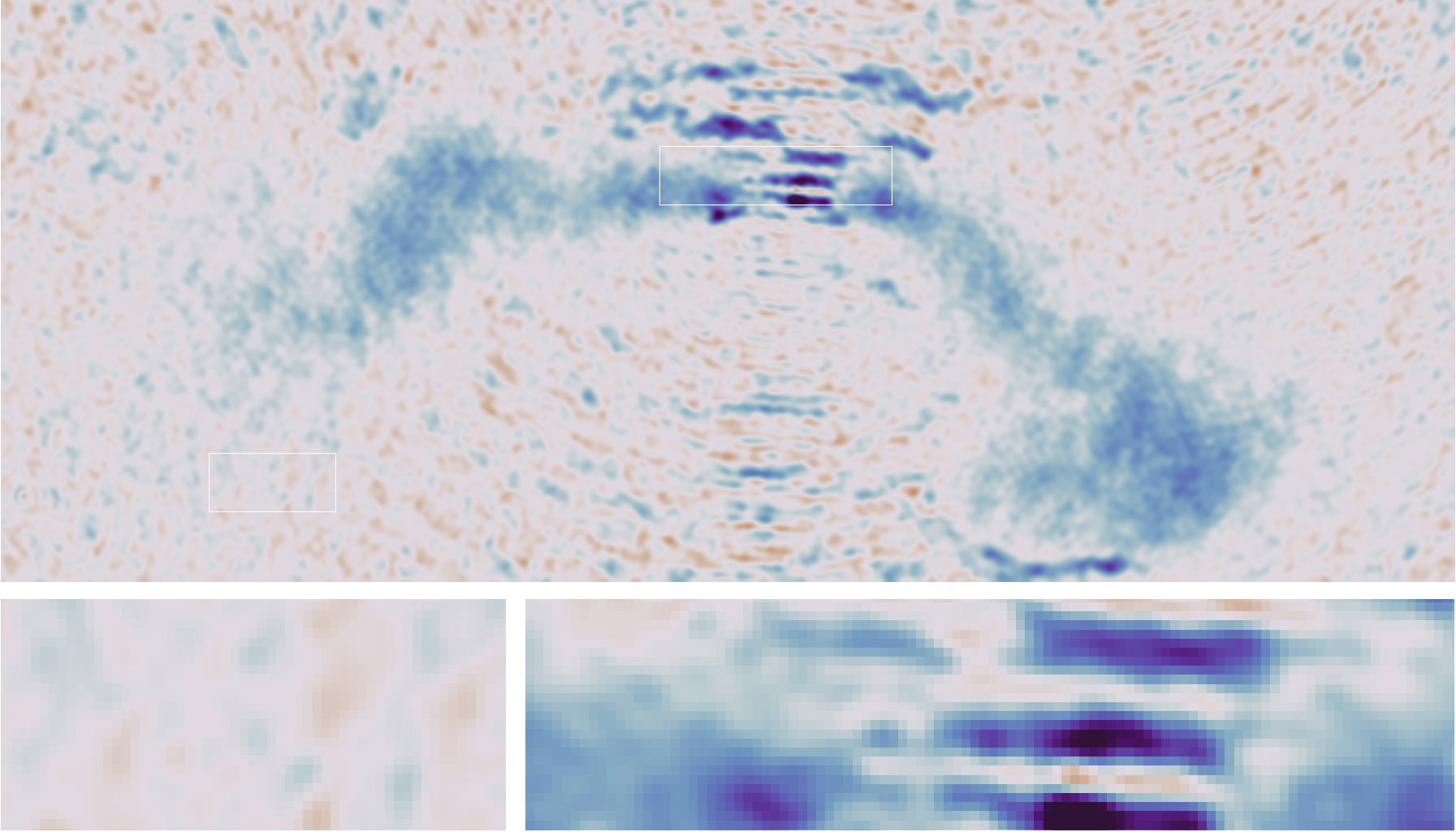} \\
(f) SARA & (g) SARA Residual \\
\includegraphics[width=0.238\textwidth]{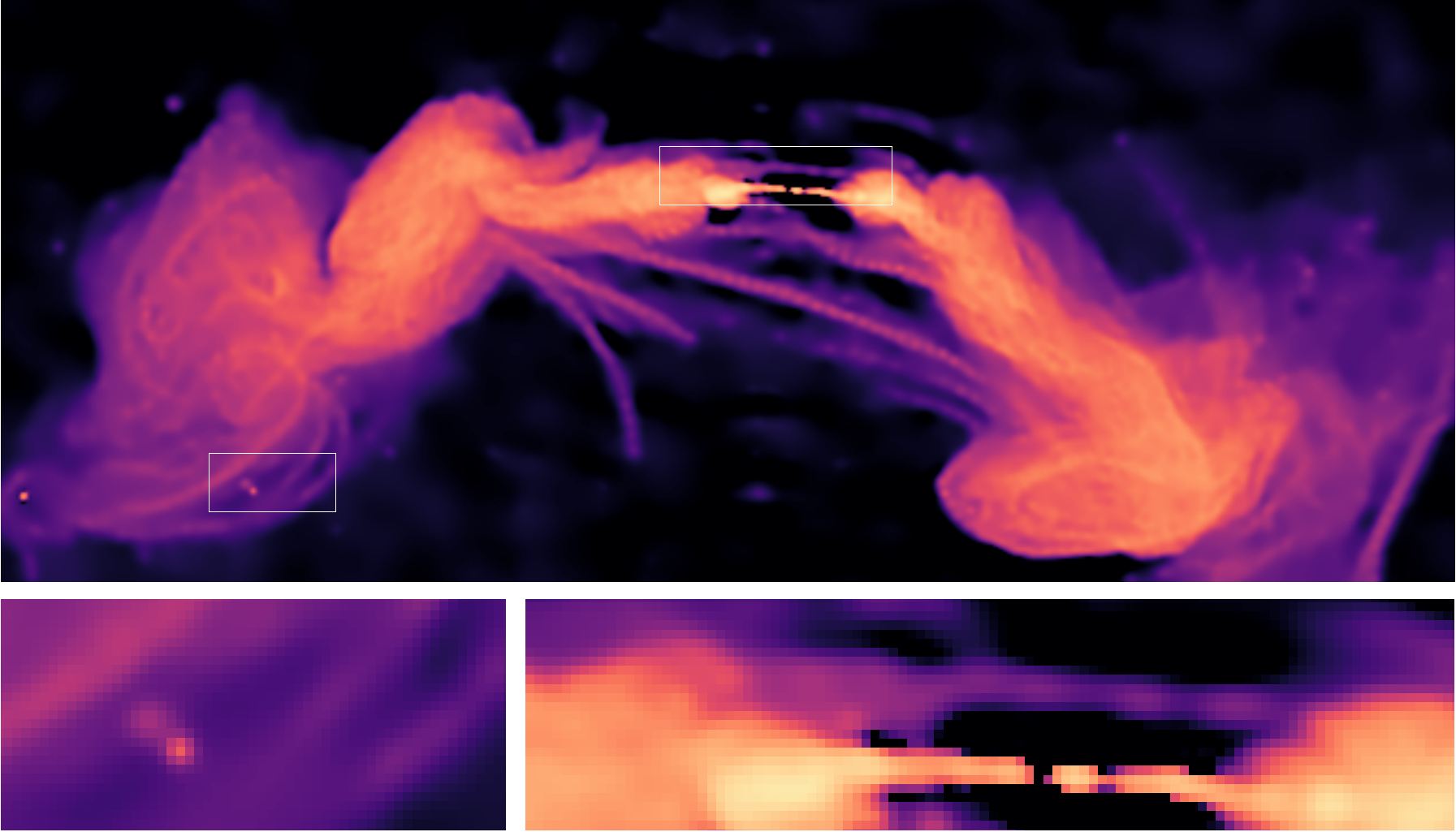} &
\includegraphics[width=0.238\textwidth]{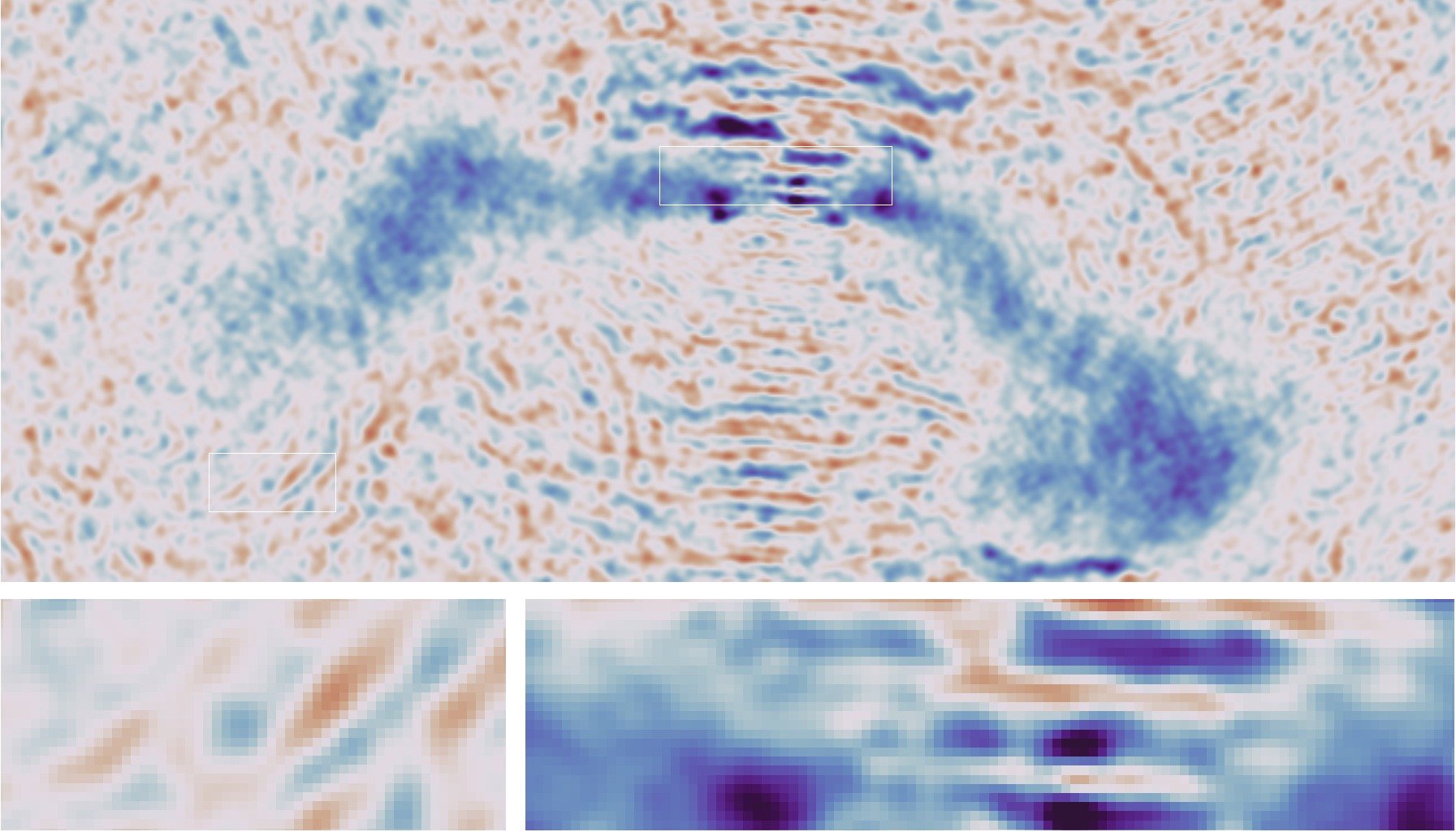} \\
(h) cAIRI\textsubscript{OAID} & (i) cAIRI\textsubscript{OAID} Residual \\
\includegraphics[width=0.238\textwidth]{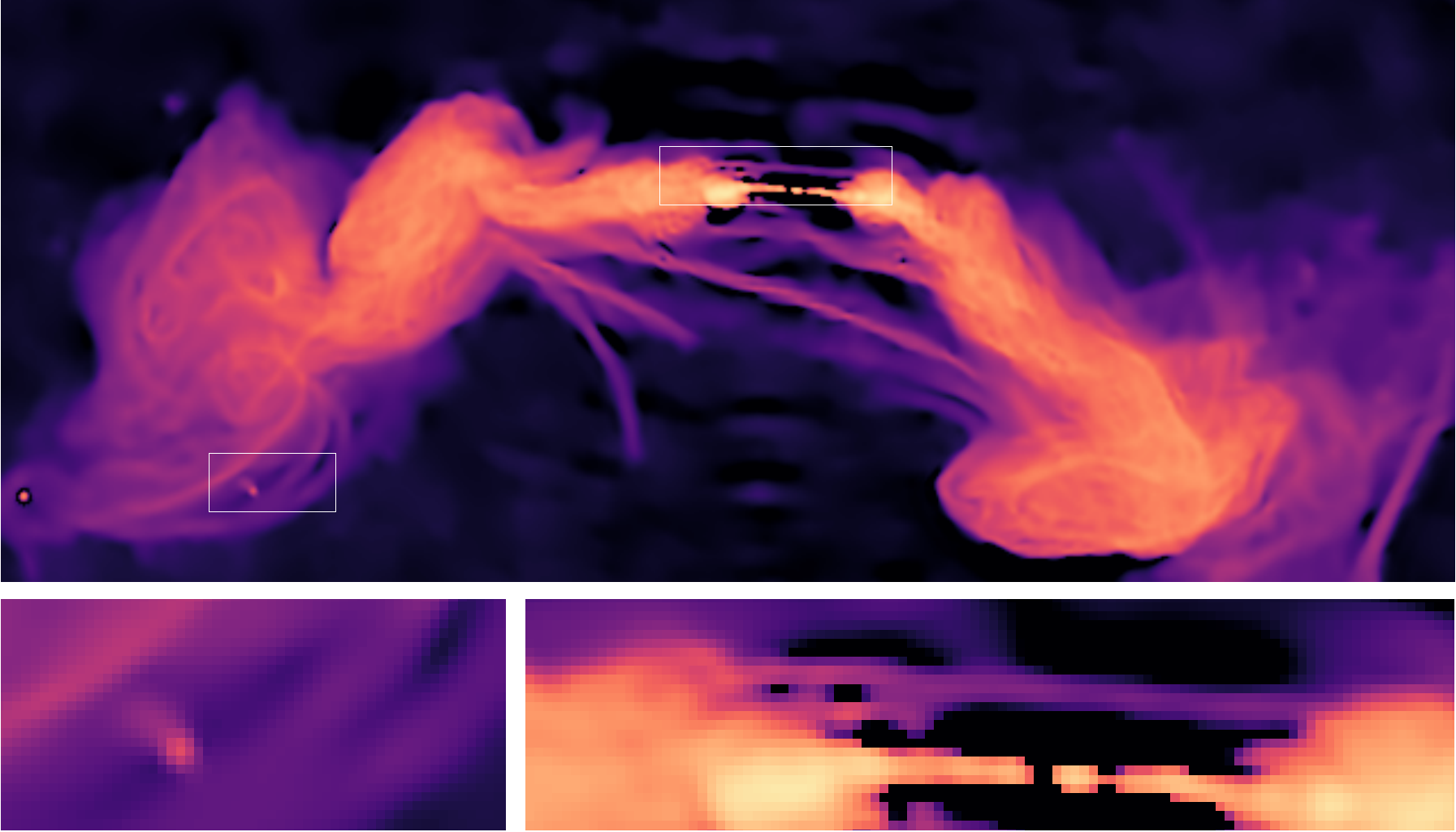} &
\includegraphics[width=0.238\textwidth]{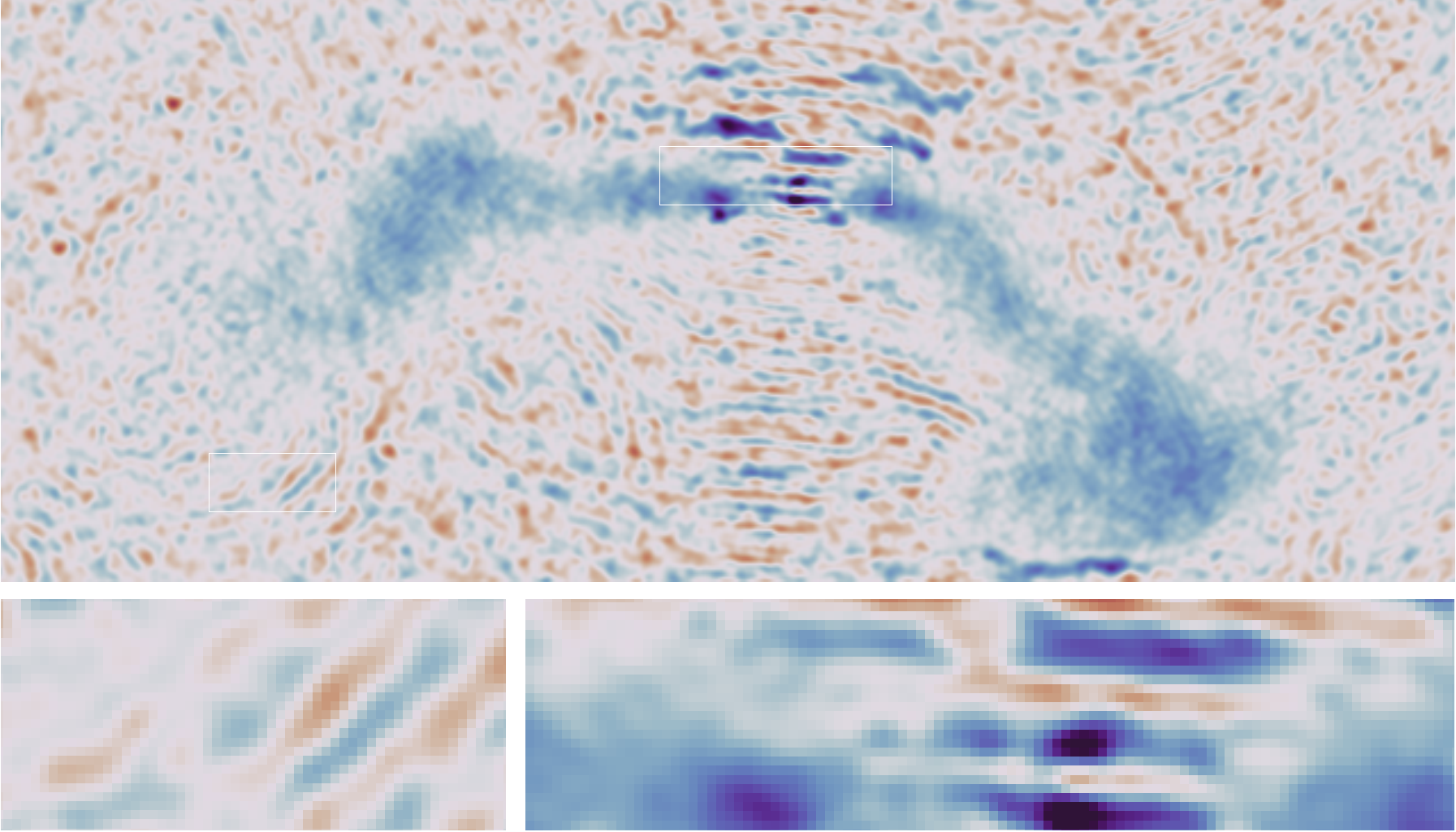} \\
(j) cAIRI\textsubscript{MRID} & (k) cAIRI\textsubscript{MRID} Residual
\end{tabular}
\caption{Imaging of a wide FoV (same as in Fig.~\ref{fig:meerkat_full_FOV}) containing the ESO 137-006 galaxy. Due to the large size (2560$\times$2560), we only display a center-crop of size 640$\times$256 over the ESO 137-006 galaxy. Below each image, we provide zooms on areas of interest corresponding to white rectangles. (a) shows the back-projected dirty image. For the rest panels, reconstructions of various methods are shown in the left column. The corresponding residual dirty images are shown in the right column.}
\label{fig:meerkat}
\end{figure}

\begin{figure} \centering
\begin{tabular}{@{\hspace{0.\tabcolsep}} c @{\hspace{0.15\tabcolsep}} c @{\hspace{0.\tabcolsep}}}
\multicolumn{2}{c}{\hspace{1\tabcolsep} \includegraphics[height=0.03\textwidth]{exp_results/meerkat_results_dr1e4/log_hor.png}} \\
\includegraphics[width=0.238\textwidth]{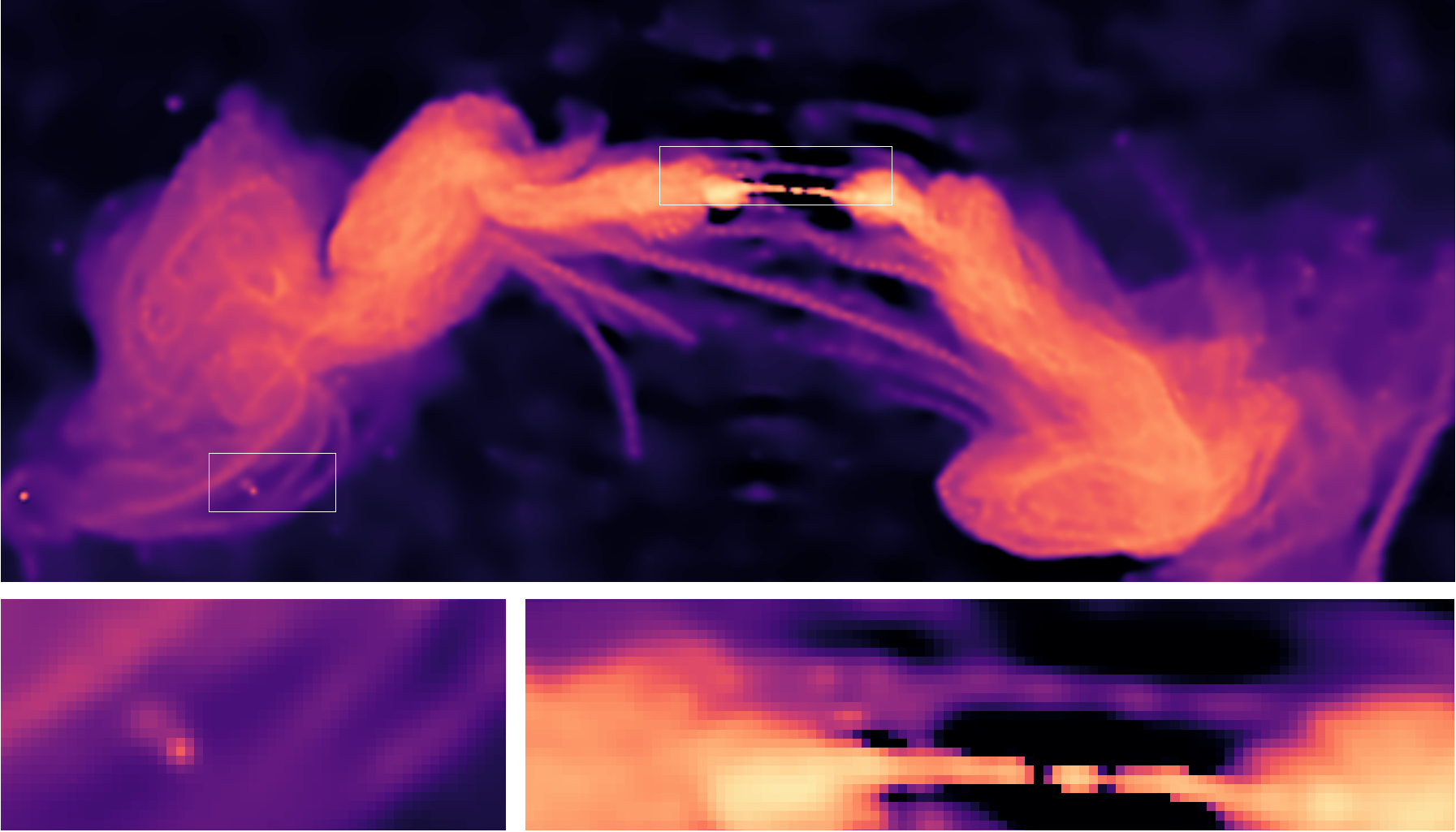} &
\includegraphics[width=0.238\textwidth]{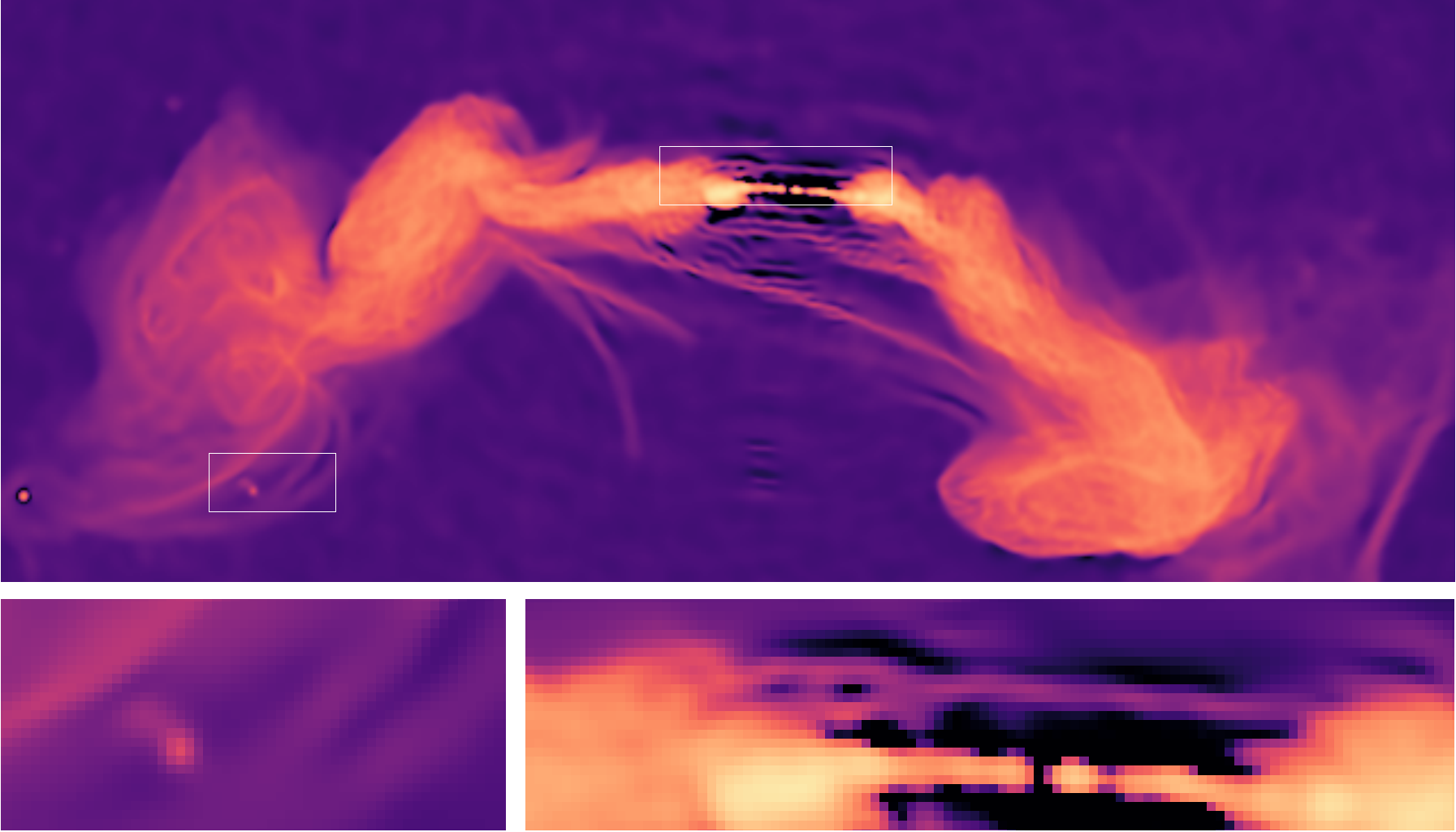} \\
(a) cAIRI\textsubscript{OAID} mean image & (b) cAIRI\textsubscript{MRID} mean image \\
\includegraphics[width=0.238\textwidth]{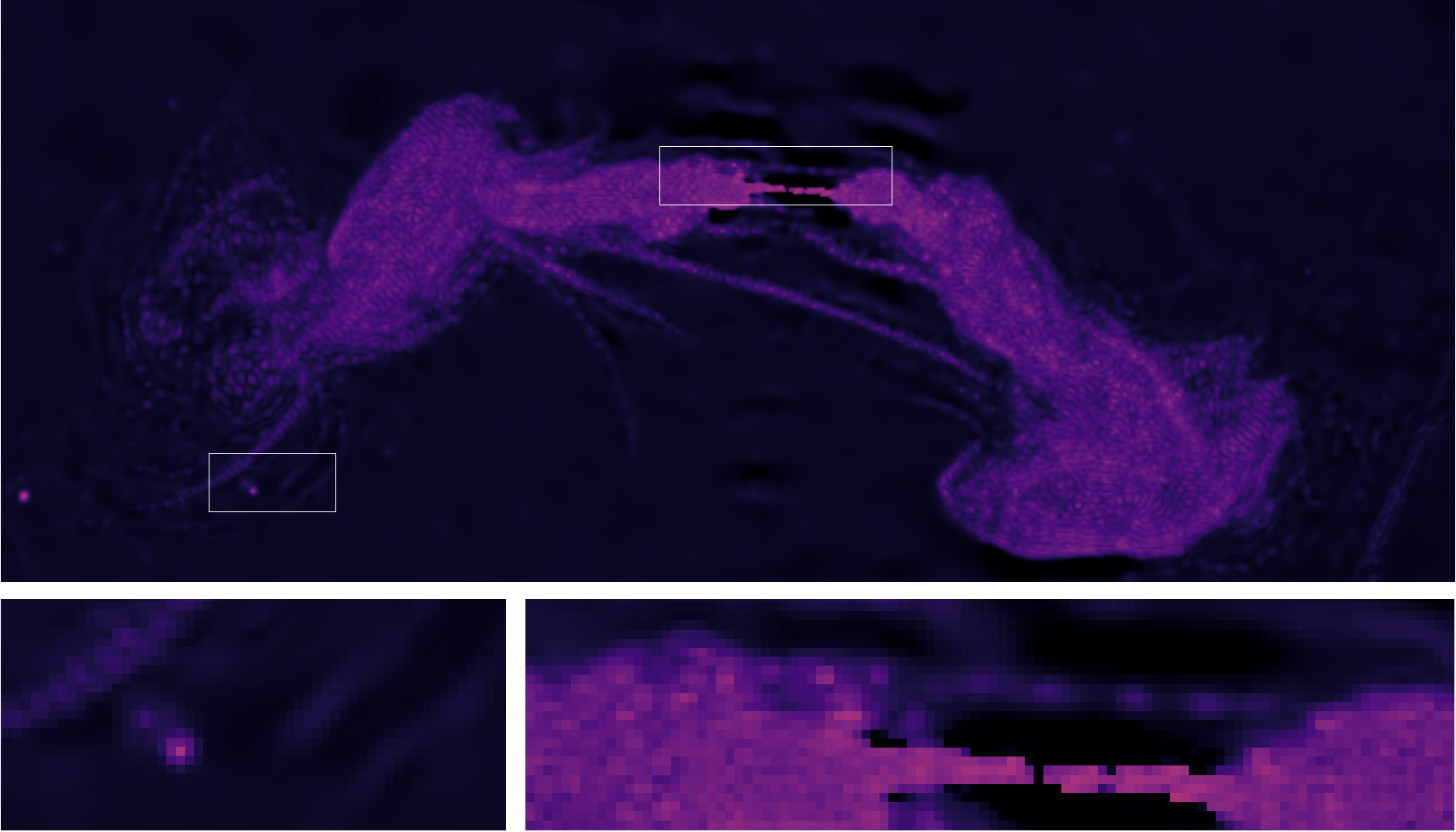} &
\includegraphics[width=0.238\textwidth]{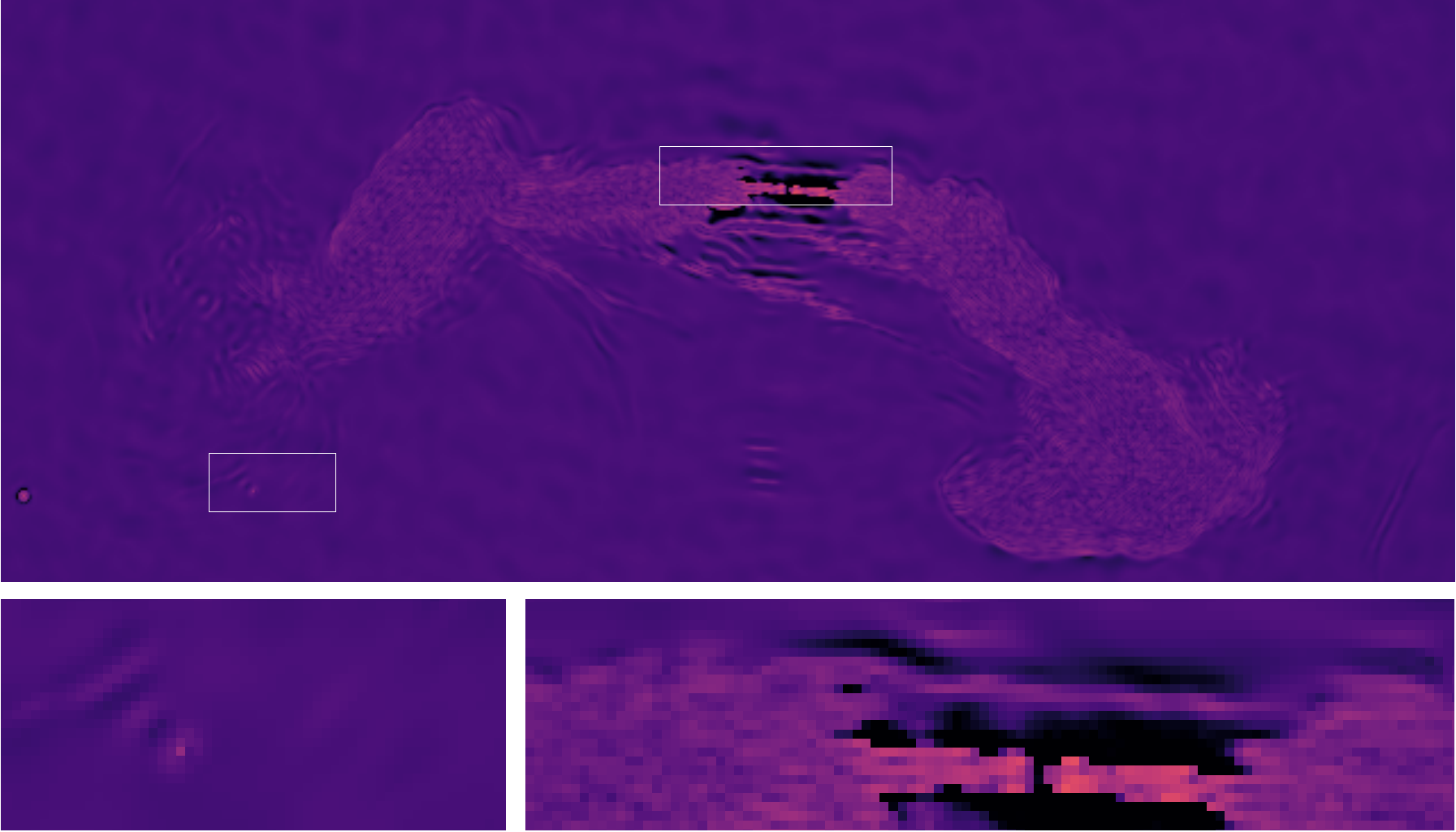} \\
(c) cAIRI\textsubscript{OAID} std. image & (d) cAIRI\textsubscript{MRID} std. image \\
\multicolumn{2}{c}{\includegraphics[height=0.03\textwidth]{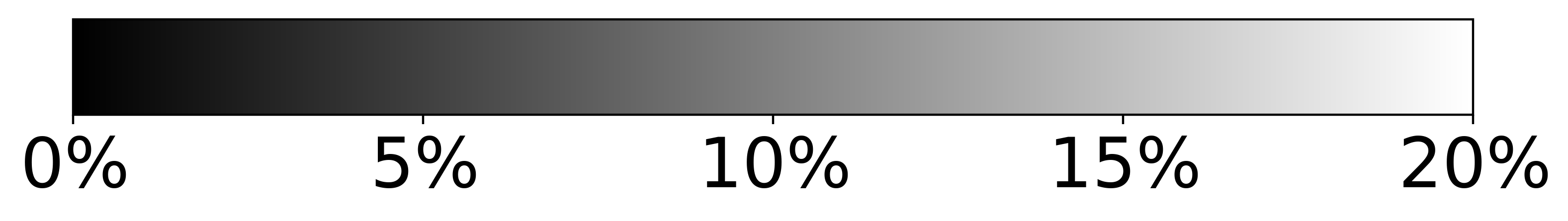}} \\
\includegraphics[width=0.238\textwidth]{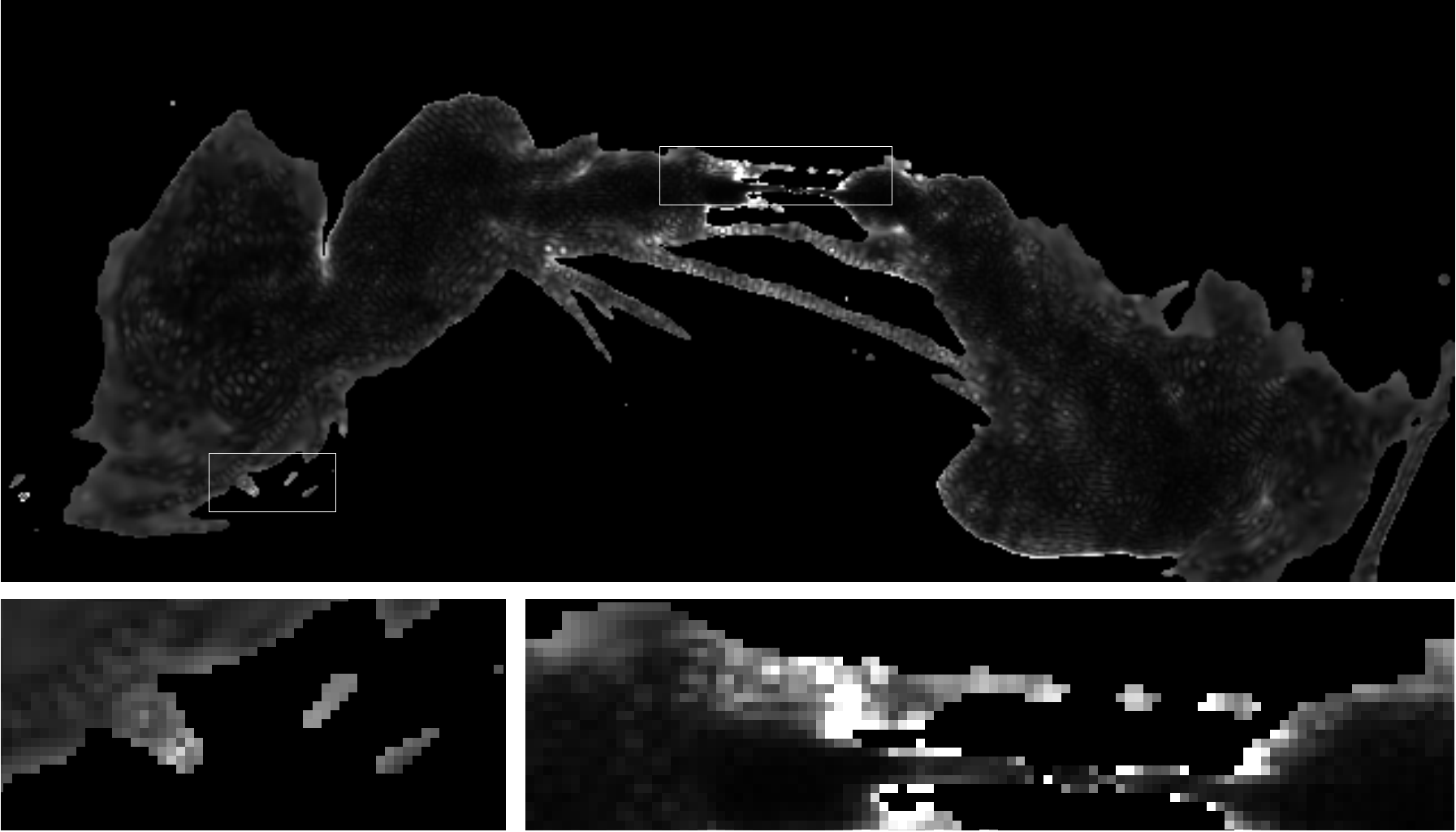} &
\includegraphics[width=0.238\textwidth]{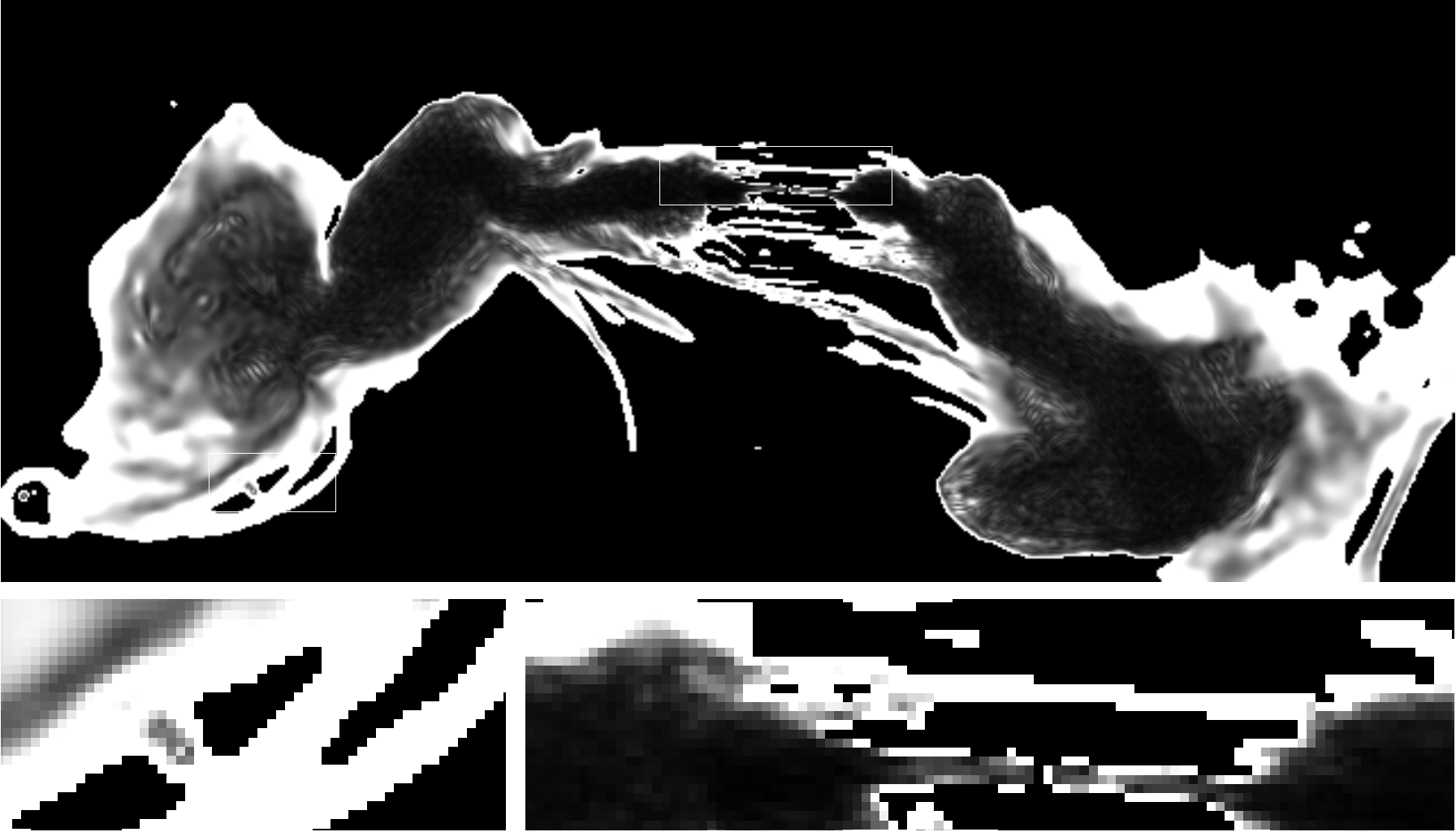} \\
(e) cAIRI\textsubscript{OAID} std/mean & (f) cAIRI\textsubscript{MRID} std/mean
\end{tabular}
\caption{Sensitivity of the MeerKAT real measurement imaging results to the training conditions of the denoiser realisation for cAIRI\textsubscript{OAID} and cAIRI\textsubscript{MRID}. Same as Fig.~\ref{fig:meerkat}, we only display a center-crop of size 640$\times$256 over the ESO 137-006 galaxy. The left column shows the statistical imaging results over 15 different reconstructions of cAIRI\textsubscript{OAID} with denoisers trained with OAID and different random status. The right column shows cAIRI\textsubscript{OAID} results with various denoisers trained with MRID. Each row shows the pixel-wise mean images, standard deviation images and relative standard deviation maps respectively.}
\label{fig:meerkat_robustness}
\end{figure}

\section{Application to real astronomical data} \label{section:real_data}

In this section, we validate our robust PnP approach  
on real observations of the
radio galaxy ESO 137-006, the loudest radio galaxy in the
Norma cluster, acquired with the MeerKAT telescope.
This field of view has recently sparked interest in the radio astronomy community for the observation of collimated synchrotron threads \citep{ramatsoku2020collimated, dabbech2022first}. 
We stress that we rely on the same shelf of denoisers that was used in the experiments on simulated data, meaning that, in line with the PnP motivation, no denoisers had to be trained specifically for the experiments in this section.

\subsection{Data acquisition}
The considered measurements correspond to a subset of the data imaged in \citet{ramatsoku2020collimated} and \citet{dabbech2022first} acquired by MeerKAT.
More specifically, we selected observations at  10 consecutive frequency channels, spanning the frequency range 963 to 971 MHz for the formation a monochromatic intensity image.
The measurement is self-calibrated for phase using the WSClean imager \citep{offringa2017optimized} and the CubiCal calibration suite \citep{kenyon2018cubical}. Full details are provided in \citet{ramatsoku2020collimated}.
We maintain the same image pixel size as in \citet{dabbech2022first} (1.68 arcsec per pixel) corresponding to a super-resolution factor around 2, but with an image size of $2560 \times 2560$ and a FoV of $1.19 \times 1.19$ square degrees.
These imaging considerations minimizes the computational load and reduces the w-effect, and is also motivated by the limited presence of objects of interest outside this window.

\subsection{Imaging results}

We show the full imaged FoV in Fig.~\ref{fig:meerkat_full_FOV} for both the baseline MS-CLEAN algorithm\footnote{The full command is: \texttt{wsclean  -size 2560 2560 -scale 1.68asec  -channel-range 3 12 -channels-out 1 -mem 80 -weight briggs 0 -super-weight 1.0 -weighting-rank-filter-size 16 -taper-gaussian 0  -grid-mode kb -kernel-size 7 -oversampling 63 -pol I -intervals-out 1 -auto-threshold 0.5 -auto-mask 2 -gain 0.1 -mgain 0.9 -multiscale -multiscale-scale-bias 0.6 -niter 5000000.}} and cAIRI\textsubscript{OAID}. More reconstruction results are displayed in Fig.~\ref{fig:meerkat}, focusing on the radio galaxy ESO 137-006.
Due to the visual similarities of the reconstructions between AIRI and cAIRI with same set of denoisers, only cAIRI results are illustrated.
The maximum intensities of the reconstructions given by various methods are around $0.05$ Jy/pixel with heuristic noise level around $7.6\times10^{-6}$, which yields a dynamic range of approximately $10^4$. Therefore, we set $x_{\text{max}}=0.05$ and $a=10^4$ in $\operatorname{rlog}_a(\cdot)$ for visualization.
Overall, the imaging quality for all algorithms aligns with the observations from experiments using simulated data (see Section~\ref{section:synthetic_resulst}). Both the pure optimization algorithms (SARA and uSARA) and the PnP algorithms significantly outperform MS-CLEAN. 
However, SARA exhibits significant wavelet artefacts, which could be explained by calibration errors in estimating $\mathrm{\bm{\Phi}}$ and inadequate estimation of the noise standard deviation for $e$ in \eqref{eq:inv_pb_gen}, on which SARA's hyperparameters rely \citep{thouvenin2023parallel}.

The artefacts induced by residual calibration errors can be seen in all algorithms, presenting as patterns distributed on arches and some filaments around the core of the galaxy.
Both cAIRI\textsubscript{OAID} and cAIRI\textsubscript{MRID} produce images with fewer artefacts than SARA and better resolution (see bottom-left zoom). 
We however underline that both reconstructions show algorithm specific artefacts: the cAIRI\textsubscript{OAID} yields reconstructions with dotted artefacts (see the filaments in Fig.~\ref{fig:meerkat}~(h)), while the cAIRI\textsubscript{MRID} yields light ringing artefacts in the reconstruction (see the filaments in Fig.~\ref{fig:meerkat}~(j)). 
The reconstruction of MS-CLEAN is smoother than others due to the convolution with the dirty beam, though it gives fainter residual image (see panel (c)).
Meanwhile, residuals in panels (e), (g), (c) and (k) of Fig.~\ref{fig:meerkat} suggest that the reconstructions with uSARA, SARA, cAIRI\textsubscript{OAID} and cAIRI\textsubscript{MRID} show less accurate data-fidelity constraints. 
However, as was observed on simulated data, lower residuals do not necessarily imply more accurate reconstructions (see Fig.~\ref{fig:3c353_dt8}), especially with the presence of calibration errors.

As in Section~\ref{subsection:robustness}, we run cAIRI with 15 different denoisers and investigate the statistics of the reconstructed images.
The results on denoisers trained on OAID show that the mean Fig.~\ref{fig:meerkat_robustness} (a) is consistent with a single sample Fig.~\ref{fig:meerkat} (h), and that only very mild variations occur between the different solutions to the problem, as suggested by the empirical standard deviation image Fig.~\ref{fig:meerkat_robustness} (c). Furthermore, the small ratio between the standard deviation and the mean Fig.~\ref{fig:meerkat_robustness} (e) confirms that the variations in the reconstructions related to the model uncertainty are small compared to the image itself.

We observe different results for denoisers trained on MRID in Fig.~\ref{fig:meerkat_robustness} (b), (d) and (f). The mean image Fig.~\ref{fig:meerkat_robustness} (b) shows a different background value than on the sample Fig.~\ref{fig:meerkat} (j), along with stronger localized artefacts. The standard deviation map Fig.~\ref{fig:meerkat_robustness} (d) suggests that different denoiser realisations can yield highly different background values in the reconstruction. The ratio map Fig.~\ref{fig:meerkat_robustness} (f) suggests that these variations dominate the signal at very faint intensities. 

This study suggests that models trained on the proposed OAID dataset are more robust than those trained on the MRID dataset. In practice, we observed that some of the networks trained on the MRID dataset showed unstable results when plugged in the PnP algorithm for real imaging, despite the non-expansive constraint of the denoisers.
We did not observe such phenomenon in the simulated experiments, and presume this might result from calibration errors in the measurement operator.

\section{Conclusion}

We have introduced variations of the AIRI PnP algorithm, towards a more general and robust PnP paradigm for RI imaging. Firstly, we have shown that the AIRI denoisers can be plugged without any alteration into a PDFB optimisation backbone, leading to cAIRI, PnP variant to the constrained SARA optimisation algorithm itself. This extends the remit of the AIRI paradigm, with the original AIRI and the new cAIRI respectively representing the PnP counterparts to uSARA and SARA. Results from simulated and real data confirm that cAIRI performs better than AIRI, while each respectively improved on their pure optimisation counterparts. All algorithms deliver better reconstruction quality than MS-CLEAN. We interpret the reported higher performance of the constrained data-fidelity approach in cAIRI over AIRI as being related to the characteristic high-dynamic ranges of interest in RI imaging, with some high intensity point sources better resolved with constrained algorithms.

Secondly, we have studied the robustness of the AIRI paradigm to strong variations in the nature of the training dataset, with denoisers trained on MRI-born images (MRID) yielding similar reconstruction quality to those trained on optical astronomy images (OAID). The capability to adapt the dynamic range of the dataset to a specific target dynamic range of observation via bespoke exponentiation procedures enables training a shelf of denoisers tailored to high dynamic ranges. In this context, we have proposed to refine the estimated target dynamic range of observation at each iteration during reconstruction, and switch denoisers accordingly on the fly. This adaptive setup was shown to improve and stabilise the reconstruction process. 

Thirdly, we analyzed the epistemic uncertainty quantification method in PnP algorithms. Our experiments show that AIRI denoisers, despite variations in parameter initialization, noise realization, and patch selection, consistently yield similar results when integrated into AIRI or cAIRI algorithms. This ensemble of solutions provides an epistemic uncertainty metric for the AIRI paradigm. It is worth pointing out that, while results show minimal relative standard deviations in reconstructed pixel intensities for OAID denoisers, higher values are reported with MRID denoisers, suggesting greater imaging robustness with the former over the latter, in particular in larger-dimensional image setting and in the presence of calibration errors.

Finally, future research directions include developing wideband and calibration functionality for the AIRI paradigm. 

\section*{Data Availability}

AIRI and cAIRI codes are available alongside the SARA codes in the \href{https://basp-group.github.io/BASPLib/}{BASPLib} code library on GitHub. BASPLib is developed and maintained by the Biomedical and Astronomical Signal Processing Laboratory (\href{https://basp.site.hw.ac.uk/}{BASP}). The denoisers trained with OAID and MRID are available on the \href{https://doi.org/10.17861/aa1f43ee-2950-4fce-9140-5ace995893b0}{Heriot-Watt University research portal}.

Images used to generate training, validation, and testing datasets are sourced as follows. Optical astronomy images are gathered from NOIRLab/NSF/AURA/H.Schweiker/WIYN/T.A.Rector (University of Alaska Anchorage). Medical images are obtained from the NYU fastMRI Initiative database \citep{zbontar2018fastmri}.

The utilised real data are observations with the MeerKAT telescope (Project ID SCI-20190418-SM-01).

\section*{Acknowledgments}
The authors thank O. Smirnov for providing the real observations with MeerKAT and A. Dabbech for helpful discussions. This work was supported by the Engineering and Physical Sciences Research Council (EPSRC), with work funded under the EPSRC grants EP/T028270/1 and EP/T028351/1, and the STFC grant ST/W000970/1. The research used Cirrus, a UK National Tier-2 HPC Service at EPCC funded by the University of Edinburgh and EPSRC (EP/P020267/1). The MeerKAT telescope is operated by SARAO, which is a facility of DST/NRF.



\bibliographystyle{mnras}
\bibliography{references_mnras} 




\bsp	
\label{lastpage}
\end{document}